\documentclass[a4paper,journal]{IEEEtran}
\IEEEoverridecommandlockouts

\usepackage{orcidlink}
\usepackage{cite}
\usepackage{bm}
\usepackage{amsmath,amssymb,amsfonts}
\usepackage{algorithmic}
\usepackage{graphicx}
\usepackage[compatibility=false]{caption}
\usepackage{subcaption}
\usepackage{textcomp}
\usepackage[table]{xcolor}
\usepackage{xcolor}
\usepackage{float}
\usepackage{multirow}
\usepackage{diagbox}
\usepackage{makecell}
\usepackage{mathrsfs}
\usepackage{MnSymbol}
\usepackage{stfloats}
\usepackage{hyperref}
\hypersetup{hidelinks}
\hypersetup{
    colorlinks=true,            
    linkcolor=red,              
    citecolor=blue,             
}
\usepackage{stfloats} 

\definecolor{Table_Pink}{RGB}{244, 177, 131} 
\definecolor{Table_Green}{RGB}{169, 209, 142} 

\def\BibTeX{{\rm B\kern-.05em{\sc i\kern-.025em b}\kern-.08em
    T\kern-.1667em\lower.7ex\hbox{E}\kern-.125emX}}

\usepackage{soul,color} 
\definecolor{myColor}{rgb}{0,0,0} 
\makeatletter
\newcommand*{\revise}{\@ifnextchar\bgroup{\revise@}{\color{myColor}}}
\newcommand*{\revise@}[1]{{\textcolor{myColor}{#1}}}
\makeatother
    
\begin{document}

\title{On the Impact of Phase Errors in Phase-Dependent Amplitudes of Near-Field RISs}

\author{\IEEEauthorblockN{Ke Wang$^{1,2}$, Chan-Tong Lam$^{2}$, and Benjamin K. Ng$^{2}$}
\IEEEauthorblockA{$^{1}$\textit{College of Information Engineering, Jiangmen Polytechnic, Jiangmen, China}\\
$^{2}$\textit{Faculty of Applied Sciences, Macao Polytechnic University, Macao S.A.R, China}\\
Emails: \{ke.wang; ctlam; bng\}@mpu.edu.mo}
}

\author{Ke~Wang\orcidlink{0000-0002-9726-1631},~\IEEEmembership{Member,~IEEE,}
        Chan-Tong~Lam\orcidlink{0000-0002-8022-7744},~\IEEEmembership{Senior Member,~IEEE,}
        ~Benjamin K.~Ng\orcidlink{0000-0001-5901-5694},~\IEEEmembership{Senior Member,~IEEE,}
        and~Yue~Liu\orcidlink{0000-0002-3292-4211},~\IEEEmembership{Member,~IEEE}
        \thanks{Manuscript received 25 June, 2025; revised 1 Nov, 2025; accepted 18 December, 2025. An earlier version of this paper was presented in part at \textit{2024 IEEE 100th Vehicular Technology Conference (VTC2024-Fall)} \cite{ken24vtcF}. This work was supported by the research funding of the Macao Polytechnic University, Macao SAR, China (Project No. RP/FCA-04/2025). Copyright (c) 2025 IEEE. Personal use of this material is permitted. However, permission to use this material for any other purposes must be obtained from the IEEE by sending a request to pubs-permissions@ieee.org.}
\thanks{Ke~Wang, Chan-Tong Lam, Benjamin K. Ng, and Yue Liu are all with the Faculty of Applied Sciences, Macao Polytechnic University, Macao SAR, 999078, China, e-mail: (kewang, ctlam, bng, yue.liu)@mpu.edu.mo.}
}

\markboth{Accepted for publication in IEEE Transactions on Vehicular Technology, 2025. DOI: 10.1109/TVT.2025.3647594.}{\MakeLowercase}

\maketitle

\begin{abstract}
\revise{This paper investigates mutual coupling between phase-dependent amplitudes (PDAs) and designed phase shifts within pixels of near-field (NF) reconfigurable intelligent surfaces (RISs) in the presence of phase errors (PEs). In contrast to existing research that treats phase shifts with errors (PSEs) and the PDAs separately, we introduce a remaining power (RP) metric to quantify the proportion of power preserved in the signals reflected by the RIS, and we prove its asymptotic convergence to theoretical values by leveraging \textit{extended Glivenko–Cantelli theorem}. Then, the RP of signals passing through RIS pixels is jointly examined under combined phase and amplitude uncertainties. In addition, we propose four pixel reflection models to capture practical conditions, and we derive approximate polynomial upper bounds for the RP with error terms by applying \textit{Taylor expansion}. Furthermore, based on \textit{Friis transmission formula} and \textit{projected aperture}, we propose a general NF channel model that incorporates the coupling between the PSEs and the PDAs. By using \textit{Cauchy–Bunyakovsky–Schwarz inequality} and \textit{Riemann sums}, we derive a closed-form upper bound on spectral efficiency, and the bound becomes tighter as the pixel area decreases. We reveal that as the RIS phase shifts approach the ends of their range, the RP under independent and identically distributed PEs is smaller than that under fully correlated PEs, whereas this relationship reverses when the phase shifts are near the middle of their range. Neglecting the PEs in the PDAs leads to an overestimation of the RIS performance gain, explaining the discrepancies between theoretical and measured results.}
\end{abstract}


\section{Introduction}

\subsection{Background and Related Works}

Reconfigurable intelligent surfaces (RISs) hold significant potential to revolutionize wireless communication technologies\cite{emil2020ris_cm}. Traditional investigations on the RIS have predominantly concentrated on far-field scenarios\cite{wu2019japb, huang2019ee}, which are suitable for long transmission distances. Recently, however, the RIS deployed in near-field (NF) environments has exhibited distinct advantages over alternative transmission enhancement methodologies\cite{liu23near,feng2024near,lu2024tutorial,lu2021near}. Given its non-uniform spherical wave and spatial non-stationarity features\cite{lu2024tutorial}, the NF RIS further highlights the propagation properties of electromagnetic waves by increasing the total reflective area and placing the communication transceivers in \textit{the Fresnel region}\cite{lu2021near}, thereby improving spectral efficiency (SE), spatial resolution, and degrees of freedom for transmission\cite{emil23elaa}.

It should be noted that we cannot consider every pixel of the NF RIS as identical, since each pixel possesses its unique transmission angle and distance\cite{liu23near,feng2024near,emil23elaa,lu2021near,lu2024tutorial}. Consequently, the hardware characteristics of each NF RIS pixel must be individually taken into account. Moreover, in practical scenarios, \textit{phase errors (PEs)} within the RIS cannot be overlooked\cite{badiu20errors}. Previous studies \cite{badiu20errors,zhou2020spectral,xing2021achievable, lu2024performance,sun2021diagnosis, taghvaee2020error, yang2023practical} primarily focused on \textit{phase shift with errors (PSEs)}, represented as $\exp(-\jmath(\phi+\Delta))$\footnote{Given that the transceiver of the system model in this paper has a single antenna, thus $\phi$ and the phase of the cascaded channel cancel each other out, and the PSE is defined as $\exp(-\jmath\Delta)$ in the rest of this work.} where $\phi$ denotes \revise{an RIS pixel phase shift}, $\Delta$ represents a PE random variable (RV) that follows specific distributions, $\exp(\cdot)$ is exponential function, and $\jmath\triangleq\sqrt{-1}$. Generally speaking, there are three types of practical PEs in the RIS, i.e., \textit{quantization errors}\cite{badiu20errors,zhou2020spectral, yang2023practical}, \textit{imperfect channel estimations}\cite{lu2024performance,badiu20errors}, and \textit{pixel hardware failures}\cite{sun2021diagnosis,taghvaee2020error, ken23vtcF, yang2023practical}. \revise{In particular, Badiu \textit{et\ al.} \cite{badiu20errors} initially revealed that the quantization and imperfect estimation errors in the RIS are respectively characterized by uniform ($\mathcal{UF}$) and \revise{von Mises ($\mathcal{VM}$) RVs}, and an RIS cascaded path is equivalent to a direct channel with Nakagami scalar fading. Zhou \textit{et\ al.} in \cite{zhou2020spectral} analyzed spectral and energy efficiencies of the RIS with \revise{$\mathcal{UF}$ RVs}, and an important finding is that the SE is constrained regardless of the pixel number becoming exceedingly large. Besides, Yang \textit{et\ al.} in \cite{yang2023practical} further explored the implications of incorporating the quantization error into the RIS system, Lu \textit{et\ al.} in \cite{lu2024performance} conducted a joint analysis of the RIS system's performance, taking into account the synergistic impacts of the estimation errors and channel aging. Moreover, Sun \textit{et\ al.} in \cite{sun2021diagnosis} proposed a method for detecting failed RIS pixels based on compressed sensing, Taghvaee \textit{et\ al.} \cite{taghvaee2020error} examined the reliability problem in the RIS by introducing an error model and a general methodology for error analysis, and Wang \textit{et\ al.} in \cite{ken23vtcF} revealed that hardware aging effects related to operation times of the RIS are significant in practical scenarios.}

Beyond the aforementioned noises exist in \revise{the PSE $\exp(-\jmath\Delta)$}, however, the actual pixel reflection coefficient exhibits its own \textit{phase-dependent amplitudes (PDAs)}\cite{abey20TC,ozturk23twc, mosleh24tvt}, which can be expressed as $\beta(\phi)$ where $\phi$ is phase shift and $\beta(\cdot)\in (0,1]$ refers to a nonlinear function\cite{abey20TC}. Specifically, Abeywickrama \textit{et\ al.} \cite{abey20TC} first introduced the approximated PDA model to mimic the equivalent circuit model of the RIS pixel, and then formulated and addressed a problem of minimizing the total transmission power through a joint optimization of the transmit beamforming vectors and the RIS phase shifts. Ozturk \textit{et\ al.} in \cite{ozturk23twc} investigated a problem of NF localization by utilizing the RIS accounting for the presence of the PDA, and they revealed that ignoring the PDA at the receiver can lead to substantial performance degradation. Moreover, Mosleh \textit{et\ al.} \cite{mosleh24tvt} showed the ergodic capacity limit of the RIS system with the PDA is directly dependent on the PDA and indirectly on the phase shift.

\subsection{Motivations and Contributions}

Given the background and related works mentioned above, the motivation for this paper is summarized in Table \ref{Table_Works}. In particular, although many previous works have explored the impacts of the PE in the PSE on the RIS-aided system\cite{badiu20errors,zhou2020spectral,xing2021achievable, lu2024performance,sun2021diagnosis, taghvaee2020error, yang2023practical, abey20TC,ozturk23twc, mosleh24tvt, ken23vtcF}, there is no research on the mutual coupling of the PSE and the PDA when the PE exists. It should be emphasized that the pixel hardware is increasingly vulnerable as it incorporates sophisticated tuning, control, and sensing systems\cite{taghvaee19fault}. \revise{Thus, from a practical perspective, PEs may also exist in the PDA. To this end, in this paper, we consider three regimes: $(i)$ perfect transfer of the PE from the PSE to the PDA, i.e., $\iota = 1$ in Cases III and IV in Table \ref{Table_Works}; $(ii)$ fully random pixel imperfections that induce i.i.d. noise at the PDA, i.e., $\iota = 0$ in Cases III and IV in Table \ref{Table_Works}; and $(iii)$ intermediate correlation levels with $\iota \in (0, 1)$. In practice, the pixel hardware failure \cite{sun2021diagnosis,taghvaee2020error, ken23vtcF, yang2023practical} progressively shifts the system from $(i)$ toward $(ii)$. Specifically, differential degradation of RIS components (e.g., varactor diodes) desynchronizes the errors between the PSE and the PDA, thereby increasing their \textit{statistical independence}. Consequently, as the RIS behavior approaches an i.i.d. noise model, the correlation coefficient $\iota$ decreases over time and approaches $0$, reflecting the emergence of random, decoupled uncertainties driven by hardware degradations\cite{ken23vtcF}.}

It is noteworthy that the omission of the above issues may be pivotal in accounting for the disparity between theoretical analyses and hardware validations \cite{pei21ristrials,tang19metasurfaces, tang2021wireless, tang2022path}. \revise{Besides, different from the existing pixel reflection models (i.e., Perfect, Cases I and II in Table \ref{Table_Works}), a practical model that explicitly captures $\mathcal{UF}$ and/or $\mathcal{VM}$ uncertainties in the PSE and/or the PDA is essential for future performance analysis and algorithm design in the NF RIS-assisted system. Therefore, the focus of this work is to determine \textit{how the PSE and the PDA, with the PE in particular, jointly affect the RIS system performance.} The contributions are summarized as follows:}

\begin{table*}[htbp]
    \centering
    \begin{tabular}{ccccc}
        \hline
        \rowcolor{Table_Green}Model & Works & Reflection Coefficient \revise{Description} & Approximated Model \\  \hline
        \rowcolor{Table_Pink}Perfect & \cite{liu23near,feng2024near,lu2024tutorial,lu2021near,emil23elaa,emil20howlarge, ozdogan20irsmodel,wu2019japb,huang2019ee,zhang2025ris} & Complete reflection w/o the PDA, the PSE or any PEs & $1\cdot\exp(-\jmath\phi)$ \\
        \rowcolor{Table_Green}Case I & \cite{badiu20errors,zhou2020spectral,xing2021achievable, lu2024performance,sun2021diagnosis, taghvaee2020error, yang2023practical, yang24twc} & Partial reflection w/ the PSE and the PDA is a constant $\beta\in(0,1]$ & $\beta\cdot\exp(-\jmath\Delta)$\\ 
        \rowcolor{Table_Pink}Case II & \cite{abey20TC,mosleh24tvt,ozturk23twc,ken23vtcF} & Partial reflection w/ the PSE and the PDA $\beta(\phi)$ is w/o any PEs & $\beta(\phi)\cdot\exp(-\jmath\Delta)$\\
       \rowcolor{Table_Green}Case III & This paper & Partial reflection w/ the PSE and the PDA $\beta(\phi)$ is w/ an single PE  & $\beta(\phi+\bar{\Delta})\cdot\exp(-\jmath\Delta)$\\
       \rowcolor{Table_Pink}Case IV & This paper &Partial reflection w/ the PSE and the PDA $\beta(\phi)$ is w/ both PEs & $\beta(\phi+\bar{\Delta}+\bar{\Theta})\cdot\exp(-\jmath(\Delta+\Theta))$\\\hline
    \end{tabular}
    \caption{Comparison between this paper and its related works. \textcolor{myColor}{Note $\beta(\cdot)\in(0,1]$ and $\phi$ is the designed pixel phase shift. $\Delta$, $\bar{\Delta}$, $\Theta$ and $\bar{\Theta}$ all denote PE RVs, $\bar{\Delta} = \iota\Delta + \sqrt{1-\iota^2}\breve{\Delta}$, where $\iota\in[0,1]$ and $\breve{\Delta}$ is i.i.d. with $\Delta$. $\Theta$ and $\bar{\Theta}$ are defined analogously.}}
    \label{Table_Works}
\end{table*}

\begin{itemize}
    \item {\revise{\textbf{Hardware and theoretical modeling}. We present the hardware configuration and schematic model of an RIS pixel, and investigate how key circuit parameters affect the PDA model. Closed-form lower and upper bounds for the ideal PDA without the PE are derived.}}
    \item {\revise{\textbf{New metric with asymptotic guarantee}. We introduce a new metric termed the RP to quantify energy conservation after a signal passes through an RIS pixel. By leveraging \textit{extended Glivenko–Cantelli theorem}, we rigorously prove the asymptotic convergence of the RP to its theoretical value, thereby establishing the statistical foundation for subsequent analysis.}}
    \item {\revise{\textbf{Unified reflection models and polynomial bounds}. We develop four practical pixel reflection models that jointly capture phase and amplitude uncertainties. Based on \textit{Taylor expansion}, we derive polynomial lower bounds for the RP with quantified approximation errors. As the RIS phase shifts approach the ends of their range, the RP under i.i.d. PEs becomes smaller than that under identical PEs, whereas this relationship reverses when the phase shifts are near the middle of their range.}}
    \item {\revise{\textbf{New NF channel and SE analysis}. Building on \textit{Friis transmission formula} and \textit{proposed reflection}, we establish a general NF line-of-sight (LoS) channel model incorporating the coupling between the PDA and the PSE. A closed-form SE expression is further derived through \textit{Cauchy–Bunyakovsky–Schwarz (CBS) inequality} and \textit{Riemann sums}}.}
\end{itemize}

\subsection{Outline, Notations, and Reproducible Research}

\revise{The remainder of this paper is organized as follows. Sec. \ref{Sec2} presents the hardware structure of the RIS pixel and its schematic model, analyzes the influence of key circuit parameters on the approximated PDA model, and derives closed-form lower and upper bounds for the ideal PDA without PEs. Sec. \ref{Sec3} introduces a unified RP-based framework and four practical pixel reflection models, along with their polynomial bounds that incorporate error terms under phase and amplitude uncertainties. Sec. \ref{Sec4} develops a general NF LoS channel model that captures the coupling between the PDA and the PSE, and provides a closed-form analysis of SE. Extensive simulation results and performance discussions are presented in Sec. \ref{Sec5}. Finally, Sec. \ref{Sec6} concludes the paper and outlines potential directions for future research.}

Notations of this paper are as follows: $\vert \cdot \vert$ denotes absolute value, $\Vert\cdot\Vert$ is $l_2$ norm, $[\cdot]^\mathsf{T}$ is transpose operation, $(p\ \mathsf{mod} \ q)$ is the remainder of the division of $p$ by $q$. Besides, $\sin(\cdot)$, $\cos(\cdot)$, $\arctan(\cdot)$, and $\arccos(\cdot)$ are respectively sine, cosine, inverse tangent, and inverse cosine functions. Moreover, $\mathcal{CN}$, $\mathcal{UF}$, $\mathcal{VM}$, \revise{$\mathbb{E}\{\cdot\}$}, $\mathsf{Im}(\cdot)$, and $\mathsf{Re}(\cdot)$ represent complex Gaussian distribution, uniform distribution, \revise{von Mises} distribution, expectation function, imaginary and real parts of a complex number, respectively. \revise{Furthermore, $\mathsf{Cov}(\cdot,\cdot)$ and $\mathsf{Var}(\cdot)$ denote covariance and variance operators, respectively.} For reproducible research, the simulation code is available at https://github.com/ken0225/On-Impact-PEs-PDAs-NF-RISs.

\section{Phase-Dependent Amplitudes in RIS Pixels}\label{Sec2}

In this section, we start by elucidating the hardware configuration of the RIS pixel. Henceforth, an analysis of the approximated PDA is proposed, especially for lower and upper bounds of the PDA in the absence of the PE. Note that we focus on the NF scenario; thus, the hardware of each pixel and its own cascaded channel should be analyzed separately.

\subsection{Hardware Structure of RIS Pixels}

\begin{figure}[htbp]
\centerline{\includegraphics[width=6cm]{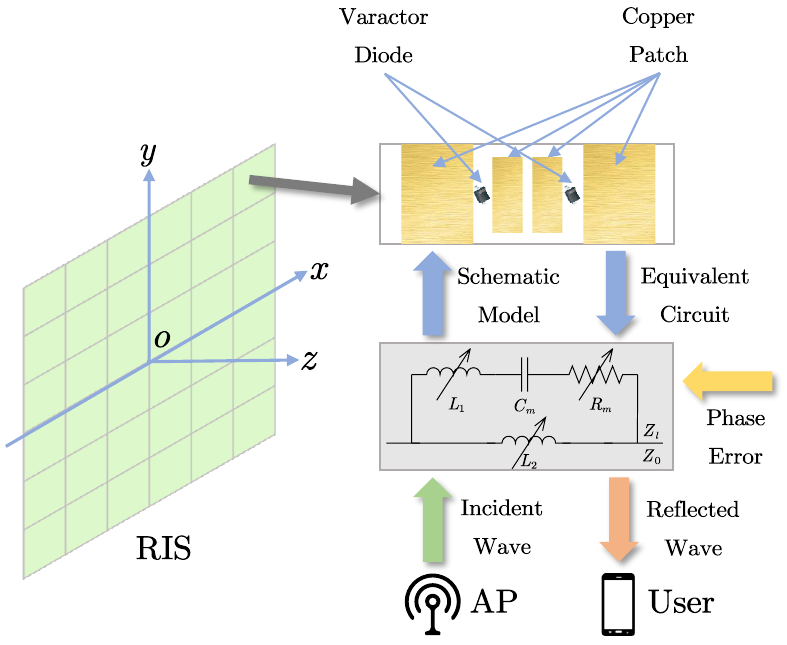}}
\caption{Top view and equivalent circuit of the RIS pixel.}
\label{Fig1_RIS_layout}
\end{figure}

A typical RIS pixel hardware contains three layers \cite{pei21ristrials,tang19metasurfaces}. As shown in Fig. \ref{Fig1_RIS_layout}, the top layer contains two pairs of copper patches, each of which is connected by a varactor diode. The middle layer is a complete metallic panel, for reflecting incoming waves and preventing energy loss, and the bottom layer consists of direct current biasing lines. A single pixel can be modeled as a parallel resonant circuit\cite{tang19metasurfaces}. Thus, without loss of generality, if an RIS has $M$ pixels, for the $m$-th pixel where $m=1,\dots, M$, we have its equivalent impedance as \cite{tang19metasurfaces,abey20TC}

\begin{equation}
    Z_m(C_m, R_m) = \frac{\jmath2\pi f_c L_1(\jmath2\pi f_c L_2 +\frac{1}{\jmath2\pi f_c C_m}+R_m)}{\jmath2\pi f_c L_1+(\jmath2\pi f_c L_2 +\frac{1}{\jmath2\pi f_c C_m}+R_m)}, 
    \label{Z_m(C_m, R_m)}
\end{equation}

\noindent where $L_1$, $L_2$, $C_m$, $R_m$, and $f_c$ are bottom layer inductance, top layer inductance, the $m$-th effective capacitance, the $m$-th effective resistance, and carrier frequency, respectively. If $Z_0$ denotes free space impedance, then the reflection coefficient of the $m$-th pixel can be obtained as

\begin{equation}
  \varsigma_m = \frac{Z_m(C_m, R_m)-Z_0}{Z_m(C_m, R_m)+Z_0}.
  \label{varsigma_m}
\end{equation}

\noindent Consequently, the final tuned phase of the $m$-th reflection coefficient $\varsigma_m$ is 

\begin{equation}\label{phi_m-arctan}
    \phi_m = \arctan \left(\frac{\mathsf{Im}(\varsigma_m)}{\mathsf{Re}(\varsigma_m)}\right).
\end{equation}

It should be emphasized that one single RIS pixel is not a beamformer but a scatterer \cite{ozdogan20irsmodel}. Therefore, a large number of pixels are arranged periodically on the top layer, and the junction capacitance of the varactor diode in each pixel is controlled by the biasing voltage. Consequently, the RIS with a total of $M$ pixels is able to reshape the incident wave\footnote{A large-area metasurface (i.e., a large number of pixels) is needed to outperform the traditional transmission enhancement techniques such as relaying\cite{emil20howlarge}. Hence $M\gg 1$.}, and each pixel has \textit{its own unique reflection coefficient.}

\subsection{Approximated Phase-Dependent Amplitudes}

\begin{figure}[htbp]
\centerline{\includegraphics[width=6cm]{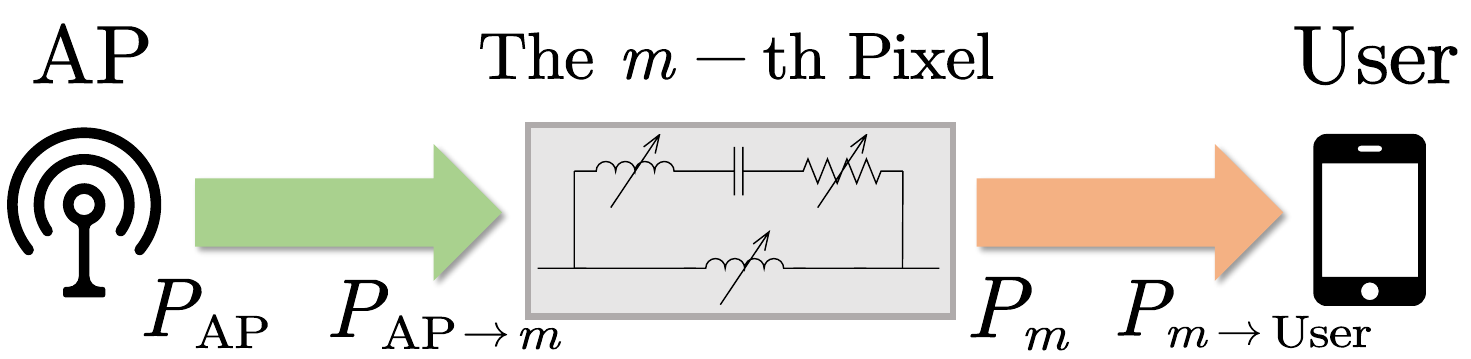}}
\caption{A downlink cascaded path of the $m$-th pixel.}
\label{Fig_Cascaded_Path}
\end{figure}

Fig. \ref{Fig_Cascaded_Path} illustrates the $m$-th downlink cascaded path of the RIS-aided communication. Note that $P_{\mathrm{AP}}$, $P_{\mathrm{AP}\rightarrow m}$, $P_m$, and $P_{m\rightarrow\mathrm{User}}$ respectively denote the transmit power by an access point (AP), the received power by the $m$-th pixel, the reflected power by the $m$-th pixel, and the received power by a user. Then we have
\revise{
\begin{equation}
    \underbrace{P_{\mathrm{AP}} \overset{(i)}{>} P_{\mathrm{AP}\rightarrow m}}_{\mathrm{Fading\ Channel}} 
    \overbrace{\underset{(ii)}{\ge}}^{\mathrm{RIS}}
    \underbrace{P_m \overset{(iii)}{>} P_{m\rightarrow\mathrm{User}}}_{\mathrm{Fading\ Channel}},
\end{equation}}

\noindent where $(i)$ and $(iii)$ are caused by large/small-scale fading and non-isotropic pixels\cite{liu23near}. But for $(ii)$, the main reasons are the PSE, the PDA and the PE\cite{badiu20errors,abey20TC}, and $(ii)$ is an equal sign when there is no PE and the PDA is always $1$. To characterize reflection amplitudes more accurately, for the $m$-th pixel, we derive an equivalent low-pass expression for the receiver power without error as

\begin{equation}\label{P_m_P_AP_m}
    P_m = \left| \beta(\phi_m)\exp(-\jmath\phi_m)\right|^2 P_{\mathrm{AP}\rightarrow m},
\end{equation}

\noindent where\footnote{Individual differences in the pixel hardware are ignored in this paper, thus $\beta_1(\cdot)=\beta_2(\cdot)=\cdots=\beta_M(\cdot)=\beta(\cdot)$.} $\beta(\phi_m)$ is the $m$-th PDA of the pixel and $\phi_m$ is the desired phase defined in \eqref{phi_m-arctan}.

\begin{figure}[htbp]
	\centering
	\begin{subfigure}{0.24\textwidth}
		\centering
		\includegraphics[width=\textwidth]{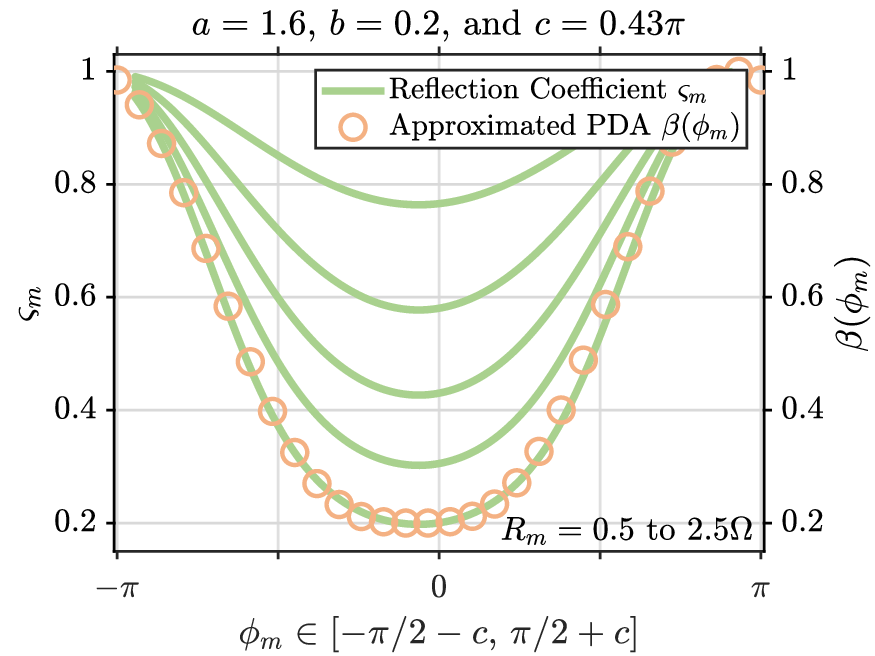}
		\caption{$\varsigma_m$ in \eqref{varsigma_m} and $\beta(\phi_m)$ in \eqref{beta(phi)_Approx_PDA}.}
        \label{Fig_Approx_PDA}
	\end{subfigure}
	\centering
	\begin{subfigure}{0.24\textwidth}
		\centering
		\includegraphics[width=\textwidth]{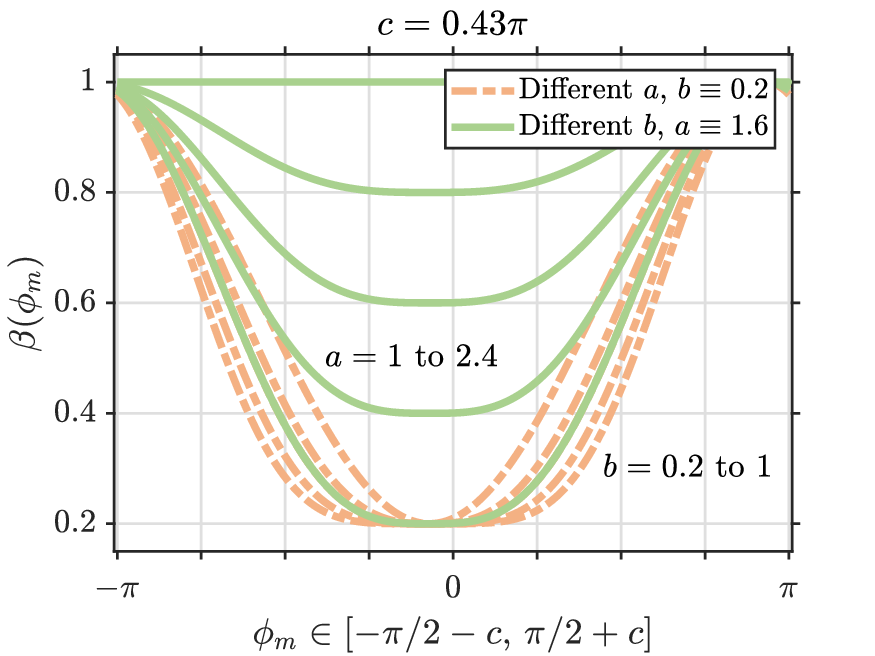}
		\caption{Different $b$ and $a$.}
		\label{Fig_Diff_b_a}
	\end{subfigure}
    \centering
	\begin{subfigure}{0.24\textwidth}
		\centering
		\includegraphics[width=\textwidth]{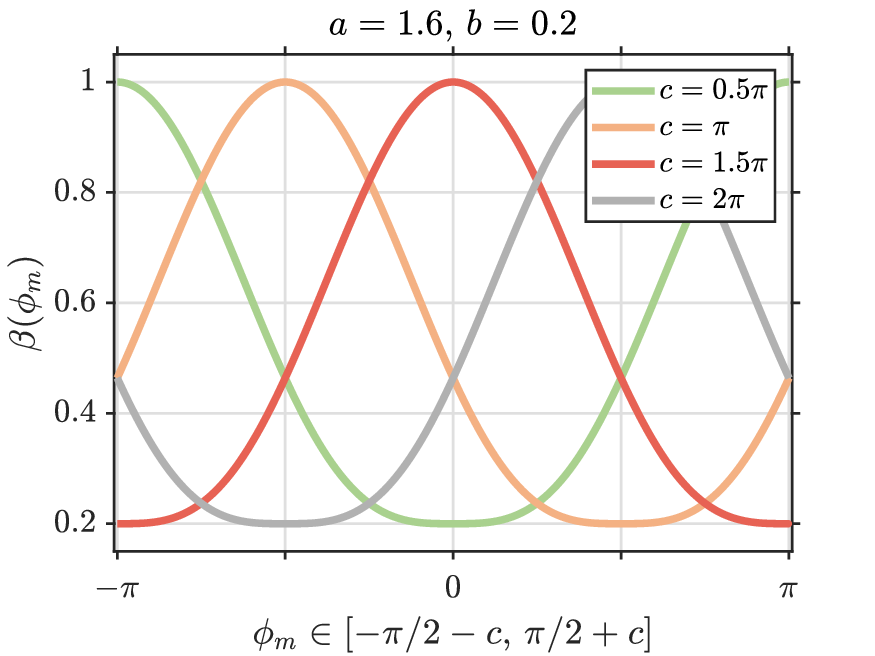}
		\caption{Different $c$ for $\beta(\phi_m)$.}
		\label{Fig_Diff_c}
	\end{subfigure}
    \centering
	\begin{subfigure}{0.24\textwidth}
		\centering
		\includegraphics[width=\textwidth]{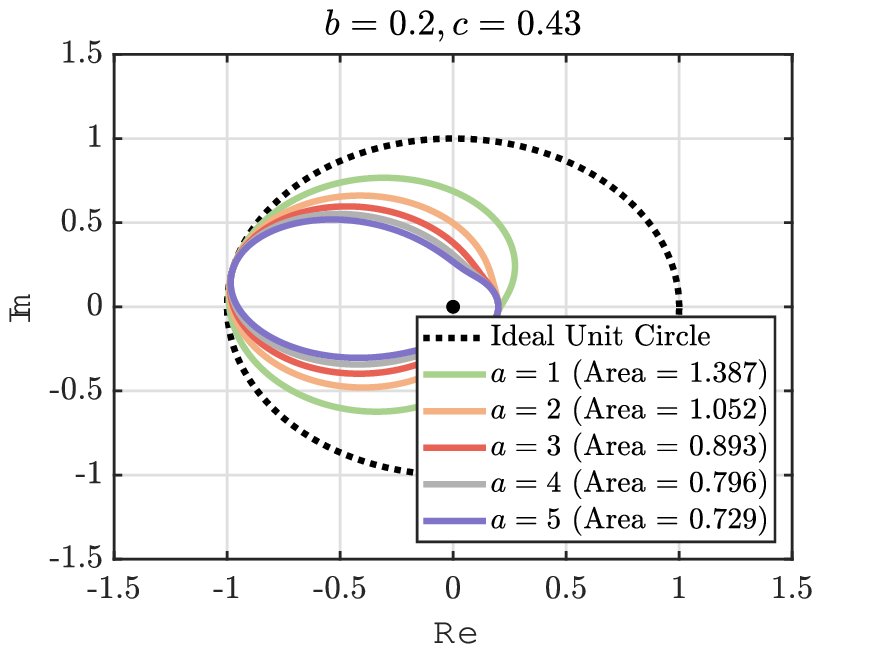}
		\caption{Different $a$ for feasible set.}
		\label{Fig_Feasible_diff_a}
	\end{subfigure}
    \centering
	\begin{subfigure}{0.24\textwidth}
		\centering
		\includegraphics[width=\textwidth]{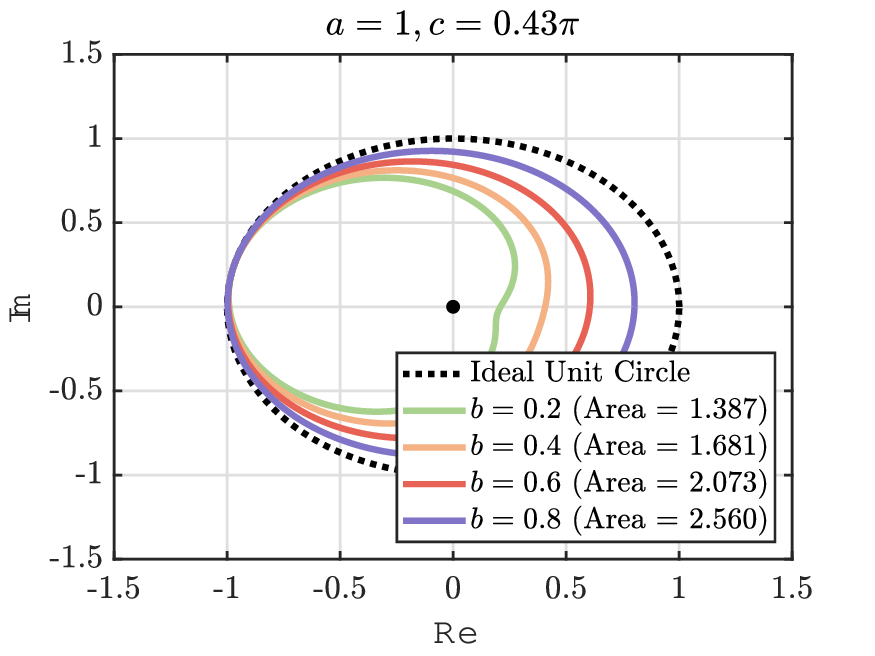}
		\caption{Different $b$ for feasible set.}
		\label{Fig_Feasible_diff_b}
	\end{subfigure}
     \centering
	\begin{subfigure}{0.24\textwidth}
		\centering
		\includegraphics[width=\textwidth]{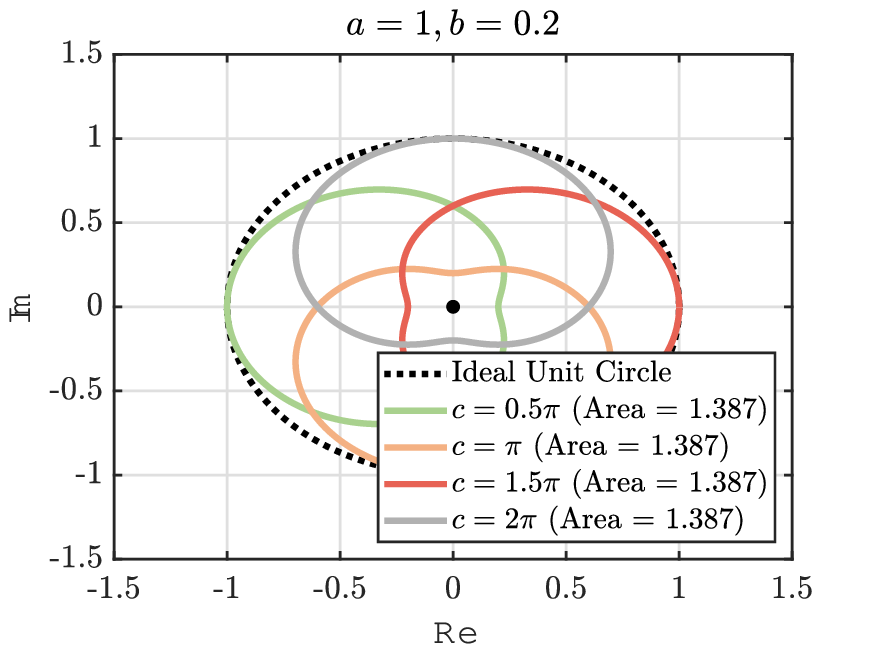}
		\caption{Different $c$ for feasible set.}
		\label{Fig_Feasible_diff_c}
	\end{subfigure}
	\caption{\textcolor{myColor}{Approximated PDA model validations and phase shift feasible set plots. Electrical parameters are from \cite{abey20TC}.}}
	\label{Fig_Sec_II}
\end{figure}

Most previous works ignored the PDA $\beta(\phi_m)$ or simply assumed it is a constant smaller than or equal one \revise{(e.g., The first two models in Table \ref{Table_Works})}. This makes sense if we consider the far-field scenario and each pixel works in the same behavior. However, if the transceiver and the RIS are in the NF case, the pixel has its own feature. In other words, given the different transmission distances between the pixels and the transceiver, we have to not only consider the different phases but also the different distances and other unique characteristics such as the PDA of each scatterer. \revise{Considering \eqref{varsigma_m} and defining $\varsigma_m = \beta(\phi_m)\exp(-\jmath \phi_m)$, an ideal equivalent phase shift model was introduced as} \cite{abey20TC}

\begin{equation}
\beta(\phi_m)=(1-b)\cdot\bigg(\frac{\sin(\phi_{m}-c)+1}{2}\bigg)^a+b,
\label{beta(phi)_Approx_PDA}
\end{equation}

\noindent where $a\ge 1$ is the steepness factor, $b\in[0,1]$ is the minimum amplitude, $c\in (0, \pi/2]$ is the horizontal distance, and all parameters are related to the specific circuit implementation. We call $\beta(\phi_m)$ an \textit{approximated phase-dependent amplitude} \revise{(e.g., Case II in Table \ref{Table_Works})}. To demonstrate that $b$ is more important than both $a$ and $c$, we plot Fig. \ref{Fig_Approx_PDA} using various parameters. It shows the relationship between phase shift  \revise{$\phi_{m}\in[-{\pi}/{2}-c,{\pi}/{2}+c]$} and reflected amplitude $\beta(\phi_{m})\in[b,1]$. When\footnote{It's worth mentioning that there is no negative phase in reality, thus for $\phi_m$, the actual phase shift should add $2k\pi$ where $k$ is a non-negative integer, and the relationship between $\phi_m$ and $\beta(\phi_m)$ in \eqref{beta(phi)_Approx_PDA} is still valid. In this paper, we use $\phi_{m}\in[-{\pi}/{2}-c, {\pi}/{2}+c]$ for the convenience of expression. It is easy to find that when $c={\pi}/{2}$, $\phi_m\in[-\pi, \pi]$.} \revise{$\phi_{m}={\pi}/{2}+c$}, the amplitude $\beta(\phi_{m})$ is maximized to $1$. This is because the reflective currents are out-of-phase with the pixel currents. However, when $\phi_{m}=-{\pi}/{2}+c$, the amplitude is minimized to $b$ since the dielectric and metallic losses increase\cite{abey20TC}. Besides, the approximated PDA model and the circuit model \eqref{varsigma_m} match well, which shows the correctness of \eqref{beta(phi)_Approx_PDA}. From Fig. \ref{Fig_Diff_b_a}, it can be observed that the minimum amplitude $b$ is more important than the other parameters\footnote{For example, when $b$ is fixed to $0.5$, the energy loss between $a=1.6$ and $2$ is just $0.2$ dB. However, if $a$ equals $2$, the energy loss between $b=1$ and $0.5$ is $3.2$ dB. More details can be found in \cite{abey20TC}.}. \revise{Fig. \ref{Fig_Diff_c} shows that $c$ does not decrease the amplitude but changes the positions of the peak and the foot. Besides, we present the designed phase shifts feasible set in the complex plane relative to the ideal unit circle under variations in $a$, $b$, and $c$. Based on Fig. \ref{Fig_Feasible_diff_a} increasing $a$ shifts and squeezes the set with a non‑monotonic area that shrinks for large $a$. Fig. \ref{Fig_Feasible_diff_b} shows increasing $b$ roughly scales it up, enlarging the area and approaching the unit circle, and Fig. \ref{Fig_Feasible_diff_c} reveals $c$ mainly causes a global rotation with nearly constant area. Thus, in order to mimic the real pixel device, choosing a suitable $b$ is of great significance. We then have a proposition as follows.}

\textbf{Proposition 2} \big(Lower and upper bounds of the ideal PDA\big): \revise{\textit{When the pixel hardware and the phase shift are both noiseless, the lower and upper bounds of $\beta(\phi_m)$ in \eqref{beta(phi)_Approx_PDA} are given by}}

\begin{equation}
    b\overset{(i)}{\le}\beta(\phi_m)\big|_{a>1}\overset{(ii)}{\le} \beta(\phi_m)\big|_{a=1}\overset{(iii)}{\le} 1.
    \label{propsition2.1}
\end{equation}

\textit{Proof}: \revise{First, when $\phi_{m}=\phi_{\mathsf{L}}$, equality holds in $(i)$. Similarly, $(ii)$ holds with equality when $\beta(\phi_m)$ attains its minimum (maximum), respectively. Moreover, since $\sin(\phi_m-c)\in[0,1]$ and $b\in[0,1]$, it is straightforward to verify $(iii)$. This completes the proof.}\ \ \ \ \ \ \ \ \ \ \ \ \ \ \ \  \ \ \ \ \ \ \ \ \ \ \ \ \ \ \ \ \ \  \ \ \ \ \ \ \ \ \ \ \ \ \ \ \ \ \ \ \ \ \ \ \ \ \ \ \ \ \ \ \ \ \ \ \ \ \ \ \ \ $\blacksquare$

\revise{Based on Fig. \ref{Fig_Sec_II}, $b$ dominates the equivalent phase shift model in \eqref{beta(phi)_Approx_PDA}.} Therefore, instead of focusing on $\beta(\phi_m)\big|_{a>1}$, we mainly study $\beta(\phi_m)\big|_{a=1}$ \revise{throughout this paper}, which can be viewed as a special case of the proposed model $\beta(\phi_m)$. Consequently, \revise{we define $\phi_{\mathsf{L}}=-\pi/2+c$, $\phi_{\mathsf{U}}=\pi/2+c$, and $\bar{\beta}(\phi_m)=\beta(\phi_m)\big|_{a=1}$ for the remainder of this work.} Since $a$ is less significant relative to other parameters, $\bar{\beta}(\phi_m)$ is indicative of the predominant characteristic of $\beta(\phi_m)$.

\section{Impacts of Phase Errors on Phase-Dependent Amplitudes}\label{Sec3}

In Sec. \ref{Sec2}, we introduce the RIS pixel circuit model $\varsigma_m$ in \eqref{varsigma_m} and its approximated equivalent model $\beta(\phi_m)$ in \eqref{beta(phi)_Approx_PDA}. In practice, however, random noise cannot be ignored\cite{badiu20errors}. The PE is caused by multiple internal and/or external reasons such as hardware degradations, channel estimations, and human-induced accidents\cite{badiu20errors,zhou2020spectral,yang24twc,taghvaee2020error,ken23vtcF}. Hence, in this section, we delve into the last four cases outlined in Table \ref{Table_Works}, with a particular focus on the implications of integrating the PE with the PDA. \revise{Normally there are three types of PE in the RIS pixel, namely \textit{quantization errors}\cite{zhou2020spectral}, \textit{hardware aging effects}\cite{ken23vtcF}, and \textit{imperfect phase estimations}\cite{badiu20errors}. The first two PE RVs follow $\mathcal{UF}$ distributions, while the last RV follow $\mathcal{VM}$ distribution. This work considers these three uncertainties}.

Suppose there is an RIS with $M$ pixels and each transceiver \revise{is equipped with} a single isotropic antenna. \revise{The total power that is emitted from the RIS\footnote{This is the power that the RIS radiates rather than the receiver receives.} can be obtained as $P_{\mathrm{RIS}}=\left\vert\sum_{m=1}^M \sqrt{P_{\mathrm{AP}\rightarrow m}}\beta(\phi_m+\bar{\Delta}_{m})\exp(-\jmath(\phi_m+\Delta_{m}))\right\vert^2$, where $\Delta_m$,  $\bar{\Delta}_m$, and $\breve{\Delta}_m$ are RVs. Note that $\bar{\Delta}_m = \iota\Delta_m + \sqrt{1-\iota^2}\breve{\Delta}_m$ and $\breve{\Delta}_m$ is i.i.d. with $\Delta_m$. The coefficient $\iota\in[0,1]$ is introduced to characterizes the correlation between the PE in the PDA (i.e., $\bar{\Delta}_m$) and in the PSE (i.e., $\Delta_m$). When $\iota=0$, $\Delta_m$ and $\bar{\Delta}_m$ are i.i.d., if $\iota=1$, $\Delta_m=\bar{\Delta}_m$. For intermediate values $\iota\in(0,1)$, $\Delta_m$ and $\bar{\Delta}_m$ exhibit partial correlation.}

\revise{Therefore, assuming that the noise RVs on different pixels (e.g., $m=1, 2$) are i.i.d., $P_{\mathrm{RIS}}$ can be rewritten as}

\begin{align}
P_{\mathrm{RIS}}=\ &
\bigg\vert\sum_{m=1}^M \sqrt{P_{\mathrm{AP}\rightarrow m}}\overbrace{\beta(\phi_m+\underbrace{\vphantom{\bar{\Delta}_{m}}\bar{\Delta}_{m}}_{\mathrm{PE}})}^{\mathrm{PDA}}\overbrace{\vphantom{\bar{\Delta}_{m}}\exp(-\jmath\underbrace{\Delta_{m}}_{\mathrm{PE}})}^{\mathrm{PSE}}\bigg\vert^2
\notag \\
\overset{(i)}{=}& P_{\mathrm{AP}\rightarrow 1} \bigg\vert\sum_{m=1}^M\beta(\phi_m+\bar{\Delta}_{m})\exp(-\jmath\Delta_{m})\bigg\vert^2
\notag\\
=\ & M^2 P_{\mathrm{AP}\rightarrow 1} \bigg\vert\frac{1}{M}\sum_{m=1}^M\beta(\phi_m+\bar{\Delta}_{m})\exp(-\jmath\Delta_{m})\bigg\vert^2
\notag\\
\revise{\xrightarrow[]{M\to\infty}}& \underbrace{\vphantom{\bigg|}M^2 P_{\mathrm{AP}\rightarrow 1}}_{\mathrm{Square}\ \mathrm{Law}} \underbrace{\bigg|\mathbb{E}\big\{\beta(\phi+\bar{\Delta})\exp(-\jmath\Delta)\big\}\bigg|^2}_{\revise{\mathrm{Remaining \ Power} \ \Gamma\ \in\ (0,1)}},
\label{P_RIS_User}
\end{align}

\noindent where $(i)$ is obtained when the pixels are all with the same transmit power $P_{\mathrm{AP}\rightarrow 1}=P_{\mathrm{AP}\rightarrow 2}=\cdots =P_{\mathrm{AP}\rightarrow M}$, respectively. \revise{As the number of reflecting pixels $M$ becomes sufficiently large, the phase shift set $\{\phi_m\}_{m=1}^M$, which is designed to be densely and uniformly distributed over the interval $[-\pi/2-c, \pi/2+c]$, form a sequence whose empirical distribution function converges to that of a continuous $\mathcal{UF}$ distribution over $[-\pi/2-c, \pi/2+c]$. By considering an extension of the \textit{Glivenko–Cantelli theorem} \footnote{\revise{If the empirical distribution of a deterministic sequence converges weakly, the function average over the sequence approaches the corresponding expectation}}\cite{Peter2006}, for the bounded continuous functions $\beta(\phi+\bar{\Delta})\exp(-\jmath\Delta)$, the final step of \eqref{P_RIS_User} is obtained.}

\begin{figure}[htbp]
	\centering
	\begin{subfigure}{0.24\textwidth}
		\centering
		\includegraphics[width=\textwidth]{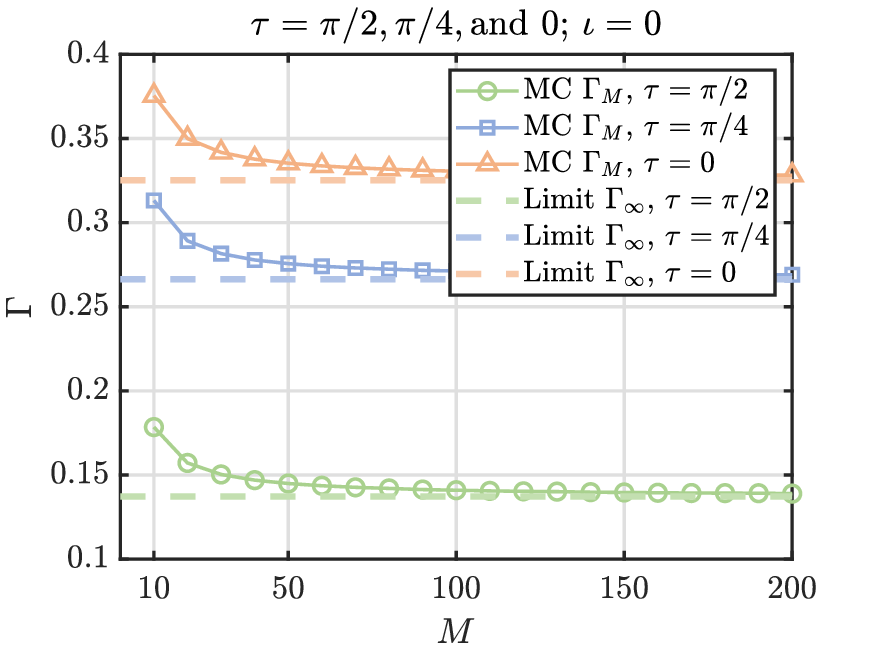}
		\caption{$\mathcal{UF}[\tau, \tau]$ and $\iota=0$.}
        \label{Fig_eq8_UF_1015}
	\end{subfigure}
    	\begin{subfigure}{0.24\textwidth}
		\centering
		\includegraphics[width=\textwidth]{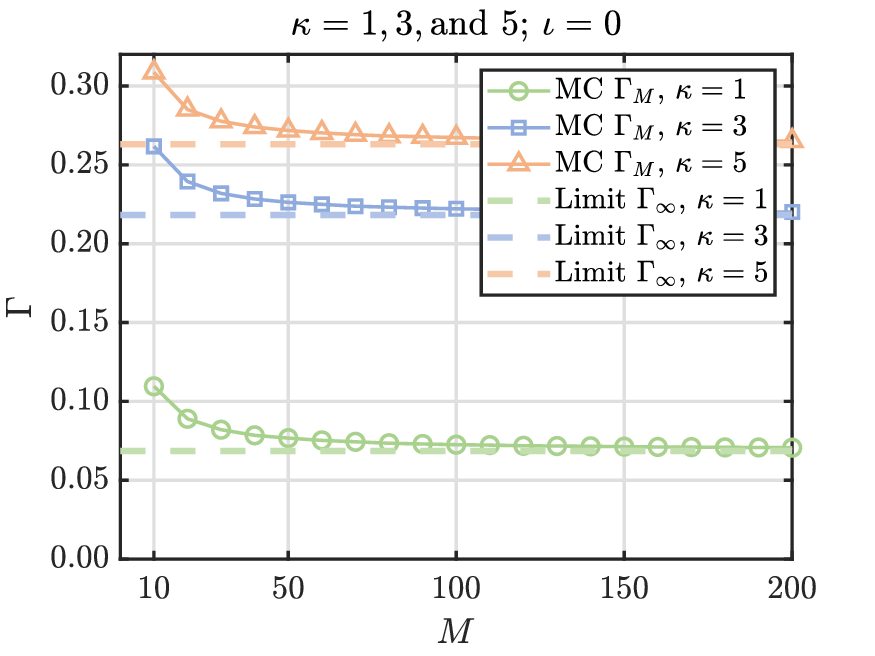}
		\caption{$\mathcal{VM}(0,\kappa)$ and $\iota=0$.}
        \label{Fig_eq8_VM_1015}
	\end{subfigure}
    	\centering
	\begin{subfigure}{0.24\textwidth}
		\centering
		\includegraphics[width=\textwidth]{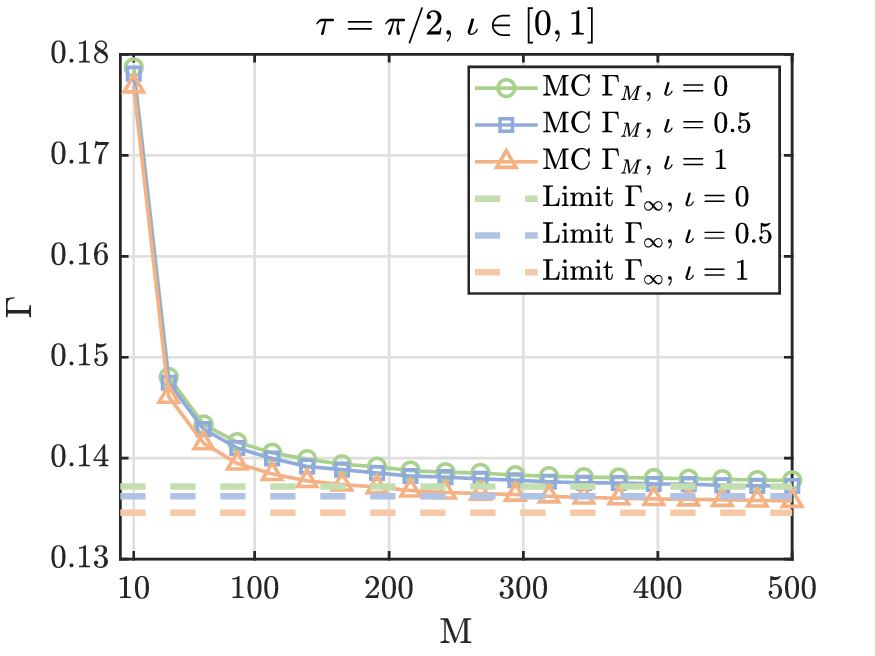}
		\caption{$\mathcal{UF}[\tau, \tau]$ and $\iota\in[0,1]$.}
        \label{Fig_eq8_UF_iota_1015}
	\end{subfigure}
    	\begin{subfigure}{0.24\textwidth}
		\centering
		\includegraphics[width=\textwidth]{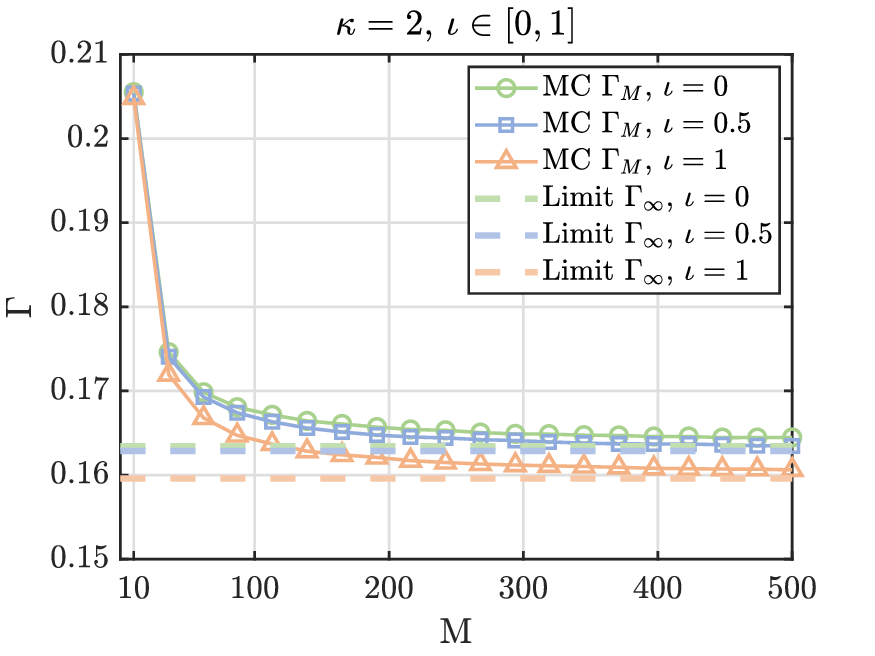}
		\caption{$\mathcal{VM}(0,\kappa)$ and $\iota\in[0,1]$.}
        \label{Fig_eq8_VM_iota_1015}
	\end{subfigure}
    	\caption{\textcolor{myColor}{Validation of \eqref{P_RIS_User}.}}
	\label{Fig_eq8_1004}
\end{figure}

It can be seen that \eqref{P_RIS_User} contains two parts. The first one, i.e., $M^2 P_{\mathrm{AP}\rightarrow m}$, reveals that the total power emitted from the RIS $P_{\mathrm{RIS}}$ increases directly proportional to $M^2$. This behavior is also called \textit{square law} in the RIS communication \cite{ozdogan20irsmodel}. \revise{Moreover, the second part, i.e., normalized \revise{\textit{remaining power} $\Gamma$}, cannot be ignored especially when we consider a practical RIS system. Previous works only considered $\vert\mathbb{E}\{\exp(-\jmath\Delta_{m})\}\vert^2$. In the rest of this paper, we \revise{define} $\Gamma=\vert\mathbb{E}\{\beta(\phi_m+\bar{\Delta}_{m})\exp(-\jmath\Delta_{m})\}\vert^2$, which represents the power that remains after a signal passes through an RIS pixel. Besides, it is worth noting that $\phi_1 \neq \phi_2 \neq \cdots\neq\phi_M$ because each pixel experiences its own cascaded fading channel, making an accurate approximation of $\Gamma$ challenging. Nevertheless, as $M$ grows large, \eqref{P_RIS_User} implies that $P_{\mathrm{RIS}}$ approaches a constant.}

\revise{To model} the quantization errors and hardware degradations of the RIS, we assume that the PEs $\delta_m$ and $\bar{\delta}_m$ at the $m$-th pixel follow a $\mathcal{UF}$ distribution \cite{badiu20errors}, i.e., $\delta_m$ and $\bar{\delta}_m\sim\mathcal{UF}[-\tau, \tau]$, where $\tau\in[0,{\pi}/{2}]$. The corresponding probability density function (PDF) is $f_{\mathcal{UF}}(\delta)=\frac{1}{2\tau}$ where $\delta\in[-\tau,\tau]$ or is zero otherwise. \revise{Hence, the reflection coefficient is $\beta(\phi_{m}+\bar{\delta}_m)\exp(-\jmath(\phi_{m}+\delta_{m}))$, where $\bar{\delta}_m = \iota\delta_m + \sqrt{1-\iota^2}\breve{\delta}_m$, $\iota\in[0,1]$ and $\breve{\delta}_m$ is i.i.d. with $\delta_m$.} For imperfect channel estimations of the RIS, we assume that the errors $\gamma_m$ and $\bar{\gamma}_m$ in the $m$-th pixel follows zero-mean $\mathcal{VM}$ distributions\cite{mardia2009directionalbook}, i.e., $\gamma_m$ and $\bar{\gamma}_m\sim\mathcal{VM}(0,\kappa)$, where $\kappa$ is concentration parameter. Accordingly, the PDF is $f_{\mathcal{VM}}(\gamma)=\frac{\exp(\kappa\cos(\gamma))}{2\pi I_0(\kappa)}$ where $\gamma\in[-\pi,\pi]$ and $I_n(\kappa)$ is the modified Bessel function of the first kind of order $n$ with $\kappa\ge 1$. \revise{Thus, the reflection coefficient is $\beta(\phi_{m}+\bar{\gamma}_m)\exp(-\jmath(\phi_{m}+\gamma_{m}))$, where $\bar{\gamma}_m = \iota\gamma_m + \sqrt{1-\iota^2}\breve{\gamma}_m$, $\iota\in[0,1]$ and $\breve{\gamma}_m$ is i.i.d. with $\gamma_m$.}


\begin{figure*}[t]
\centering
\includegraphics[width=0.85\textwidth]{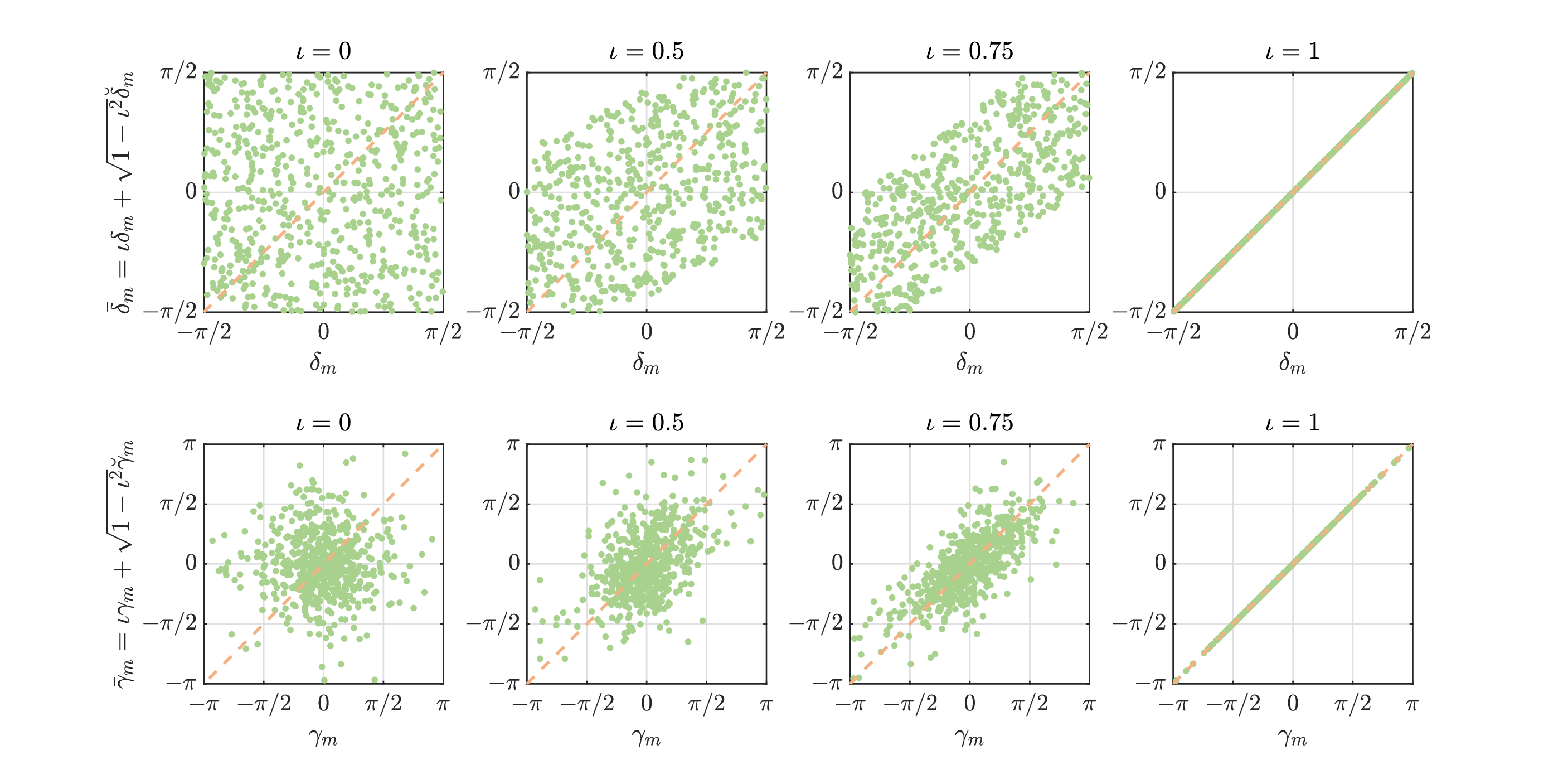}
\caption{\textcolor{myColor}{Scatter plots of $\mathcal{UF}$ and $\mathcal{VM}$ RVs for different correlation coefficients $\iota\in [0, 0.5, 0.75, 1]$.}}
\label{Fig_UF_VM_iota_0_1}
\end{figure*}

\revise{It can be observed from Figs. \ref{Fig_eq8_UF_1015} and \ref{Fig_eq8_VM_1015} that, when $\iota=0$, as $M$ increases from $1$ to $200$, Monte Carlo (MC) result $\Gamma_M=\big\vert\frac{1}{M}\sum_{m=1}^M\beta(\phi_m+\bar{\Delta}_{m})\exp(-\jmath\Delta_{m})\big\vert^2$ converges asymptotically to analytical (AN) limit $\Gamma_{\infty}=\big\vert\mathbb{E}\big\{\beta(\phi+\bar{\Delta})\exp(-\jmath\Delta)\big\vert^2$. Besides, as the PE becomes more severe (i.e., $\tau$ increases from $0$ to $\pi/2$ and $\kappa$ increases from $1$ to $5$), $\Gamma_M$ and $\Gamma_{\infty}$ decrease accordingly. Besides, from Figs. \ref{Fig_eq8_UF_iota_1015} and \ref{Fig_eq8_VM_iota_1015}, if $\iota=1$, the same convergence behavior is observed, i.e., as $M$ grows from $1$ to $500$, the MC result $\Gamma_M$ approaches its AN limit $\Gamma_{\infty}$. More importantly, both $\Gamma_M$ and $\Gamma_{\infty}$ with $\iota=1$ are smaller than these with $\iota=0$. This implies if an RIS has a sufficiently large number of pixels (e.g., $M>200$ when $\iota=0$ and $M>500$ when $\iota=1$) and the average phase approaches zero, then fully coupled noise (i.e., $\iota=1$) is more detrimental than i.i.d. noise (i.e., $\iota=0$). The top row of Fig. \ref{Fig_UF_VM_iota_0_1} plots the characteristics of $\bar{\delta}_m=\iota\delta_m+\sqrt{1-\iota^2}\breve{\delta}_m$ when i.i.d. RVs $\delta$ and $\breve{\delta}_m$ follow a $\mathcal{UF}$ distribution with $\tau=\pi/2$. As the correlation coefficient $\iota$ increases from $0$ to $1$, the scatter points transition from occupying a square region to converging along the diagonal, fully aligning at $\iota=1$. In the third figure of this row, the overall trend of the green dots is flatter than the orange line. This is because the best-fit line for the green dots is not the line $\bar{\delta}_m={\delta}_m$, but rather $\bar{\delta}_m=0.75{\delta}_m$. The bottom row of Fig. \ref{Fig_UF_VM_iota_0_1} presents the same construction method and i.i.d. RVs $\gamma$ and $\breve{\gamma}_m$ follow a $\mathcal{VM}$ distribution. As $\iota$ increases, the correlation strengthens and points cluster toward the diagonal, yet the central concentration of the $\mathcal{VM}$ and sparse tails result in higher density at the center. In the rest of this section, we study the last four cases in Table \ref{Table_Works} as follows\footnote{\revise{In Cases II, III, and IV, we assume $\phi_1=\phi_2=\cdots=\phi_m$ so that the effect of uncertainties can be examined under a fixed phase shift.}}.}

\subsection{Case I: When the PDA is A Constant}

When the PDA of the $m$-th pixel is a constant for all $M$ pixels and \revise{only the PSE contains the PE}, we have the \revise{RP} $\Gamma$ as

\begin{equation}
    \Gamma = \underbrace{\boxed{\left\vert\mathbb{E}\big\{\beta\exp(-\jmath\Delta_m)\big\}\right\vert^2}}_{\mathrm{Case \ I}},
    \label{Gamma_Case_I}
\end{equation}

\noindent where $\revise{\Delta_m}=\delta_m$ or $\gamma_m$. Consequently, the impact of noise on $\Gamma$ can be described by the following two propositions.

\textbf{Proposition 3.1} \big($\beta\exp(-\jmath\delta_{m})$, the PDA is a constant and only the PSE contains \revise{$\mathcal{UF}$ \revise{RVs}}\big): \textit{When the $m$-th PDA is a constant $\beta$, and the PSE includes $\delta_{m}\sim\mathcal{UF}[-\tau, \tau]$ where \revise{$\tau\in [0,{\pi}/{2}]$}, then the \revise{RP} $\Gamma$ in \eqref{Prop_31} can be obtained as\footnote{Except for special instructions, we assume in the remaining part of this paper that $\Gamma_{(\cdot)}=\Gamma_{(\cdot)}(\phi_m)$.}}

\begin{align}\label{Prop_31}
 &\Gamma_{(\mathrm{3.1})}= \ \left\vert\mathbb{E}\big\{\beta\exp(-\jmath\delta_{m})\big\}\right\vert^2
 \notag\\
=\ & \beta^2 \left(1-\frac{1}{3}\tau^2+\frac{2}{45}\tau^4-\frac{1}{360}\tau^6+\frac{1}{14400}\tau^8\right)+\revise{\mathcal{O}\left(\tau^{10}\right)},
\end{align}

\noindent \textit{where $\mathcal{O}(\cdot)$ denotes higher‑order terms.}

\textit{Proof}: Please see Appendix A in \cite{ken24vtcF}.  \ \ \ \ \ \ \ \ \ \ \ \ \ \  \ \ \ \ \ \ \ \  \ \ \ \ \ \ \ \ $\blacksquare$

\begin{figure}[htbp]
	\centering
	\begin{subfigure}{0.24\textwidth}
		\centering
		\includegraphics[width=\textwidth]{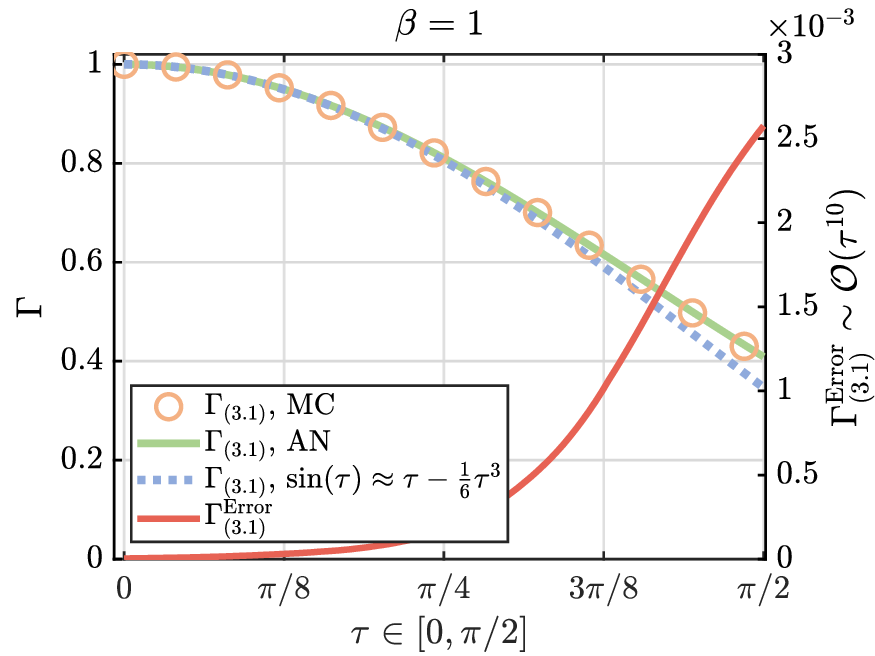}
		\caption{Validation of Proposition 3.1.}
        \label{Fig_Prop_31_1012}
	\end{subfigure}
	\centering
	\begin{subfigure}{0.24\textwidth}
		\centering
		\includegraphics[width=\textwidth]{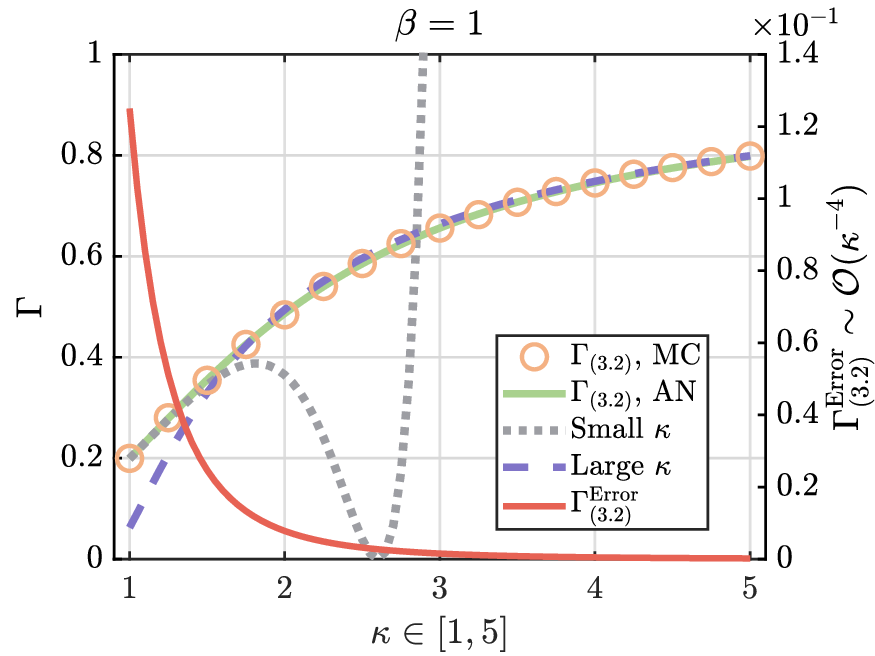}
		\caption{Validation of Proposition 3.2.}
		\label{Fig_Prop_32_1012}
	\end{subfigure}
	\caption{\textcolor{myColor}{Validation of Case I.}}
	\label{Fig_Sec_II_Case_I}
\end{figure}

Fig. \ref{Fig_Prop_31_1012} reveals that $ \Gamma_{(\mathrm{3.1})}$ needs at least the first three terms of the Taylor series expansion of $\sin(\tau)$. \revise{The error term\footnote{For clarity, we smooth the error curves without affecting their accuracy.} is $\Gamma_{(3.1)}^{\mathrm{Error}}=\big\vert\Gamma_{(3.1)}^{\mathrm{MC}}-\Gamma_{(3.1)}^{\mathrm{AN}}\big\vert\sim\mathcal{O}(\tau^{10})$ and becomes significant when $\tau$ approaches $\pi/2$.}

\begin{figure}[htbp]
	\centering
	\begin{subfigure}{0.24\textwidth}
		\centering
		\includegraphics[width=\textwidth]{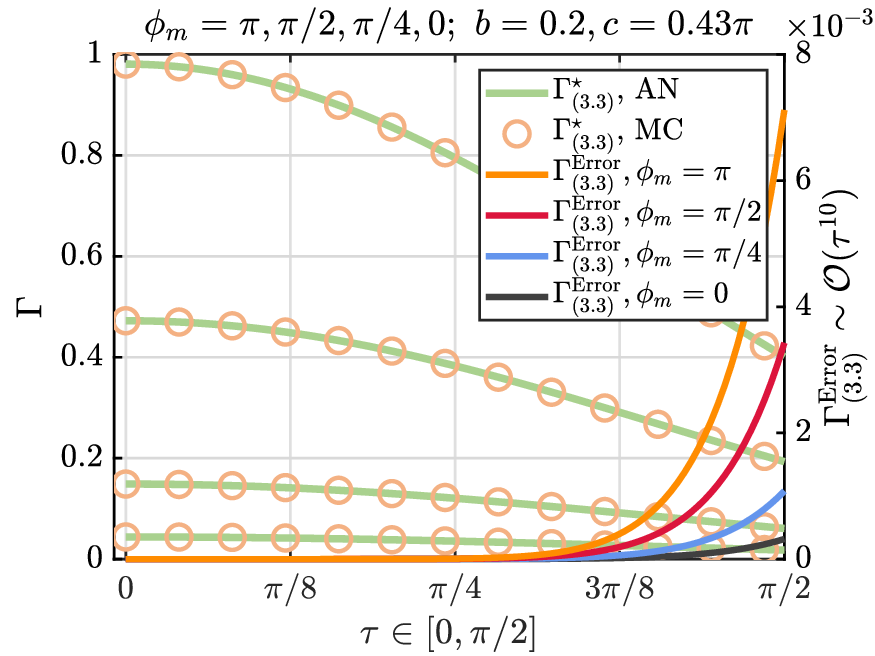}
		\caption{Validation of $\Gamma_{(3.3)}^\star$.}
        \label{Fig_Prop_33_Diff_phi_m}
	\end{subfigure}
	\centering
	\begin{subfigure}{0.24\textwidth}
		\centering
		\includegraphics[width=\textwidth]{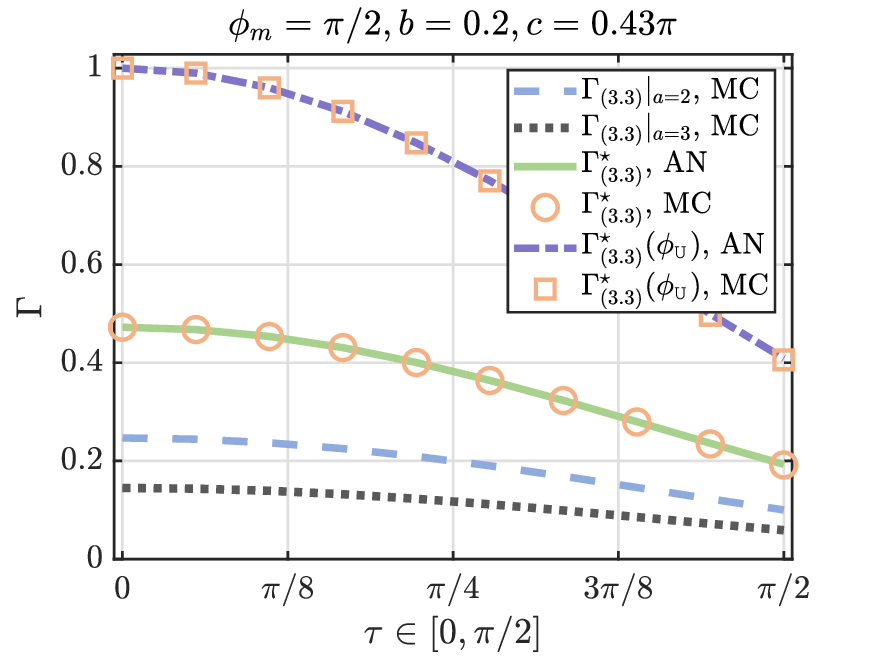}
		\caption{Validation of Prop. 3.3.}
		\label{Fig_Prop_33_Comparison}
	\end{subfigure}
	\caption{\textcolor{myColor}{Validation of Prop. 3.3.}}
	\label{Fig_Sec_II_Case_II_Prop_33}
\end{figure}

\textbf{Proposition 3.2} \big($\beta\exp(-\jmath\gamma_{m})$, the PDA is a constant and only the PSE contains \revise{$\mathcal{VM}$ noises}\big): \textit{When the $m$-th PDA is a constant $\beta$, and the PSE includes $\gamma_{m}\sim\mathcal{VM}(0,\kappa)$ where $\kappa \ge 1$, then the \revise{RP} $\Gamma$ in \eqref{Gamma_Case_I} can be obtained as}

\begin{align}\label{Prop_32}
\Gamma_{(\mathrm{3.2})}&=\left\vert\mathbb{E}\big\{\beta\exp(-\jmath\gamma_{m})\big\}\right\vert^2=\beta^2 \rho^2,
\end{align}

\noindent \textit{where}

\begin{align}
\rho=\begin{cases}
\frac{\kappa}{2}\left(1-\frac{\kappa^2}{8}+\frac{\kappa^4}{48}-\frac{11\kappa^6}{3072}\right)\revise{+\mathcal{O}(\kappa^9)}&\text{for small $\kappa$}
\\
1-\frac{1}{2\kappa}-\frac{1}{8\kappa^2}-\frac{1}{8\kappa^3}\revise{+\mathcal{O}(\kappa^{-4})}&\text{for large $\kappa$}.
\end{cases}
\label{rho}
\end{align}

\textit{Proof}: Please see Appendix \ref{Proof_Prop_31_32}.  \ \ \ \ \ \ \ \ \ \ \ \ \ \  \ \ \ \ \ \ \ \  \ \ \ \ \ \ \ \ $\blacksquare$

From Fig. \ref{Fig_Prop_32_1012}, we can see that if $\kappa\in[1,1.6]$, $\frac{\kappa}{2}\left(1-\frac{\kappa^2}{8}+\frac{\kappa^4}{48}-\frac{11\kappa^6}{3072}\right)$ is a suitable approximation of $\rho$, otherwise we should use $1-\frac{1}{2\kappa}-\frac{1}{8\kappa^2}-\frac{1}{8\kappa^3}$. \revise{For small $\kappa$, $\rho^2$ incurs a truncation error of $\Gamma_{(3.2)}^{\mathrm{Error}}\sim\mathcal{O}(\kappa^{9})$. In contrast, for large $\kappa$, the asymptotic expansion yields an error of $\Gamma_{(3.2)}^{\mathrm{Error}}\sim\mathcal{O}(\kappa^{-4})$. To ensure correctness and reproducibility, $\mathcal{VM}$ RVs are generated using $\mathsf{CircStat}$ \cite{berens2009circstat} and $\mathsf{vmrand}$\cite{best1979vm} in this paper.}

\subsection{Case II: When the PDA is without Phase Errors}

When the PDA of the $m$-th pixel is $\beta(\phi_m)$ in \eqref{beta(phi)_Approx_PDA} without any errors and the PSE is with the PE, we have the \revise{RP} $\Gamma$ as

\begin{equation}
    \Gamma = \underbrace{\boxed{\left\vert\mathbb{E}\big\{\beta(\phi_m)\exp(-\jmath\Delta_m)\big\}\right\vert^2}}_{\mathrm{Case \ II}},
    \label{Gamma_Case_II}
\end{equation}

\noindent where $\Delta=\delta_m$ or $\gamma_m$. Hence, the impact of noise on $\Gamma$ can be described by the following two propositions.

\textbf{Proposition 3.3} \big($\beta(\phi_m)\exp(-\jmath\delta_m)$, the PDA is ideal and the PSE contains \revise{$\mathcal{UF}$ RVs}\big): \textit{When the $m$-th PDA is $\beta(\phi_m)$ and without \revise{any noises}, and the PSE includes $\delta_{m}\sim\mathcal{UF}[-\tau, \tau]$ where $x\in \big[0,{\pi}/{2}\big]$, then the upper bounds of the \revise{RP} $\Gamma$ in \eqref{Gamma_Case_II} can be obtained as}

\begin{equation}
    \Gamma_{(3.3)} \overset{(i)}{\le} \Gamma_{(3.3)}^{\star}< \Gamma_{(3.3)}^{\star}(\phi_{\mathsf{U}}),
    \label{Prop_33}
\end{equation}

\noindent \textit{where $\Gamma_{(3.3)}=\big|\mathbb{E}\big\{\beta(\phi_m)\exp(-\jmath\delta_m)\big\}\big|^2$, $\Gamma_{(3.3)}^\star=\big|\mathbb{E}\big\{\bar{\beta}(\phi_m)\exp(-\jmath\delta_m)\big\}\big|^2=\zeta\big(1-\frac{1}{3}\tau^2+\frac{2}{45}\tau^4-\frac{1}{360}\tau^6+\frac{1}{14400}\tau^8\big)\revise{+\mathcal{O}\big(\tau^{10}\big)}$ and $\zeta = \frac{(1-b)^2}{4}\sin^2(\phi_{m}-c)+\frac{(1-b^2)}{2}\sin(\phi_{m}-c)+\frac{(1+b)^2}{4}$. \revise{$\Gamma_{(3.3)}^{\star}(\phi_{\mathsf{U}})=\big(1-\frac{1}{3}\tau^2+\frac{2}{45}\tau^4-\frac{1}{360}\tau^6+\frac{1}{14400}\tau^8\big)$.} Moreover, the equality in $(i)$ is satisfied when\footnote{This condition also applies to Propositions 3.4 to 3.10.} $a=1$. $\Gamma_{(3.3)}^\star\sim\mathcal{O}(\tau^{10})$.}

\textit{Proof}: Please see Appendix B in \cite{ken24vtcF}.  \ \ \ \ \ \ \ \ \ \ \ \ \ \  \ \ \ \ \ \ \ \  \ \ \ \ \ \ \ \ $\blacksquare$

It is not difficult to find that $b^2\le\zeta\le 1$, and $b^2$ and $1$ can be obtained respectively when $\phi_{m}=\phi_{\mathsf{L}}$ and $\phi_{\mathsf{U}}$. Besides, Fig. \ref{Fig_Prop_33_Diff_phi_m} reveals an important insight that different $\phi_m$ are affected differently by the same error. In particular, if $\phi_m=\pi$, the RP decreases from $1$ to $0.4$ when $\tau$ increase from $0$ to ${\pi}/{2}$. However, when $\phi_m={\pi}/{4}$, the RP decrease can be ignored even $\tau$ approaches to ${\pi}/{2}$. \revise{The error term is $\Gamma_{(3.3)}^{\mathrm{Error}}=\big\vert\Gamma_{(3.3)}^{\mathrm{MC}}-\Gamma_{(3.3)}^{\mathrm{AN}}\big\vert\sim\mathcal{O}(\tau^{10})$}. Fig. \ref{Fig_Prop_33_Comparison} illustrates that when $\phi_m=\phi_{\mathsf{U}}$, the RIS pixel always has full reflections, i.e., $\beta(\phi_m)$ is maximized to $1$. However, in practice, $\phi_{\mathsf{U}}$ cannot align $M$ paths to the receiver. Therefore, $\Gamma_{(3.3)}^\star(\phi_{\mathsf{U}})$ represents the upper bound on RP resulting from maximizing amplitude, rather than aiming to enhance the overall performance of the system. A well-designed $\phi_m$ in $\Gamma_{(3.3)}^\star$ denotes a trade-off between maximizing the PDA and aligning cascaded channels.

\textbf{Proposition 3.4} \big($\beta(\phi_m)\exp(-\jmath\gamma_m)$, the PDA is ideal and the PSE contains \revise{$\mathcal{VM}$ RVs}\big): \textit{When the $m$-th PDA is $\beta(\phi_m)$ and without any errors, and the PSE includes $\gamma_{m}\sim\mathcal{VM}(0,\kappa)$ where $\kappa \ge 1$, then the upper bounds of the \revise{RP} $\Gamma$ in \eqref{Gamma_Case_II} can be obtained as}

\begin{equation}
    \Gamma_{(3.4)} \ {\le} \  \Gamma_{(3.4)}^{\star}< \Gamma_{(3.4)}^{\star}(\phi_{\mathsf{U}}),
    \label{Prop_34}
\end{equation}

\noindent \textit{where $\Gamma_{(3.4)}=\big|\mathbb{E}\big\{\beta(\phi_m)\exp(-\jmath\gamma_m)\big\}\big|^2$, $\Gamma_{(3.4)}^\star=\big|\mathbb{E}\big\{\bar{\beta}(\phi_m)\exp(-\jmath\gamma_m)\big\}\big|^2=\zeta\rho^2$, \revise{$\Gamma_{(3.4)}^{\star}(\phi_{\mathsf{U}})=\rho^2$,} and $\rho$ and $\zeta$ are the same as they in Props. 3.2 and 3.3, respectively.}

\textit{Proof}: The proof is the same as Props. 3.1 to 3.3 and is omitted here.  \ \ \ \ \ \ \ \ \ \ \ \ \ \  \ \ \ \ \ \ \ \ \ \ \ \ \ \ \ \ \ \ \ \ \ \ \ \ \ \ \  \ \ \ \ \ \ \ \ \ \ \ \ \ \ \ \ $\blacksquare$

\begin{figure}[htbp]
	\centering
	\begin{subfigure}{0.24\textwidth}
		\centering
		\includegraphics[width=\textwidth]{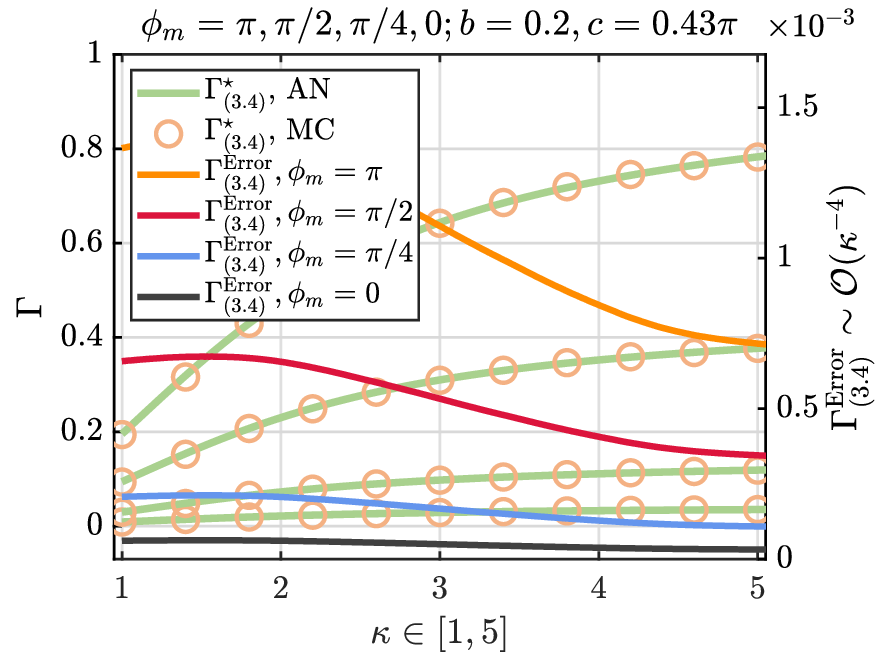}
		\caption{Validation of $\Gamma_{(3.4)}^\star$.}
        \label{Fig_Prop_34_Diff_phi_m}
	\end{subfigure}
	\centering
	\begin{subfigure}{0.24\textwidth}
		\centering
		\includegraphics[width=\textwidth]{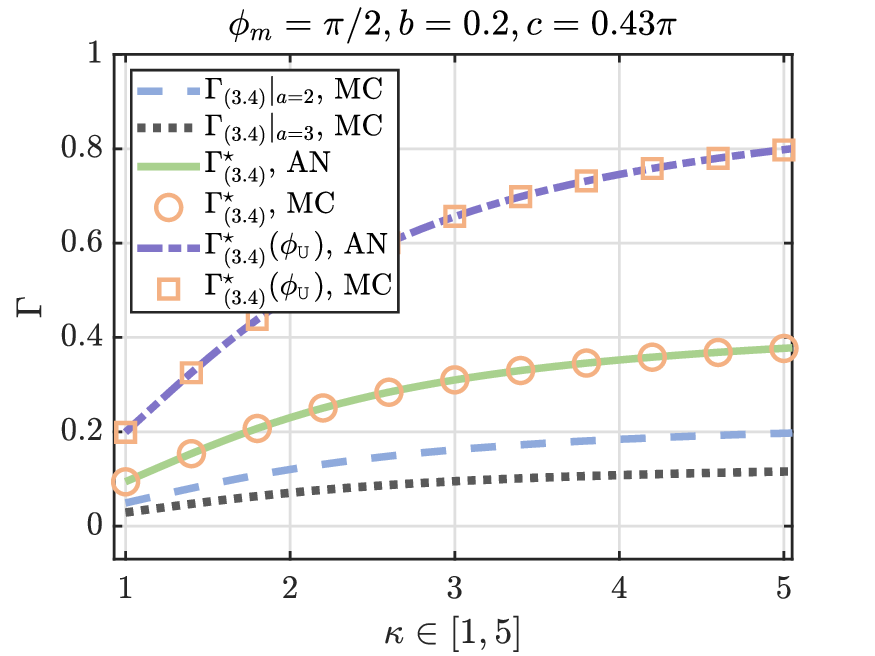}
		\caption{Validation of Prop. 3.4.}
		\label{Fig_Prop_34_Comparison}
	\end{subfigure}
	\caption{\textcolor{myColor}{Validation of Prop. 3.4.}}
	\label{Fig_Sec_II_Case_II_Prop_34}
\end{figure}

When the concentration parameter $\kappa$ increases, the distribution of directions becomes more concentrated around the mean direction $0$, and the RP grows. \revise{The error term is $\Gamma_{(3.4)}^{\mathrm{Error}}=\big\vert\Gamma_{(3.4)}^{\mathrm{MC}}-\Gamma_{(3.4)}^{\mathrm{AN}}\big\vert\sim\mathcal{O}(\kappa^{10})$}. When $\kappa=0$, the $\mathcal{VM}$ distribution reduces to a $\mathcal{UF}$ distribution. It can be observed from Fig. \ref{Fig_Prop_34_Diff_phi_m} that when $\kappa=1$, the RP is from $0$ to $0.2$ when $\phi_m=0$ to $\pi$. When $\kappa = 5$ and $\phi_m=\pi$, the RP achieves to $0.8$. However, even $\kappa = 5$, $\Gamma_{(3.4)}^\star$ is still limited as $\phi_m=0$. Therefore, the PE is no longer important when $\phi_m$ approaches $0$, because the amplitude available for unitize at this time is already at its lowest value. Besides, Fig. \ref{Fig_Prop_34_Comparison} shows the correctness of \eqref{Prop_34}.

\subsection{Case III: When the PDA is with Single Error}

When the PDA of the $m$-th pixel is the approximated model \eqref{beta(phi)_Approx_PDA} with $\bar{\Delta}_m$ and the PSE is with $\Delta_m$, we have the RP $\Gamma$ as

\begin{equation}
    \Gamma = \underbrace{\boxed{\left\vert\mathbb{E}\big\{\beta(\phi_m+\bar{\Delta}_m)\exp(-\jmath\Delta_m)\big\}\right\vert^2}}_{\mathrm{Case \ III}},
    \label{Gamma_Case_III}
\end{equation}

\noindent where $\Delta_m=\delta_m$ or $\gamma_m$, $\bar{\Delta}_m=\delta_m$ or $\gamma_m$, and $\Delta_m$ and $\bar{\Delta}_m$ are the same or i.i.d. Hence, the impact of noise on $\Gamma$ can be described by the following four propositions.

\textbf{Proposition 3.5} \big($\beta(\phi_m+\delta_m)\exp(-\jmath\delta_{m})$, \revise{$\iota=1$}\big): \textit{When $\iota=1$, the $m$-th PDA is $\beta(\phi_m+\delta_m)$, and the PDA and the PSE all contain $\delta_{m}\sim\mathcal{UF}[-\tau, \tau]$ where $\tau\in [0, {\pi}/{2}]$, then the upper bound of \revise{the RP} $\Gamma$ in \eqref{Gamma_Case_III} can be obtained as}

\begin{equation}
    \Gamma_{(3.5)} \ {\le} \  \Gamma_{(3.5)}^{\star}=(\eta_1+\eta_2)^2+\eta_3
    < \Gamma_{(3.5)}^{\star}(\phi_{\mathsf{U}}),
    \label{Prop_35}
\end{equation}

\noindent \textit{where $\Gamma_{(\mathrm{3.5})}=\big\vert\mathbb{E}\{\beta(\phi_m+\delta_m)\exp(-\jmath\delta_{m})\}\big\vert^2$, $\Gamma_{(3.5)}^{\star}=\big\vert\mathbb{E}\{\bar{\beta}(\phi_m+\delta_m)\exp(-\jmath\delta_{m})\}\big\vert^2$, $\eta_1=\frac{(1-b)}{2}\sin(\phi_{m}-c)\big(1-\frac{1}{3}\tau^2+\frac{1}{15}\tau^4-\frac{2}{315}\tau^6\big)+\mathcal{O}
(\tau^8)$, $\eta_2=\frac{1+b}{2}\big(1-\frac{1}{6}\tau^2+\frac{1}{120}\tau^4-\frac{1}{5040}\tau^6\big)+\mathcal{O}
(\tau^8)$, $\eta_3=\frac{(1-b)^2}{4}\big(\frac{1}{3}\tau^2-\frac{1}{15}\tau^4+\frac{2}{315}\tau^6\big)^2 \cos^2(\phi_{m}-c)+\mathcal{O}
(\tau^{12})$, \revise{ $\Gamma_{(3.5)}^{\mathrm{Error}}=\big\vert\Gamma_{(3.5)}^{\mathrm{MC}}-\Gamma_{(3.5)}^{\mathrm{AN}}\big\vert\sim\mathcal{O}(\tau^{8})$}. $\Gamma_{(3.5)}^{\star}(\phi_{\mathsf{U}})=({\eta_1}/ \sin(\phi_{m}-c)+\eta_2)^2$.}

\textit{Proof}: Please see Appendix C in \cite{ken24vtcF}. \ \ \ \ \ \ \ \ \ \ \ \ \ \  \ \ \ \ \ \ \ \  \ \ \ \ \ \ \ \ $\blacksquare$

\begin{figure}[htbp]
	\centering
	\begin{subfigure}{0.24\textwidth}
		\centering
		\includegraphics[width=\textwidth]{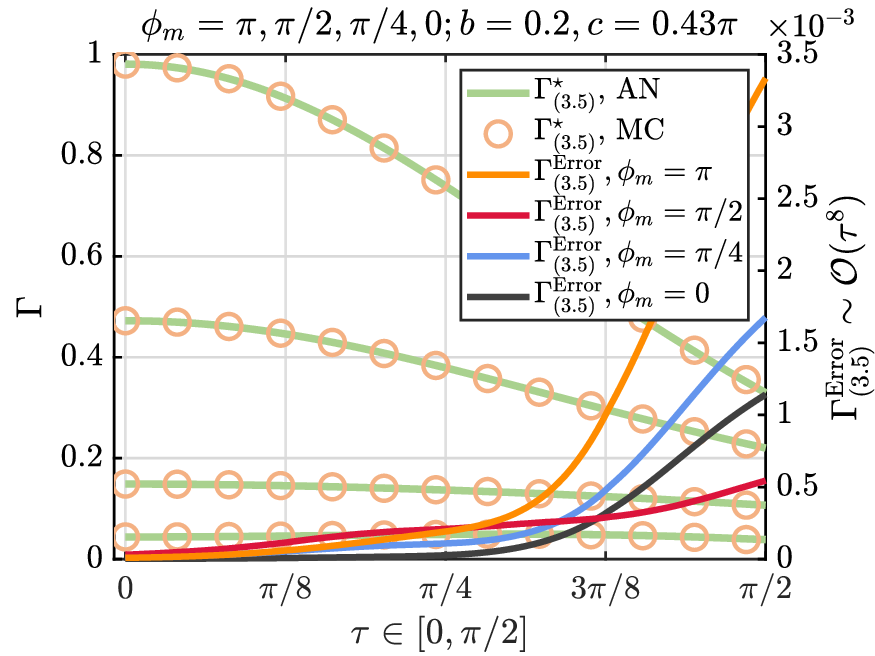}
		\caption{Validation of $\Gamma_{(3.5)}^\star$.}
        \label{Fig_Prop_35_Diff_phi_m}
	\end{subfigure}
	\centering
	\begin{subfigure}{0.24\textwidth}
		\centering
		\includegraphics[width=\textwidth]{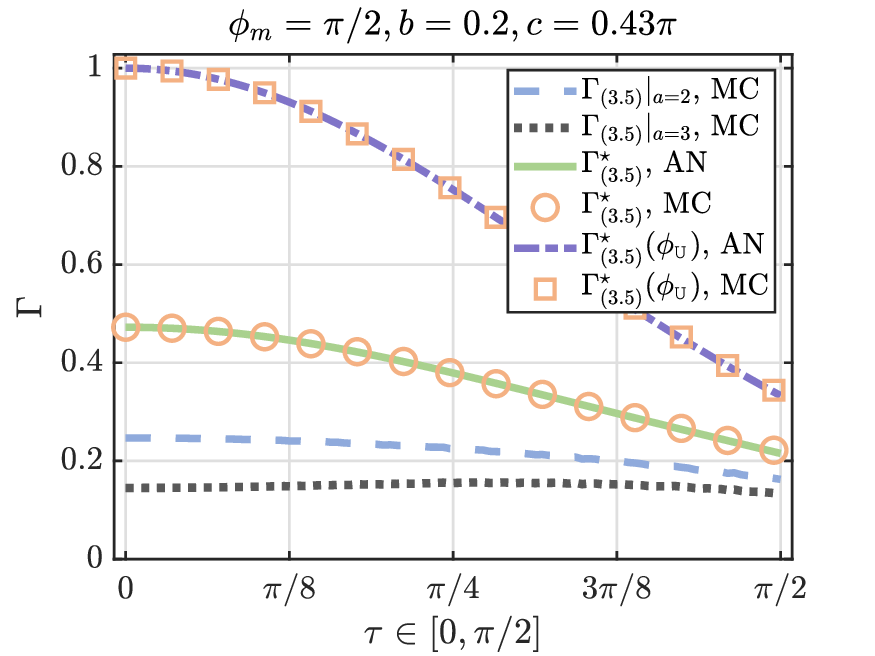}
		\caption{Validation of Prop. 3.5.}
		\label{Fig_Prop_35_Comparison}
	\end{subfigure}
	\caption{\textcolor{myColor}{Validation of Prop. 3.5.}}
	\label{Fig_Sec_II_Case_III_Prop_35}
\end{figure}

From Fig. \ref{Fig_Prop_35_Diff_phi_m}, it can be seen that the approximated AN expression of $\Gamma_{(3.5)}^\star$ is correct. Besides, when the amplitude is large (e.g., $\phi_m=\pi$), the impact of the PE $\delta_m$ on the PDA is more significant. Besides, the AN and the MC results exhibits a noticeable deviation when $\phi_m=\pi$ and $\tau={\pi}/2$ since the approximation expression of $\Gamma_{(3.5)}^\star$ utilizes a Taylor expansion at $x=0$. If the amplitude of PDA is large, then the approximation will be somewhat inaccurate when $\tau={\pi}/2$. Fig. \ref{Fig_Prop_35_Comparison} shows that $\Gamma$ of $a=2$ and $3$ are all small and they are not influenced by the PE $\delta_m$ very seriously. Moreover, $\Gamma_{(3.5)}^{\star}(\phi_{\mathsf{U}})$ performs best since \revise{$\phi_{\mathsf{U}}={\pi}/{2}+c$} offers the highest $\beta(\phi_m)$, but this strategy may lead to the mismatch on the phases of the cascaded and/or direct links.

\revise{\textbf{Remark I} \big(In Fig. \ref{Fig_Prop_35_Diff_phi_m}, why $\phi_m={\pi}/{2}$ has the smallest error when $\phi_m$ is close to $\pi/2$?\big): \textit{In Fig.\ref{Fig_Prop_35_Diff_phi_m}, the error is smallest when $\phi_m$ is close to $\pi/2$, because the PDA is less sensitive to the PE near this point. Specifically, in the PDA model, at $\phi_m \approx \pi/2$, the derivative $\partial\beta(\phi_m)/\partial\phi_m$ is small, resulting in milder changes in it when adding $\delta_m$. Thus, the Taylor expansion approximation based on small $\tau$ is more accurate, with reduced impact from higher-order terms, leading to the minimal error.}}

\textbf{Proposition 3.6} \big($\beta(\phi_m+\gamma_m)\exp(-\jmath\gamma_{m})$, \revise{$\iota=1$}\big): \textit{When $\iota=1$, the $m$-th PDA is $\beta(\phi_m+\gamma_m)$, and the PDA and the PSE all contain $\gamma_{m}\sim\mathcal{VM}(0,\kappa)$ where $\kappa \ge 1$, then the upper bounds of the \revise{RP} $\Gamma$ in \eqref{Gamma_Case_III} can be obtained as}

\begin{equation}
    \Gamma_{(3.6)} \ {\le} \  \Gamma_{(3.6)}^{\star}= (\eta_1+\eta_2)^2+\eta_3
    < \Gamma_{(3.6)}^{\star}(\phi_{\mathsf{U}}),
    \label{Prop_36}
\end{equation}

\noindent \revise{\textit{where $\Gamma_{(\mathrm{3.6})}=\big\vert\mathbb{E}\{\beta(\phi_m+\gamma_m)\exp(-\jmath\gamma_{m})\}\big\vert^2$, $ \Gamma_{(3.6)}^{\star}=\big\vert\mathbb{E}\{\bar{\beta}(\phi_m+\gamma_m)\exp(-\jmath\gamma_{m})\}\big\vert^2$, $\eta_1=\frac{1+b}{2}\rho$, $\eta_2=\frac{1-b}{4}\sin(\phi_m-c)(1+\bar{\rho})$, and $\eta_3=\frac{(1-b)^2}{16}\cos^2(\phi-c)(1-\bar{\rho})^2$. \revise{$\Gamma_{(\mathrm{3.6})}^{\star}(\phi_{\mathsf{U}})=\left(\eta_1+\frac{(1-b)(1+\bar{\rho})}{4}\right)^2$.} Note $\bar{\rho}$ is defined as}}
\revise{
\begin{align}
    \bar{\rho} =\begin{cases}
\frac{\kappa^2}{8}\left(1-\frac{\kappa^2}{6}+\frac{11\kappa^4}{384}\right)+\mathcal{O}(\kappa^6)&\text{for small $\kappa$}
\\
1-\frac{2}{\kappa}+\frac{1}{\kappa^2}-\frac{1}{4\kappa^3}+\mathcal{O}(\kappa^{-4})&\text{for large $\kappa$}.
\end{cases}
\label{bar_rho}
\end{align}}

\textit{Proof}: The proof is similar to that of Prop. 3.2 and is omitted here.

\begin{figure}[htbp]
	\centering
	\begin{subfigure}{0.24\textwidth}
		\centering
		\includegraphics[width=\textwidth]{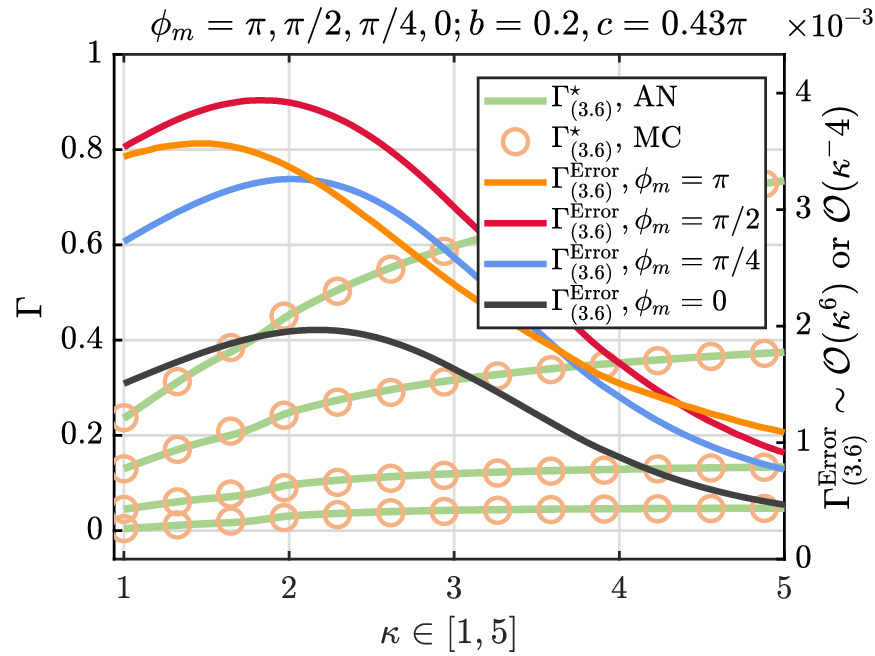}
		\caption{Validation of $\Gamma_{(3.6)}^\star$.}
        \label{Fig_Prop_36_Diff_phi_m}
	\end{subfigure}
	\centering
	\begin{subfigure}{0.24\textwidth}
		\centering
		\includegraphics[width=\textwidth]{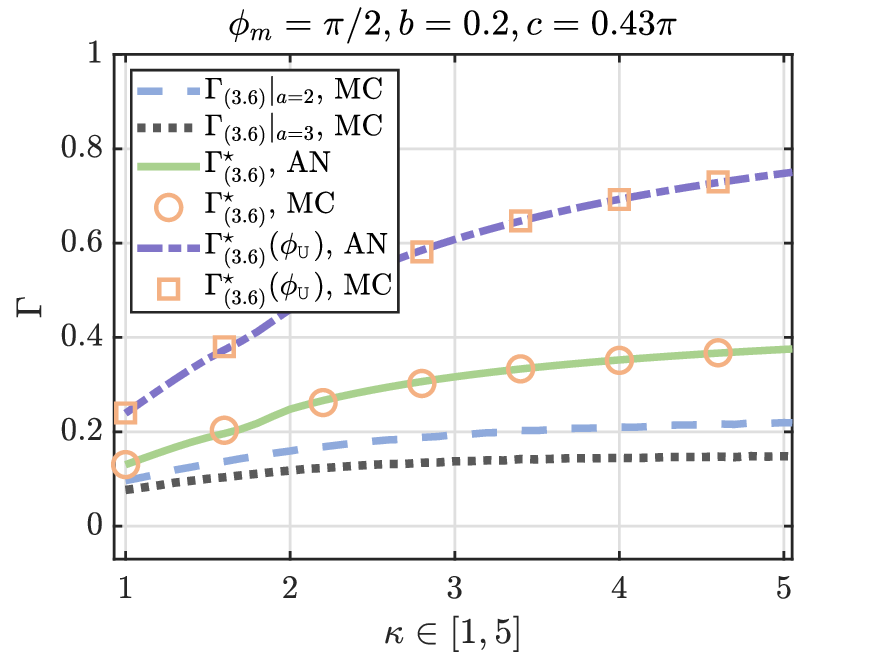}
		\caption{Validation of Prop. 3.6.}
		\label{Fig_Prop_36_Comparison}
	\end{subfigure}
	\caption{\textcolor{myColor}{Validation of Prop. 3.6.}}
	\label{Fig_Sec_II_Case_III_Prop_36}
\end{figure}

Note that \revise{$\Gamma_{(3.6)}^{\mathrm{Error}}=\big\vert\Gamma_{(3.6)}^{\mathrm{MC}}-\Gamma_{(3.6)}^{\mathrm{AN}}\big\vert\sim\mathcal{O}(\kappa^{6})$ or $\kappa^{-4}$. Fig. \ref{Fig_Prop_36_Diff_phi_m} shows the correctness of the approximation of $\Gamma_{(3.6)}^{\star}$. The error first increases and then decreases due to \eqref{bar_rho}.} Fig. \ref{Fig_Prop_36_Comparison} shows the correctness of \eqref{Prop_36}. 

\textbf{Proposition 3.7} \big($\beta(\phi_m+\bar{\delta}_m)\exp(-\jmath\delta_{m})$, \revise{$\iota=0$}\big): \textit{When $\iota=0$, the $m$-th PDA is $\beta_m(\phi_m+\bar{\delta}_{m})$, and the PDA and the PSE respectively contain $\bar{\delta}_{m}\sim\mathcal{UF}[-\tau, \tau]$ and $\delta_{m}\sim\mathcal{UF}[-\tau, \tau]$, where $\tau\in \big[0,{\pi}/{2}\big]$ and $\bar{\delta}$ and ${\delta}$ are i.i.d, then the upper bounds of the \revise{RP} $\Gamma$ in \eqref{Gamma_Case_III} can be obtained as}

\begin{equation}
    \Gamma_{(3.7)}\ {\le} \  \Gamma_{(3.7)}^{\star}=\eta_1^2\eta_2^2
    < \Gamma_{(3.7)}^{\star}(\phi_{\mathsf{U}}),
    \label{Prop_37}
\end{equation}

\noindent \textit{where $\Gamma_{(\mathrm{3.7})}=\big\vert\mathbb{E}\big\{\beta(\phi_m+\bar{\delta}_m)\exp(-\jmath\delta_{m})\big\}\big\vert^2$, $\Gamma_{(3.7)}^{\star}=\big\vert\mathbb{E}\big\{\bar{\beta}(\phi_m+\bar{\delta}_m)\exp(-\jmath\delta_{m})\big\}\big\vert^2$, $\eta_1=\frac{(1-b)}{2}\sin(\phi_m-c)\eta_2+\frac{(1+b)}{2}$, and $\eta_2=1-\frac{1}{6}\tau^2+\frac{1}{120}\tau^4+\mathcal{O}(\tau^6)$. \revise{$\Gamma_{(\mathrm{3.7})}^{\star}(\phi_{\mathsf{U}})=\left(\frac{1-b}{2}\eta_2+\frac{1+b}{2}\right)^2\eta_2^2$.}}

\textit{Proof}: The proof is the same as Propositions 3.1, 3.3 and 3.5 and it is omitted here. \ \ \ \ \ \ \ \ \ \ \ \ \ \  \ \ \ \ \ \ \ \  \ \ \ \ \ \ \ \  \ \ \  \ \ \ \ \ \ \ \  \ \ \ \ \ \ \ \  \ \ \ \ \ \ \  \ \ \ \ \ \ \ \ $\blacksquare$

\begin{figure}[htbp]
	\centering
	\begin{subfigure}{0.24\textwidth}
		\centering
		\includegraphics[width=\textwidth]{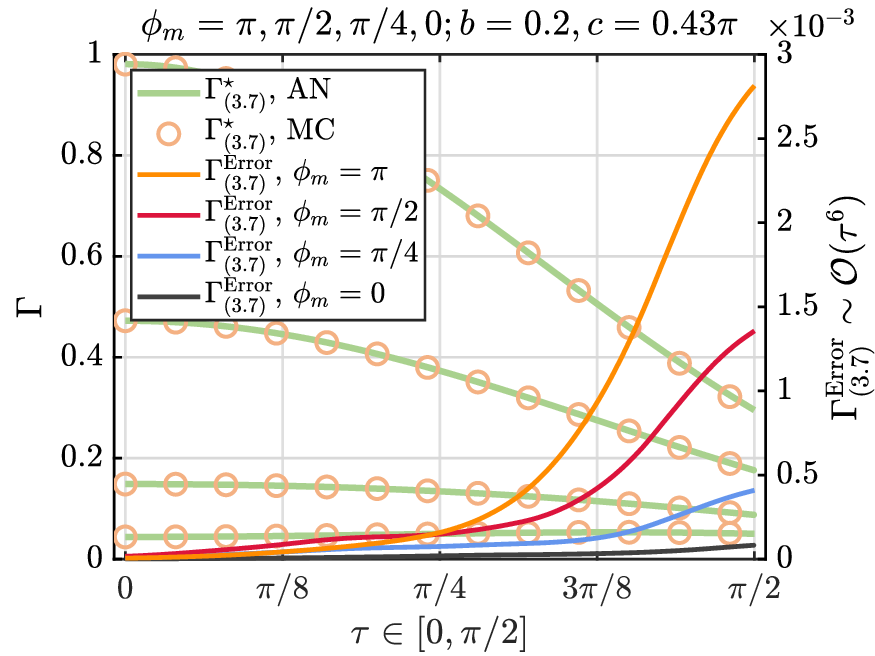}
		\caption{Validation of $\Gamma_{(3.7)}^\star$.}
        \label{Fig_Prop_37_Diff_phi_m}
	\end{subfigure}
	\centering
	\begin{subfigure}{0.24\textwidth}
		\centering
		\includegraphics[width=\textwidth]{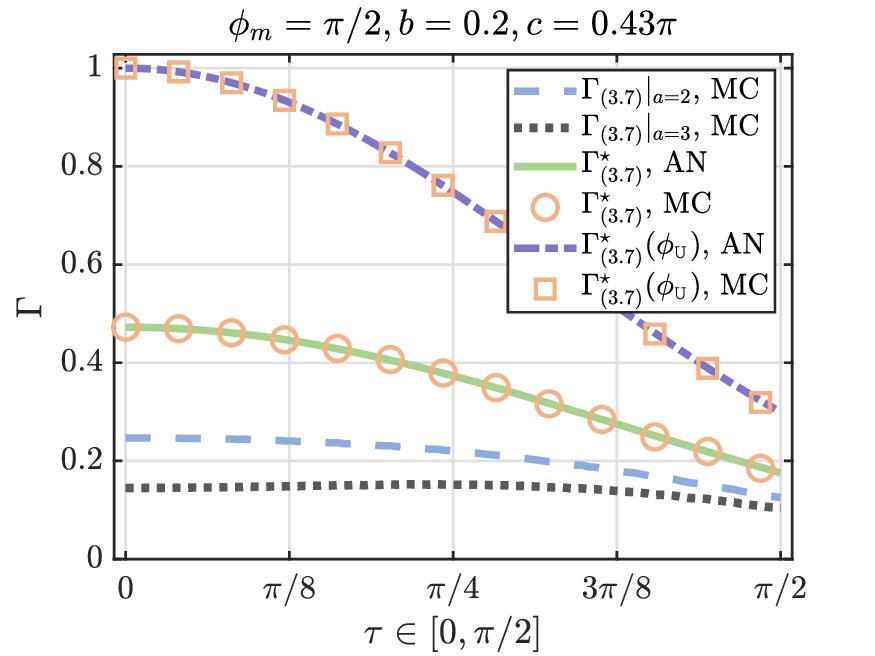}
		\caption{Validation of Prop. 3.7.}
		\label{Fig_Prop_37_Comparison}
	\end{subfigure}
	\caption{\textcolor{myColor}{Validation of Prop. 3.7.}}
	\label{Fig_Sec_II_Case_III_Prop_37}
\end{figure}

Note that \revise{ $\Gamma_{(3.7)}^{\mathrm{Error}}=\big\vert\Gamma_{(3.7)}^{\mathrm{MC}}-\Gamma_{(3.7)}^{\mathrm{AN}}\big\vert\sim\mathcal{O}(\tau^{6})$}. Fig. \ref{Fig_Prop_37_Diff_phi_m} shows the correctness of the approximation of $\Gamma_{(3.7)}^{\star}$. Fig. \ref{Fig_Prop_37_Comparison} shows the correctness of \eqref{Prop_37}.

\begin{figure}[htbp]
	\centering
	\begin{subfigure}{0.24\textwidth}
		\centering
		\includegraphics[width=\textwidth]{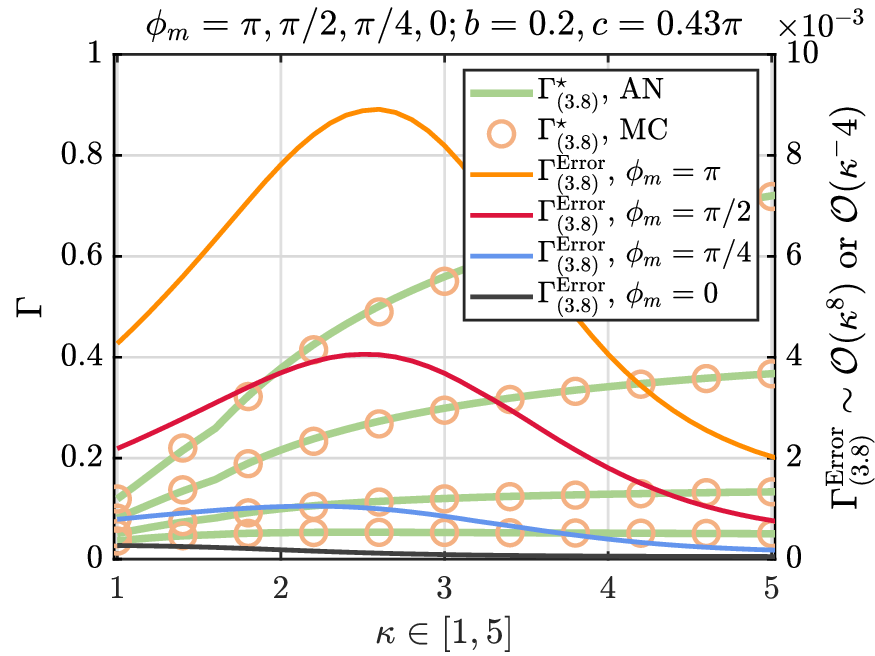}
		\caption{Validation of $\Gamma_{(3.8)}^\star$.}
        \label{Fig_Prop_38_Diff_phi_m}
	\end{subfigure}
	\centering
	\begin{subfigure}{0.24\textwidth}
		\centering
		\includegraphics[width=\textwidth]{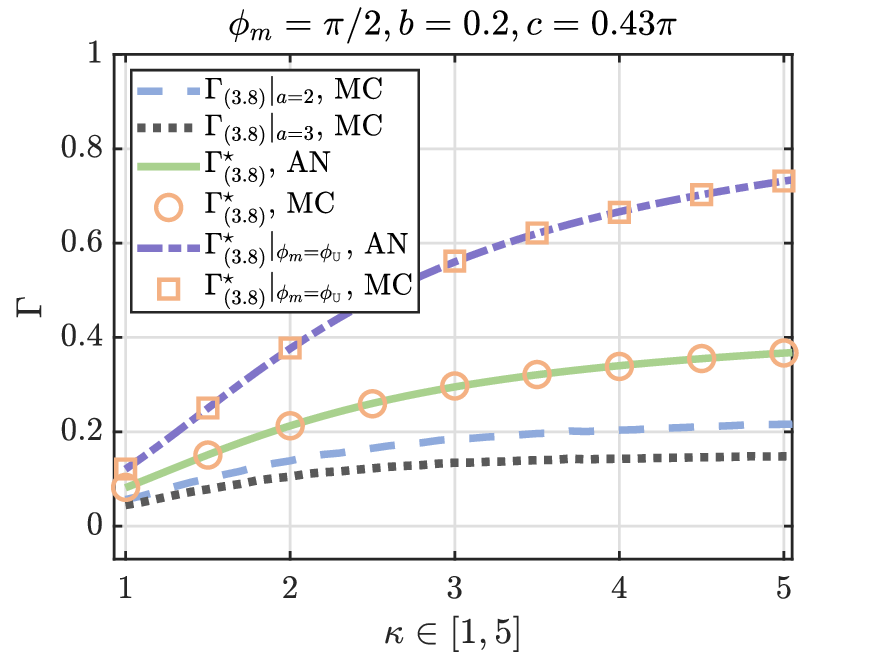}
		\caption{Validation of Prop. 3.8.}
		\label{Fig_Prop_38_Comparison}
	\end{subfigure}
	\caption{\textcolor{myColor}{Validation of Prop. 3.8.}}
	\label{Fig_Sec_II_Case_III_Prop_38}
\end{figure}

\textbf{Proposition 3.8} \big($\beta(\phi_m+\bar{\gamma}_m)\exp(-\jmath\gamma_{m})$, \revise{$\iota=0$}\big): \textit{When the $m$-th PDA is $\beta_m(\phi_m+\bar{\gamma}_{m})$, the PDA and the PSE respectively contain $\bar{\gamma}_{m}\sim\mathcal{VM}(0,\kappa)$ and $\gamma_{m}\sim\mathcal{VM}(0,\kappa)$ where $\kappa \ge 1$, and $\bar{\gamma}_m$ and ${\gamma}_m$ are i.i.d., then the upper bounds of the RP $\Gamma$ in \eqref{Gamma_Case_III} can be obtained as}

\begin{equation}
    \Gamma_{(3.8)}\ {\le} \  \Gamma_{(3.8)}^{\star}=\eta^2\rho^2
    < \Gamma_{(3.8)}^{\star}(\phi_{\mathsf{U}}),
    \label{Prop_38}
\end{equation}

\noindent \textit{where $\Gamma_{(\mathrm{3.8})}=\big\vert\mathbb{E}\big\{\beta(\phi_m+\bar{\gamma}_m)\exp(-\jmath\gamma_{m})\big\}\big\vert^2$, $\Gamma_{(3.8)}^{\star}=\big\vert\mathbb{E}\big\{\bar{\beta}(\phi_m+\bar{\gamma}_m)\exp(-\jmath\gamma_{m})\big\}\big\vert^2$, $\eta=\frac{(1-b)}{2}\sin(\phi_m-c)\rho+\frac{(1+b)}{2}$, \revise{and $\Gamma_{(3.8)}^{\star}(\phi_{\mathsf{U}})=\big(\frac{1-b}{2}\rho+\frac{1+b}{2}\big)^2\rho^2$.}}

\textit{Proof}: Consider Appendices \ref{Proof_Prop_31_32}, the proof is obvious.

Note that \revise{$\Gamma_{(3.8)}^{\mathrm{Error}}=\big\vert\Gamma_{(3.8)}^{\mathrm{MC}}-\Gamma_{(3.8)}^{\mathrm{AN}}\big\vert\sim\mathcal{O}(\kappa^{6})$ or $\mathcal{O}(\kappa^{-4}$)}. Fig. \ref{Fig_Prop_38_Diff_phi_m} shows the correctness of the approximation of $\Gamma_{(3.8)}^{\star}$. \revise{The error first increases and then decreases due to \eqref{rho}. Fig. \ref{Fig_Prop_38_Comparison} shows the correctness of \eqref{Prop_38}.}

\begin{figure}[htbp]
	\centering
	\begin{subfigure}{0.24\textwidth}
		\centering
		\includegraphics[width=\textwidth]{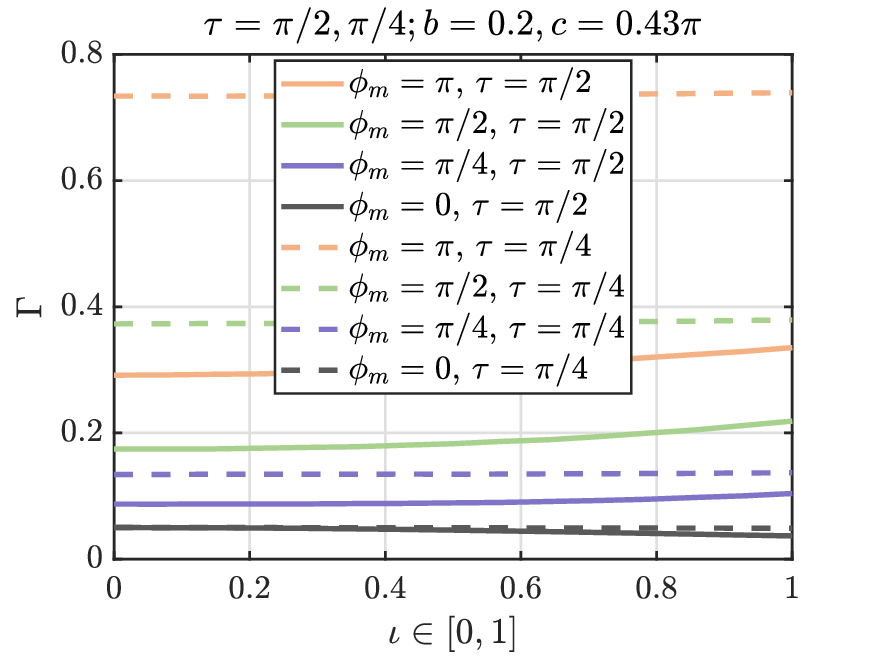}
		\caption{$\Gamma$ with $\mathcal{UF}$ RVs.}
        \label{Fig_iota_vs_Gamma_UF}
	\end{subfigure}
	\centering
	\begin{subfigure}{0.24\textwidth}
		\centering
		\includegraphics[width=\textwidth]{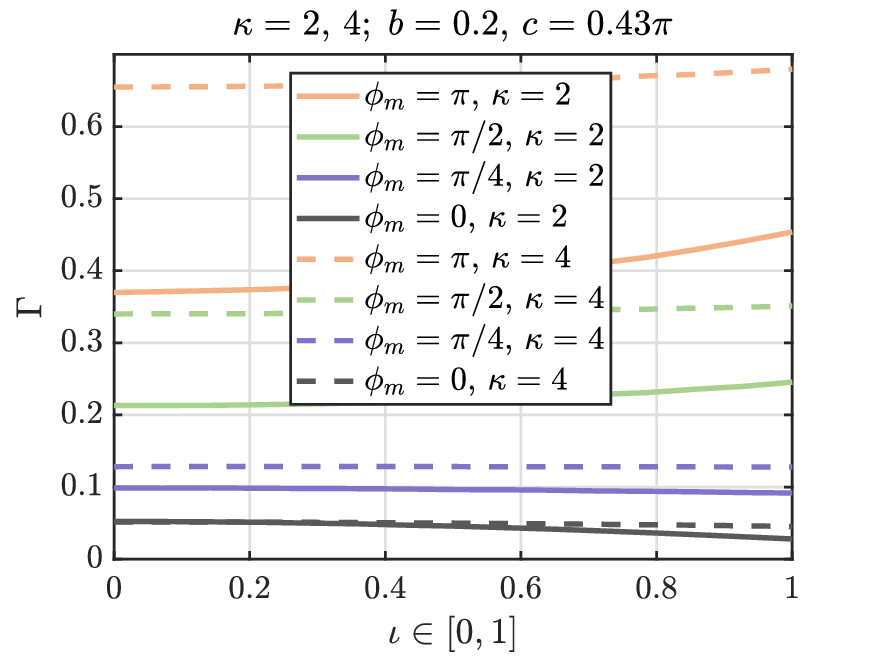}
		\caption{$\Gamma$ with $\mathcal{VM}$ RVs.}
		\label{Fig_iota_vs_Gamma_VM}
	\end{subfigure}
	\caption{\textcolor{myColor}{$\iota$ vs. $\Gamma$ considering $\mathcal{UF}$ and $\mathcal{VM}$ RVs.}}
	\label{Remark_iota_vs_Gamma}
\end{figure}

\revise{\textbf{Remark II} \big(If $\phi_m$ approaches to $0$, why $\Gamma$ decreases when $\iota$ increases?\big): \textit{As observed in Figs. \ref{Fig_iota_vs_Gamma_UF} and \ref{Fig_iota_vs_Gamma_VM}, the RP $\Gamma$ decreases as $\phi_m$ approaches $0$. For instance, $\Gamma$ is slightly lower at $\phi_m\approx 0$ than at $\phi_m\approx \pi/4$. This is because when $\phi_m\approx 0$, the system operates near the linear region of the phase response function $\beta(\cdot)$. As the correlation coefficient $\iota$ increases from $0$ toward $1$, the amplitude and phase noise become increasingly coupled. This coupling induces a negative covariance between the amplitude and the phase. We provide a strict proof of $\Gamma(0)|_{\iota=0}>\Gamma(0)|_{\iota=1}$ in Appendix \ref{Proof_Remark_II}. Fig. \ref{Fig_Remark_2_UF_VM} verify this remark through MC simulations of vector field distributions under $\mathcal{UF}$ and $\mathcal{VM}$ RVs, respectively. In particular, when $\phi_m=0$, length of the synthesized vector $|\mathsf{P}_{\iota=1}|<|\mathsf{P}_{\iota=0}|$ and the inequality is reversed when $\phi_m=\pi/2$.}}

\begin{figure*}[t]
\centering
\includegraphics[width=0.85\textwidth]{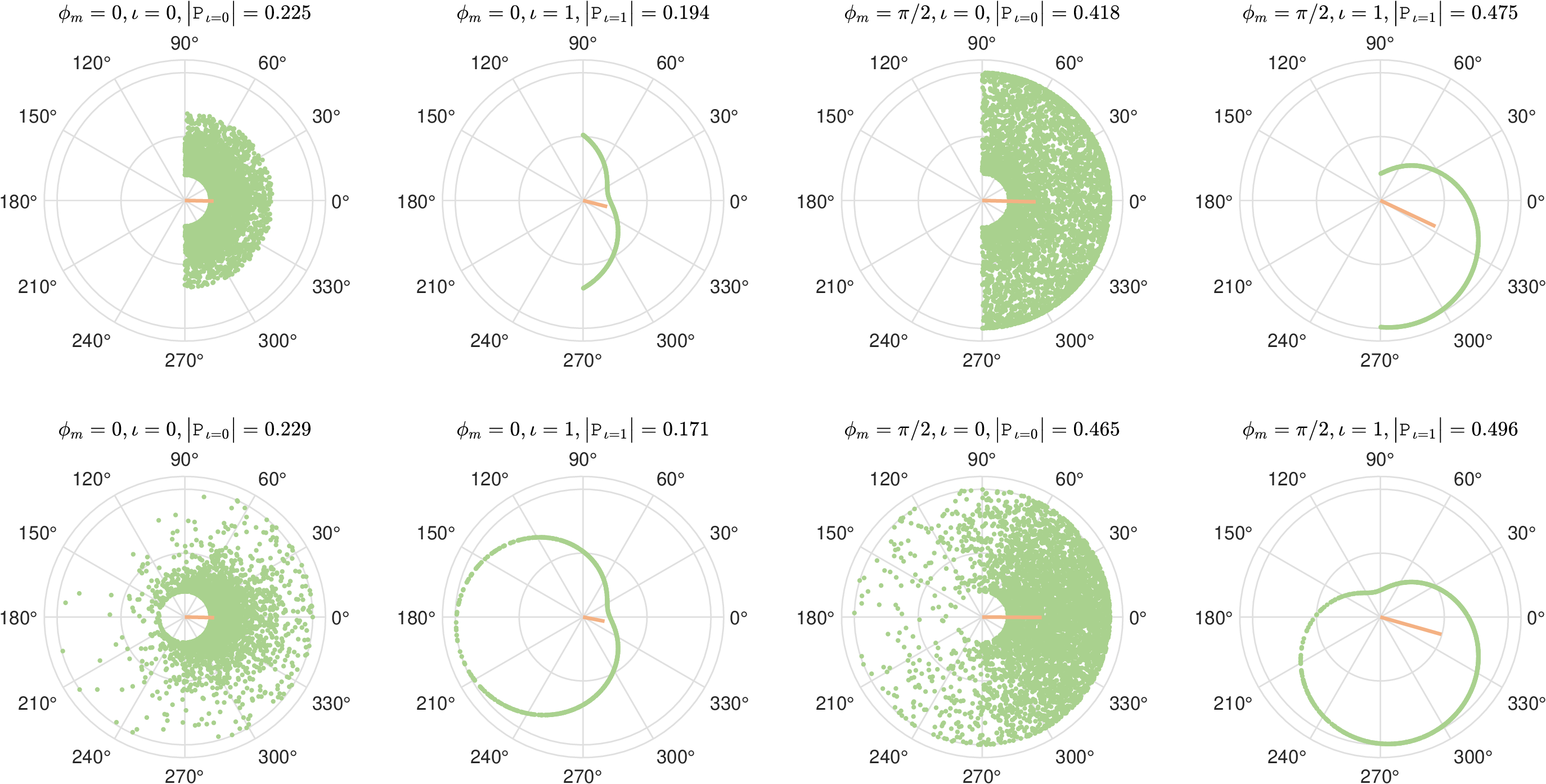}
\caption{\textcolor{myColor}{Vector field distribution with $\mathcal{UF}$ and $\mathcal{VM}$ RVs to verify Remark II. Note that $\mathsf{P}=\mathsf{P}(\bar{\Delta},\Delta)$ as Appendix \ref{Proof_Remark_II} defines.}}
\label{Fig_Remark_2_UF_VM}
\end{figure*}

\subsection{Case IV: When the PDA is with Both Errors}

When the PDA contains both $\bar{\Delta}_m$ and $\bar{\Theta}_m$ and the PSE also contains both $\Delta_m$ and $\Theta_m$, we have the RP $\Gamma$ as

\begin{equation}
    \Gamma = \underbrace{\boxed{\left\vert\mathbb{E}\big\{\beta(\phi_m+\bar{\Delta}_m+\bar{\Theta}_m)\exp(-\jmath(\Delta_m+\Theta_m)\big\}\right\vert^2}}_{\mathrm{Case \ IV}},
    \label{Gamma_Case_IV}
\end{equation}

\noindent where $\bar{\Delta}_m$ and $\Delta_m$ are $\mathcal{UF}$ RVs, $\bar{\Delta}_m=\Delta_m$ or they are i.i.d. $\bar{\Theta}_m$ and $\Theta_m$ are $\mathcal{VM}$ RVs, $\bar{\Theta}_m=\Theta_m$ or they are i.i.d. Therefore, the impact of noise on $\Gamma$ can be described by the following two propositions.

\textbf{Proposition 3.9} \big($\beta(\phi_m+\delta_m+\gamma_m)\exp(-\jmath(\delta_m+\gamma_{m}))$, the PDA and the PSE are all with $\delta_m$ and $\gamma_m$, $\iota=1$\big): \textit{When $\iota=1$, the $m$-th PDA is $\beta(\phi_m+\delta_m+\gamma_m)$, and the PDA and the PSE all contain $\delta_m\sim\mathcal{UF}[-\tau,\tau]$ where \revise{$\tau\in [0,{\pi}/{2}]$} and $\gamma_m\sim\mathcal{VM}(0,\kappa)$ where $\kappa \ge 1.6$, then the upper bounds of the \revise{RP} $\Gamma$ in \eqref{P_RIS_User} can be obtained as}

\begin{align}\label{Prop_3.9}
&\Gamma_{(\mathrm{3.9})}{\le} \  \Gamma_{(\mathrm{3.9})}^{\star}\\
=&\ \left(\frac{(1-b)\sin(\phi_m-c)}{2}(\eta_1-\eta_4)+\frac{1+b}{2}\eta_5\right)^2
\notag \\
+&\frac{(1-b)^2\cos^2(\phi_m-c)}{4}(\eta_2+\eta_3)^2
< \Gamma_{(\mathrm{3.9})}^{\star}(\phi_{\mathsf{U}}),
\end{align}

\noindent where $\Gamma_{(\mathrm{3.9})}=\big\vert\mathbb{E}\big\{\beta(\phi_m+\delta_m+\gamma_m)\exp(-\jmath(\delta_m+\gamma_{m}))\big\}\big\vert^2 $, $\Gamma_{(\mathrm{3.9})}^{\star}=\big\vert\mathbb{E}\big\{\bar{\beta}(\phi_m+\delta_m+\gamma_m)\exp(-\jmath(\delta_m+\gamma_{m}))\big\}\big\vert^2$, $\eta_1 = \big(1 - \frac{\tau^2}{3} + \frac{\tau^4}{15}-\frac{2\tau^6}{315}+\mathcal{O}(\tau^8)\big)\big(1 - \frac{1}{8\kappa} - \frac{1}{64\kappa^2} - \frac{1}{128\kappa^3}+\mathcal{O}(\kappa^{-4})\big)$, $\eta_2 = \big(\frac{\tau^2}{3} - \frac{\tau^4}{15}+\frac{2\tau^6}{315}+\mathcal{O}(\tau^8)\big)\big(1 - \frac{1}{8\kappa} - \frac{1}{64\kappa^2} - \frac{1}{128\kappa^3}+\mathcal{O}(\kappa^{-4})\big)$, $\eta_3 = \big(1 - \frac{\tau^2}{3} + \frac{\tau^4}{15}-\frac{2\tau^6}{315}+\mathcal{O}(\tau^8)\big)\big(\frac{1}{8\kappa} + \frac{1}{64\kappa^2} + \frac{1}{128\kappa^3}+\mathcal{O}(\kappa^{-4})\big)$, $\eta_4 =\big(\frac{\tau^2}{3} - \frac{\tau^4}{15}+\frac{2\tau^6}{315}+\mathcal{O}(\tau^8)\big) \big(\frac{1}{8\kappa} + \frac{1}{64\kappa^2} + \frac{1}{128\kappa^3}+\mathcal{O}(\kappa^{-4})\big)$, and $\eta_5 = \big(1-\frac{\tau^2}{6}+\frac{\tau^4}{120}+\mathcal{O}(\tau^6)\big)\big(1 - \frac{1}{2\kappa} - \frac{1}{8\kappa^2}-\frac{1}{8\kappa^3}+\mathcal{O}(\kappa^{-4})\big)$.

\textit{Proof}: Please see Appendix \ref{Proof_Prop_39}. \ \ \ \  \ \  \ \ \ \ \ \ \ \  \ \ \ \ \ \ \  \ \ \ \ \ \ \ \ $\blacksquare$

\begin{figure}[htbp]
	\centering
	\begin{subfigure}{0.24\textwidth}
		\centering
		\includegraphics[width=\textwidth]{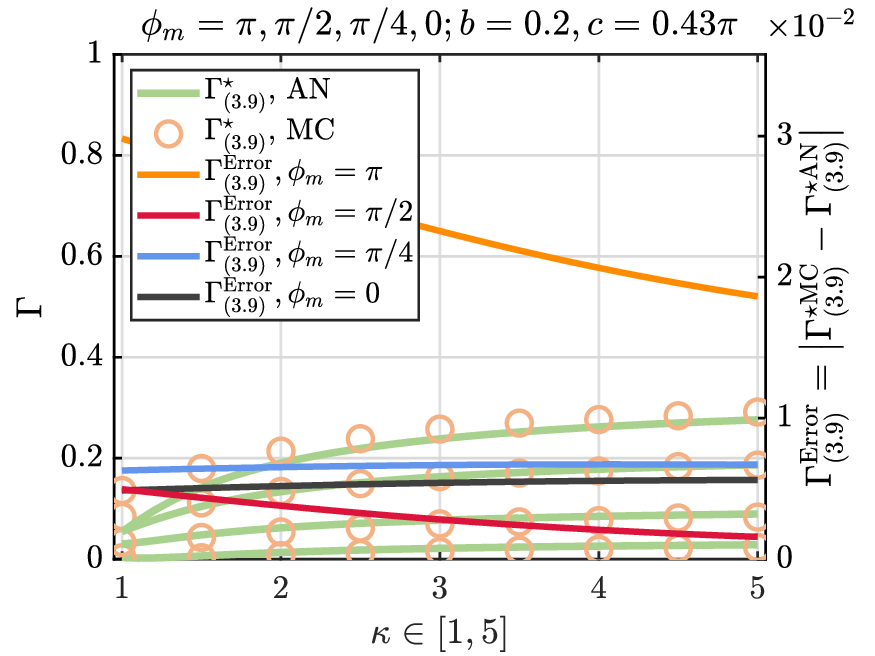}
		\caption{Validation of $\Gamma_{(3.9)}^\star$.}
        \label{Fig_Prop_39_diff_phi}
	\end{subfigure}
	\centering
	\begin{subfigure}{0.24\textwidth}
		\centering
		\includegraphics[width=\textwidth]{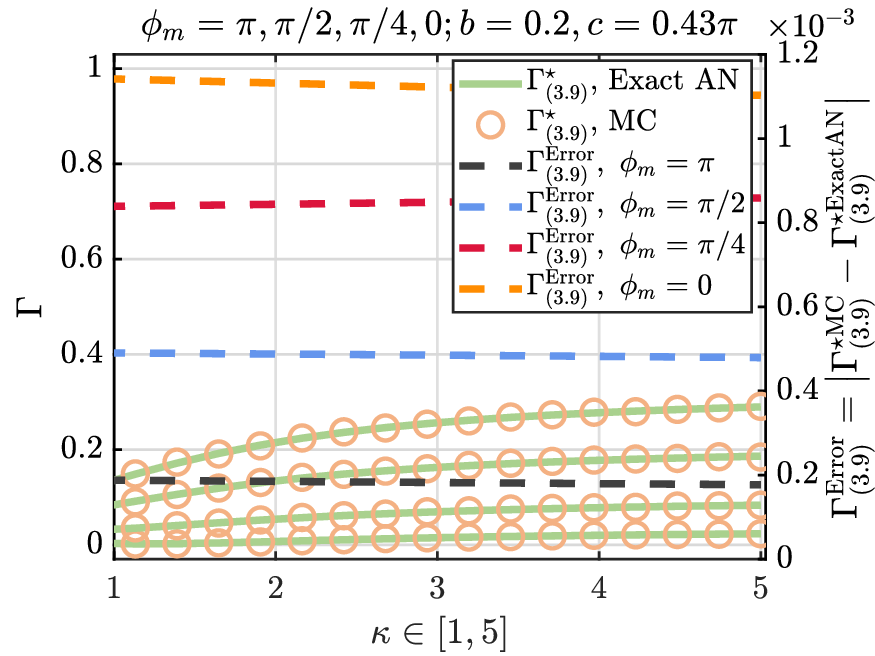}
		\caption{Exact expression of $\Gamma_{(3.9)}^\star$.}
		\label{Fig_Prop_39_diff_Exact}
	\end{subfigure}
	\caption{\textcolor{myColor}{Validation of Prop. 3.9.}}
	\label{Fig_Sec_II_Case_III_Prop_39}
\end{figure}

\revise{Fig. \ref{Fig_Prop_39_diff_phi} verifies the accuracy of the approximation for $\Gamma_{(3.9)}^{\star}$. Fig. \ref{Fig_Prop_39_diff_Exact} validates the exact expression of Prop. 3.9 by directly performing numerical integration of $\Gamma_{(3.9)}^{\star}$. Given that $\Gamma_{(3.9)}^{\mathrm{Error}}$ is on the order of $10^{-2}$, consider use direct numerical integration rather than an approximation}

\textbf{Proposition 3.10} \big($\beta(\phi_m+\bar{\delta}_m+\bar{\gamma}_m)\exp(-\jmath(\delta_m+\gamma_{m}))$, the PDA are with $\bar{\delta}_m$ and $\bar{\gamma}_m$, and the PSE are with $\delta_m$ and $\gamma_m$, $\iota=0$\big): \textit{When $\iota=0$, the $m$-th PDA is $\beta_m(\phi_m+\bar{\delta}_m+\bar{\gamma}_m)$, and it  contains $\bar{\delta}_m\sim\mathcal{UF}[-\tau,\tau]$ and $\bar{\gamma}_m\sim\mathcal{VM}(0,\kappa)$. The PSE contains $\delta_m\sim\mathcal{UF}[-\tau,\tau]$ and $\gamma_m\sim\mathcal{VM}(0,\kappa)$. Note that $\tau\in \big[0,{\pi}/{2}\big]$ and $\kappa\ge 1$. If $\bar{\delta}_m$ and $\delta_m$ are i.i.d., $\bar{\gamma}_m$ and $\gamma_m$ are i.i.d, then the upper bounds of the RP $\Gamma$ in \eqref{P_RIS_User} can be obtained as}

\begin{align}\label{Prop_3.10}
\Gamma_{(\mathrm{3.10})}\ {\le} \ \Gamma_{(\mathrm{3.10})}^{\star}\approx&\ \left(\frac{1-b}{2}\sin(\phi_m-c)\eta+\frac{1+b}{2}\right)^2\eta^2
\notag \\
< & \ \Gamma_{(\mathrm{3.10})}^{\star}(\phi_{\mathsf{U}}),
\end{align}

\noindent \textit{where $\Gamma_{(\mathrm{3.10})}=\big\vert\mathbb{E}\big\{\beta(\phi_m+\bar{\delta}_m+\bar{\gamma}_m)\exp(-\jmath(\delta_m+\gamma_{m}))\big\}\big\vert^2$, $\Gamma_{(\mathrm{3.10})}^{\star}=\big\vert\mathbb{E}\big\{\bar{\beta}(\phi_m+\bar{\delta}_m+\bar{\gamma}_m)\exp(-\jmath(\delta_m+\gamma_{m}))\big\}\big\vert^2$, and $\eta=\big(1-\frac{1}{6}\tau^2+\frac{1}{120}\tau^4\big)\rho$. \revise{$\Gamma_{(\mathrm{3.10})}^{\star}(\phi_{\mathsf{U}})=\left(\frac{1-b}{2}\eta+\frac{1+b}{2}\right)^2\eta^2$.}}

\textit{Proof}: Please see Appendix \ref{Proof_Prop_310}. \ \ \ \  \ \  \ \ \ \ \ \ \ \  \ \ \ \ \ \ \  \ \ \ \ \ \ \ \ $\blacksquare$

\begin{figure}[htbp]
	\centering
	\begin{subfigure}{0.24\textwidth}
		\centering
		\includegraphics[width=\textwidth]{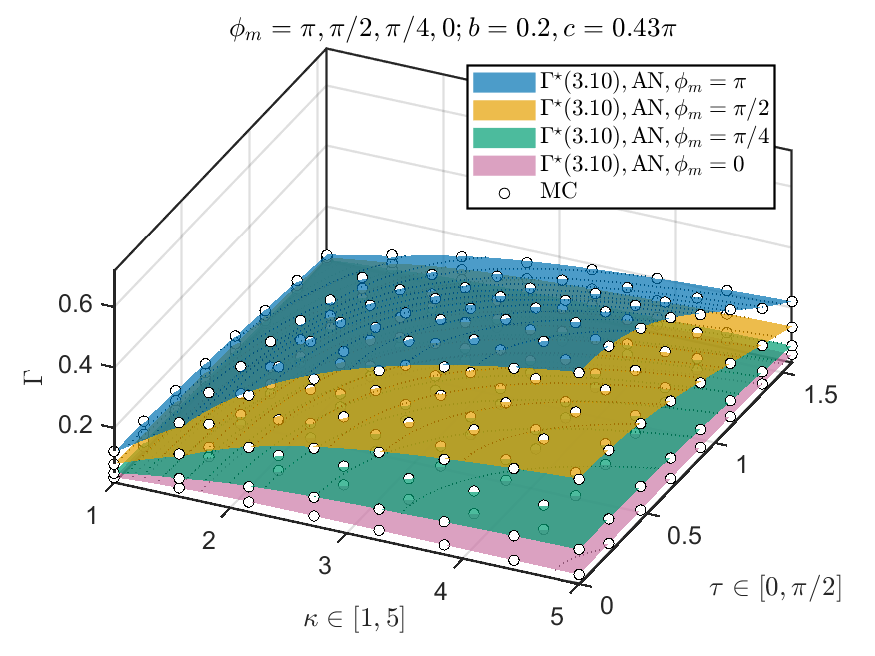}
		\caption{Validation of $\Gamma_{(3.10)}^\star$.}
        \label{Fig_Prop_310_Diff_phi_m}
	\end{subfigure}
	\centering
	\begin{subfigure}{0.24\textwidth}
		\centering
		\includegraphics[width=\textwidth]{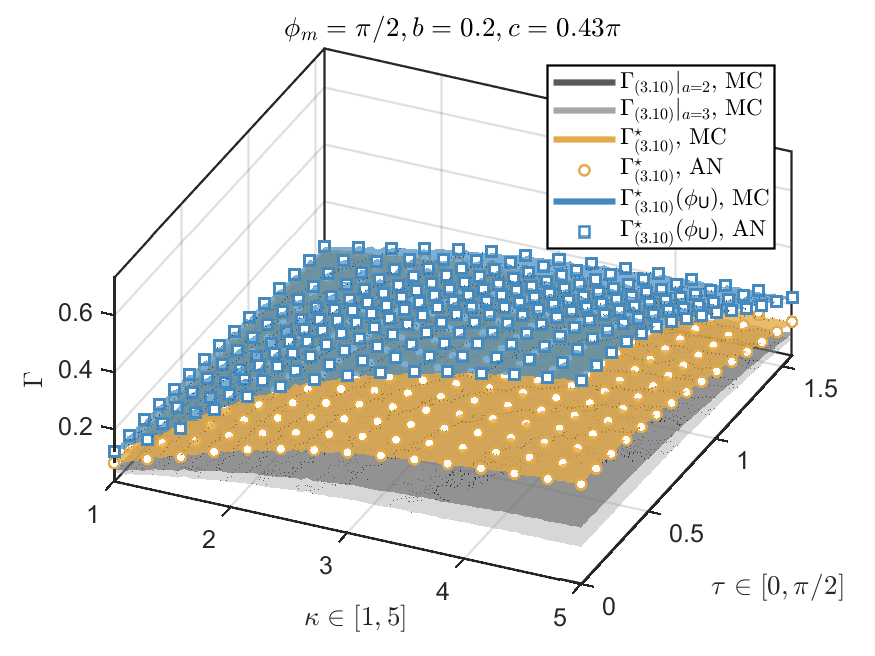}
		\caption{Validation of Prop. 3.10.}
		\label{/Fig_Prop_310_Comparison}
	\end{subfigure}
	\caption{\textcolor{myColor}{Validation of Prop. 3.10.}}
	\label{Fig_Sec_II_Case_III_Prop_310}
\end{figure}

\begin{figure}[htbp]
	\centering
	\begin{subfigure}{0.24\textwidth}
		\centering
		\includegraphics[width=\textwidth]{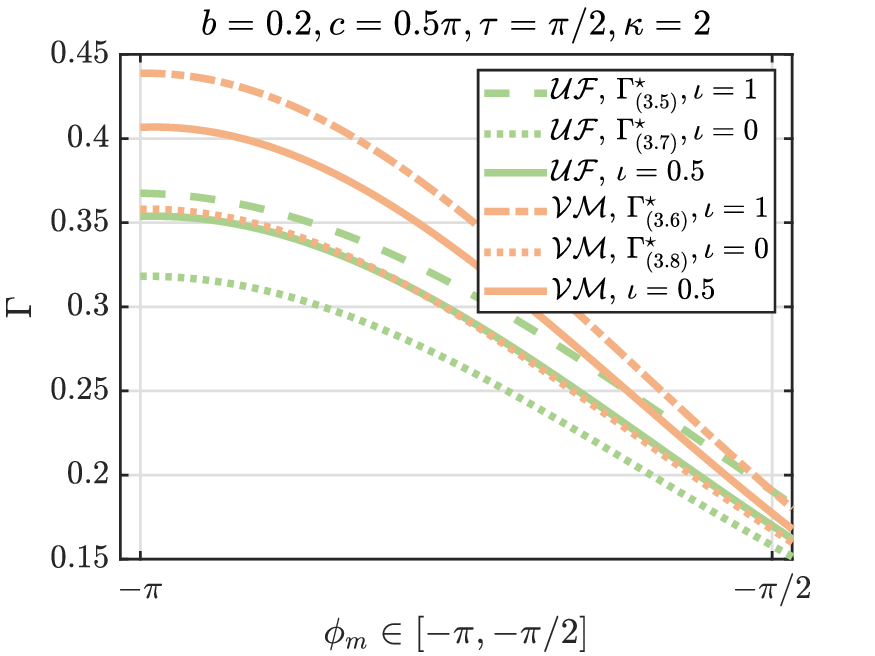}
		\caption{$\phi_m\in[-\pi,-\pi/2]$.}
        \label{Fig_iota_Prop_35373638_m_pi_m_pi2}
	\end{subfigure}
	\centering
	\begin{subfigure}{0.24\textwidth}
		\centering
		\includegraphics[width=\textwidth]{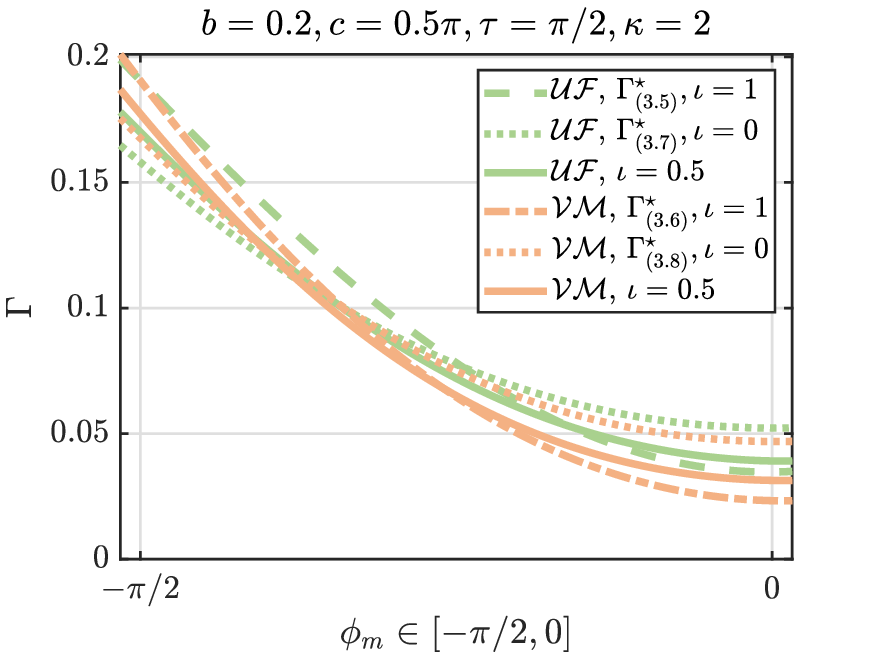}
		\caption{$\phi_m\in[-\pi/2,0]$.}
		\label{Fig_iota_Prop_35373638_m_pi2_0}
	\end{subfigure}
	\caption{\textcolor{myColor}{RP comparison under different $\iota$.}}
	\label{Fig_compare_Prop}
\end{figure}

\revise{Figs. \ref{Fig_Prop_310_Diff_phi_m} and \ref{/Fig_Prop_310_Comparison} present 3D plots of the MC and AN results for Prop. 3.10. Figs. \ref{Fig_iota_Prop_35373638_m_pi_m_pi2} and \ref{Fig_iota_Prop_35373638_m_pi2_0} illustrate that as the RIS phase shifts approach the ends of their range, the RP under i.i.d. PEs in both the PSE and PDA cases becomes smaller than that under identical PEs, whereas this relationship reverses when the phase shifts are near the middle of their range. Therefore, when the RIS is new and its components are highly correlated (i.e., $\iota=1$), the phase shift set $\{\phi_m\}_{m=1}^M$ should be configured near the ends of their range to achieve a higher $\Gamma$. Conversely, after long-term use with reduced correlation (i.e., $\iota=0$), $\{\phi_m\}_{m=1}^M$ should be set close to the middle of the range.}

\section{Channel Modeling and Spectral Efficiency Analysis}\label{Sec4}

In this section, we first present a LoS channel model of the NF RIS-aided communication considering the PDA, the PSE, and the PE. Subsequently, a tight closed-form SE upper bound for the model is derived.

\subsection{Line-of-Sight Channel Near-Field Modeling}

Consider an RIS-aided system comprising an AP, a user, and an RIS. The transceivers both equip single isotropic antennas. The direct link is blocked, and the RIS with $M$ pixels is a planar surface on a rectangular grid spaced $d_x$ and $d_y$ apart in the $xoy$-plane of a three-dimensional (3D) Cartesian coordinate system where $o$ denotes the origin point, and the area and the geometric center of the RIS are respectively $d_x\sqrt{M}\times d_y\sqrt{M}$ and $[0, h_y, 0]^\mathsf{T}$, where $h_y$ is the height of the RIS. Furthermore, it is assumed that the hardware structure and area of each pixel are the same, i.e., the area is $d_x d_y$. The position of the AP and the user are $D_{\mathrm{AP}}=[x_{\mathrm{AP}}, y_{\mathrm{AP}},z_{\mathrm{AP}}]^\mathsf{T}$ and $D_{\mathrm{User}}=[x_{\mathrm{User}}, y_{\mathrm{User}},z_{\mathrm{User}}]^\mathsf{T}$, respectively. The coordinate of the RIS is $D_{m_1 m_2}=[\Psi(m_1, d_x, \sqrt{M}), \Psi(m_2, d_y, \sqrt{M})+h_y, 0]^\mathsf{T}$, where $m_1$ and $m_2 \in \big\{ ((\sqrt{M}+1)\ \mathsf{mod}\ 2)-\big\lfloor\frac{\sqrt{M}}{2}\big\rfloor,\dots,\big\lfloor\frac{\sqrt{M}}{2}\big\rfloor\big\}$, $\Psi(m_1, d_x, \sqrt{M})=d_x(m_1-0.5((\sqrt{M}+1)\ \mathsf{mod}\ 2))$, and $\Psi(m_2, d_y, \sqrt{M})=d_y(m_2-0.5((\sqrt{M}+1)\ \mathsf{mod}\ 2))$. Moreover, by using an applicable mapping function $\mathcal{M}(m_1, m_2)=m$ where $m=1,\dots,M$, we can replace the notation $D_{m_1 m_2}$ with $D_m=[x_m, y_m, 0]^\mathsf{T}$ to guarantee a more concise expression. Additionally, for the AP and the user, it is assumed that the \textit{visibility region} \cite{carvalho20non-stat} of the RIS always consists of total $M$ pixels, i.e., the transceiver can observe the entire RIS.

Therefore, the distances between the AP and the $m$-th pixel and between the $m$-th pixel and the user can be respectively obtained as $d_{\mathrm{AP}\rightarrow m} = \Vert D_{m} - D_{\mathrm{AP}} \Vert$ and $d_{m\rightarrow\mathrm{User}} = \Vert D_{\mathrm{User}} - D_{m} \Vert$. Accordingly, the time delays are $\varpi_{\mathrm{AP}\rightarrow m} = d_{\mathrm{AP}\rightarrow m}/\nu$ and $\varpi_{\mathrm{m}\rightarrow \mathrm{User}} = d_{\mathrm{m}\rightarrow \mathrm{User}}/\nu$, respectively, where $\nu$ denotes the speed of light. Using \textit{Friis transmission formula} \cite{friis1946IRE}, for the $m$-th AP-to-pixel path $g_m$, we have\footnote{Although spatial correlations and polarization effects can adversely impact system performance\cite{emil23elaa}, these factors are primarily associated with the interval between two pixels and the RIS position\cite{emil2020near}. As such, they are not directly related to the PE, the PSE, or the PDA, which are the focus of this study. Consequently, this paper does not delve into the implications of the spatial correlation and the polarization.}

\begin{equation}\label{AP-to-pixel_path}
    \frac{P_{\mathrm{AP}\rightarrow m}}{P_{\mathrm{AP}}}=\bigg(\frac{\lambda}{4\pi d_{\mathrm{AP}\rightarrow m}}\bigg)^2
    G_{\mathrm{AP}}
    G_{\mathrm{AP}\rightarrow m},
\end{equation}

\noindent where $P_{\mathrm{AP}\rightarrow m}$ is the received power of the $m$-th RIS pixel, which is part of the power that is emitted from the AP, i.e., $P_{\mathrm{AP}}$. Besides, $G_{\mathrm{AP}}$ and $G_{\mathrm{AP}\rightarrow m}$ are the AP antenna gain and the $m$-th pixel gain from the direction of the AP. For $G_{\mathrm{AP}\rightarrow m}$, we consider a generic directional gain pattern\cite{balanis16book}, i.e., $\vartheta_{\mathrm{AP}\rightarrow m}(\theta_{\mathrm{AP}\rightarrow m},\psi_{\mathrm{AP}\rightarrow m})=\bar{r}\cos^{2r}(\theta_{\mathrm{AP}\rightarrow m})$ if $\theta_{\mathrm{AP}\rightarrow m}\in[0,{\pi}/{2})$ and $\psi_{\mathrm{AP}\rightarrow m}\in[0,2\pi]$, or $0$ otherwise. Note that $\theta_{\mathrm{AP}\rightarrow m}=\arccos\left(\frac{z_{\mathrm{AP}}}{d_{\mathrm{AP}\rightarrow m}}\right)\in[0,{\pi}/{2})$ and $\psi_{\mathrm{AP}\rightarrow m}=\arctan\left(\frac{y_{\mathrm{AP}}-y_m}{x_{\mathrm{AP}}-x_m}\right)\in[0,2\pi]$. Note that the directional gain pattern should be considered in the NF scenario\cite{feng2024near}, thus each pixel has its own effective aperture. Consider the law of power conservation\cite{emil2024introduction}, we have

\begin{align}
    &\int_0^{2\pi}\partial\psi_{\mathrm{AP}\rightarrow m}\int_0^{\frac{\pi}{2}}\bar{r}\cos^{2r}(\theta_{\mathrm{AP}\rightarrow m})\sin(\theta_{\mathrm{AP}\rightarrow m})
    \notag \\
    =\ &\bar{r}2\pi\left[-\frac{\cos^{(2r+1)}(\theta_{\mathrm{AP}\rightarrow m})}{2r+1}\right]_0^{\frac{\pi}{2}}
    =\ \bar{r}\frac{2\pi}{2r+1} = 4\pi,
\end{align}

\noindent thus $\bar{r}=2(2r+1)$. In this paper, we assume cosine pattern, i.e., \textit{projected aperture}\cite{feng2024near}, thus $r=\frac{1}{2}$ and $\bar{r}=4$. Consequently, $\vartheta_{\mathrm{AP}\rightarrow m}(\theta_{\mathrm{AP}\rightarrow m},\psi_{\mathrm{AP}\rightarrow m})=4\cos(\theta_{\mathrm{AP}\rightarrow m})$. Similarly, for the $m$-th pixel-to-user path $h_m$, we have

\begin{equation}\label{pixel-to-user_path}
    \frac{P_{m \rightarrow \mathrm{User}}}{P_{m}}=\left(\frac{\lambda}{4\pi d_{m \rightarrow \mathrm{User}}}\right)^2
    G_{\mathrm{User}}
    G_{m \rightarrow \mathrm{User}},
\end{equation}

\noindent where $P_{m \rightarrow \mathrm{User}}$ is the received power of the user, which is part of the power that is emitted from the $m$-th RIS pixel, i.e., $P_{m}$. It is worth noting that the pixel may contain the PSE, the PDA, and the PEs, thus $P_{\mathrm{AP}\rightarrow m} \geq P_m$. Besides,  $G_{\mathrm{User}}$ and $G_{m \rightarrow \mathrm{User}}$ are respectively the user antenna gain and the $m$-th pixel gain from the direction of the user, and $\vartheta_{m \rightarrow \mathrm{User}}(\theta_{m \rightarrow \mathrm{User}},\psi_{m \rightarrow \mathrm{User}})=4\cos(\theta_{m \rightarrow \mathrm{User}})$ where $\theta_{m \rightarrow \mathrm{User}}=\arccos\left(\frac{z_{\mathrm{User}}}{d_{m\rightarrow \mathrm{User}}}\right)\in[0,\frac{\pi}{2})$ and $\psi_{m \rightarrow \mathrm{User}}=\arctan\left(\frac{y_{\mathrm{User}}-y_m}{x_{\mathrm{User}}-x_m}\right)\in[0,2\pi]$. Besides, based on Fig. \ref{Fig_Cascaded_Path}, we have $\Gamma_m=\frac{P_m}{P_{\mathrm{AP}\rightarrow m}}$, where $\Gamma_m$ denotes the RP for the $m$-th pixel with the PSE, the PDA, and the PEs.

\begin{figure*}[!b] 
\centering
\begin{minipage}{0.95\textwidth}
\noindent\rule{\linewidth}{0.6pt}
\begin{equation}
\label{P_m_User_to_P_AP}
    \frac{P_{m \rightarrow \mathrm{User}}}{P_{\mathrm{AP}}} = \overbrace{\frac{P_{\mathrm{AP}\rightarrow m}}{P_{\mathrm{AP}}}}^{|g_m|^2}
    \overbrace{\frac{P_m}{P_{\mathrm{AP}\rightarrow m}}}^{\Gamma_m}
    \overbrace{\frac{P_{m \rightarrow \mathrm{User}}}{P_{m}}}^{|h_m|^2}
    =\underbrace{\bigg(\frac{\lambda}{4\pi d_{\mathrm{AP}\rightarrow m}}\bigg)^2
    G_{\mathrm{AP}}
    G_{\mathrm{AP}\rightarrow m}}_{A^2_{\mathrm{AP}\rightarrow m}}
    \Gamma_m 
    \underbrace{\bigg(\frac{\lambda}{4\pi d_{m \rightarrow \mathrm{User}}}\bigg)^2
    G_{\mathrm{User}}
    G_{m \rightarrow \mathrm{User}}}_{A^2_{m \rightarrow \mathrm{User}}},
\end{equation}
\end{minipage}
\end{figure*}

Therefore, the $m$-th end-to-end power gain from the AP to the user can be obtained as \eqref{P_m_User_to_P_AP} in the next page, where $G_{\mathrm{AP}\rightarrow m}=\frac{4\pi}{\lambda^2}d_xd_y\vartheta_{\mathrm{AP}\rightarrow m}=\frac{16\pi}{\lambda^2}d_xd_y\cos(\theta_{\mathrm{AP}\rightarrow m})$ and $G_{m \rightarrow \mathrm{User}}=\frac{4\pi}{\lambda^2}d_xd_y\vartheta_{m \rightarrow \mathrm{User}}=\frac{16\pi}{\lambda^2}d_xd_y\cos(\theta_{m \rightarrow \mathrm{User}})$. Since the antennas in the transceiver are isotropic, then $G_{\mathrm{AP}}=G_{\mathrm{User}}=1$. Thus the $m$-th LoS channel of the AP-to-pixel and the pixel-to-user paths are respectively $g_{m} = A_{\mathrm{AP}\rightarrow m}\exp(-\jmath 2\pi f_c \varpi_{\mathrm{AP}\rightarrow m})$ and $h_{m} = A_{m \rightarrow \mathrm{User}}\exp(-\jmath 2\pi f_c \varpi_{m \rightarrow \mathrm{User}})$, where $f_c$ is carrier frequency, 

\begin{equation}
    A_{\mathrm{AP}\rightarrow m} = \frac{\lambda}{4\pi d_{\mathrm{AP}\rightarrow m}}\sqrt{G_{\mathrm{AP}\rightarrow m}},
    \label{A_AP_m}
\end{equation}

\noindent and 

\begin{equation}
A_{m \rightarrow \mathrm{User}}=\frac{\lambda}{4\pi d_{m \rightarrow \mathrm{User}}}\sqrt{G_{m \rightarrow \mathrm{User}}}.
    \label{A_m_User}
\end{equation}

Therefore, for the $m$-th channel, if the model is without any PEs, the PSE, or the PDA, the baseband equivalent of the received passband signal can be obtained as $g_m h_m \exp(-\jmath\phi_m)$ where $\phi_m=-2\pi f_c(\varpi_{\mathrm{AP}\rightarrow m}+\varpi_{m \rightarrow \mathrm{User}})+2k\pi$ and $k$ is a non-negative integer. Consider the RIS with $M$ pixels, and the PSE and the PDA coexist, then we have a total baseband equivalent channel as

\begin{align}\label{y}
    y  =  hs\sqrt{P_{\mathrm{AP}}} + \omega,
\end{align}

\noindent where $h=\sum_{m}g_{m}\beta(\phi_m+\bar{\Delta}_{m})\exp(-\jmath(\phi_m+\Delta_{m}))h_{m}$, $\bar{\Delta}_{m}$ and $\Delta_{m}$ are the PEs, $s$ is a unit-power signal symbol, and $\omega\sim\mathcal{CN}(0, \sigma^2)$ denotes additive white Gaussian noise with the variance $\sigma^2$, then the SE at the user side is given by

\begin{equation}
\mathsf{SE} = \mathbb{E}\bigg\{\log_2 \bigg(1+\frac{P_{\mathrm{AP}} \vert h\vert^2}{\sigma^2}\bigg)\bigg\}.
\end{equation}

\revise{\textbf{Remark III}: Note that for the RIS with $M$ pixels, $\Gamma_m$ generally differ because each scatterer exhibits distinct characteristics. Consequently, the term $\beta(\phi_m+\bar{\Delta}_{m})$ of $h$ in \eqref{y} cannot be factored outside the summation, and a closed-form expression for $\Gamma_m$ is generally unavailable. However, based on the last step of \eqref{P_RIS_User}, when the number of pixels is sufficiently large (e.g., larger than $200$), the RP obtained in the previous section can be used to evaluate the SE. Accordingly, in the next section, we derive the SE expression together with its lower and upper bounds.}

\subsection{Spectral Efficiency Analysis}\label{Sec4_B}

It can be obtained that\cite{emil23elaa}

\begin{align}
\mathsf{SE} = \mathbb{E}\bigg\{\log_2 \bigg(1+\frac{P_{\mathrm{AP}}\vert h\vert^2}{\sigma^2}\bigg)\bigg\}
\approx\log_2 \bigg(1+\frac{P_{\mathrm{AP}}\mathbb{E}\{\vert h\vert^2\}}{\sigma^2}\bigg).
\end{align}

\noindent Based on \eqref{beta(phi)_Approx_PDA} and recall $\phi_{\mathsf{U}}={\pi}/{2}+c$ and \revise{$\phi_{\mathsf{L}}=-{\pi}/{2}+c$}, we have $\mathsf{L} < \mathsf{H} < \mathsf{U}$,
where 

\begin{align}
\mathsf{L} =\ &  \mathbb{E}\bigg\{\bigg\vert \sum_{m=1}^M g_{m}\beta(\phi_{\mathsf{L}}+\Delta_{m})\exp(-\jmath(\phi_{m}+\Delta_{m}))h_{m}\bigg\vert^2\bigg\}
\notag \\
\approx\ & \bigg( \sum_{m=1}^M A_{\mathrm{AP}\rightarrow m}
\sqrt{\Gamma_{(\cdot)}(\phi_\mathsf{L})}
A_{m \rightarrow \mathrm{User}}\bigg)^2
\notag \\
=\ &\Gamma_{(\cdot)}(\phi_\mathsf{L})\bigg( \sum_{m=1}^M A_{\mathrm{AP}\rightarrow m}
A_{m \rightarrow \mathrm{User}}\bigg)^2, 
\label{L}
\end{align}

\begin{align}
\mathsf{H} =\ &  \mathbb{E}\bigg\{\bigg\vert \sum_{m=1}^M g_{m}\beta(\phi_{m}+\Delta_{m})\exp(-\jmath(\phi_{m}+\Delta_{m}))h_{m}\bigg\vert^2\bigg\}
\notag \\
\approx\ & \bigg( \sum_{m=1}^M A_{\mathrm{AP}\rightarrow m}
\sqrt{\Gamma_{(\cdot)}(\phi_m)}
A_{m \rightarrow \mathrm{User}}\bigg)^2,
\end{align}

\noindent and 

\begin{align}
\mathsf{U} =\ &  \mathbb{E}\bigg\{\bigg\vert \sum_{m=1}^M g_{m}\beta(\phi_{\mathsf{U}}+\Delta_{m})\exp(-\jmath(\phi_{m}+\Delta_{m}))h_{m}\bigg\vert^2\bigg\}
\notag \\
\approx\ & \bigg( \sum_{m=1}^M A_{\mathrm{AP}\rightarrow m}
\sqrt{\Gamma_{(\cdot)}(\phi_\mathsf{U})}
A_{m \rightarrow \mathrm{User}}\bigg)^2
\notag \\
=\ &\Gamma_{(\cdot)}(\phi_\mathsf{U})\bigg( \sum_{m=1}^M A_{\mathrm{AP}\rightarrow m}
A_{m \rightarrow \mathrm{User}}\bigg)^2.
\label{U}
\end{align}

\begin{figure*}[!b] 
\centering
\begin{minipage}{0.95\textwidth}
\noindent\rule{\linewidth}{0.6pt}
\begin{equation}
\label{non-ideal_gain}
\bigg| \sum_{m=1}^M A_{\mathrm{AP}\rightarrow m}A_{m \rightarrow \mathrm{User}}\bigg|^2
   =\frac{z_{\mathrm{AP}}z_{\mathrm{User}}}{\pi^2}
   \bigg( \sum_{m=1}^M \frac{d_x d_y}{(d_{\mathrm{AP}\rightarrow m}d_{m \rightarrow \mathrm{User}})^{\frac{3}{2}}}\bigg)^2
   =\frac{d_x^2 d_y^2 z_{\mathrm{AP}}z_{\mathrm{User}}}{\pi^2}
   \bigg( \sum_{m=1}^M \frac{1}{(d_{\mathrm{AP}\rightarrow m}d_{m \rightarrow \mathrm{User}})^{\frac{3}{2}}}\bigg)^2
\end{equation}
\begin{equation}
\label{non-ideal_gain_v2}
\bigg| \sum_{m=1}^M A_{\mathrm{AP}\rightarrow m}A_{m \rightarrow \mathrm{User}}\bigg|^2\le\ \frac{d_x^2 d_y^2 z_{\mathrm{AP}}z_{\mathrm{User}}}{\pi^2}
   \sum_{m=1}^M\mathsf{S}_m \sum_{m=1}^M\mathsf{T}_m
   \le\ \frac{z_{\mathrm{AP}}z_{\mathrm{User}}}{\pi^2}
\sum_{m=1}^M\mathsf{S}_m^{\mathrm{Upper}}\sum_{m=1}^M\mathsf{T}_m^{\mathrm{Upper}}
\end{equation}
\end{minipage}
\end{figure*}

Using Props. 3.1 to 3.10, we can obtain $\Gamma_{(\cdot)}(\phi_\mathsf{L})$ and $\Gamma_{(\cdot)}(\phi_\mathsf{U})$ considering different noises situations. Next, we first compute $\vert \sum_{m} A_{\mathrm{AP}\rightarrow m}A_{m \rightarrow \mathrm{User}}\vert^2$, then $\vert \sum_{m} A_{\mathrm{AP}\rightarrow m}\sqrt{\Gamma_{(\cdot)}}A_{m \rightarrow \mathrm{User}}\vert^2$ can be obtained easily. Note that $A_{\mathrm{AP}\rightarrow m}$ in \eqref{A_AP_m} and $A_{m \rightarrow \mathrm{User}}$ in \eqref{A_m_User} only contain non-negative components, we have \eqref{non-ideal_gain} in the next page. Based on \textit{CBS inequality} \cite{gradshteyn2014table}, we have \eqref{non-ideal_gain_v2} in the next page, where $\mathsf{S}_m = \big((x_m-x_{\mathrm{AP}})^2+(y_m-y_{\mathrm{AP}})^2+z^2_{\mathrm{AP}}\big)^{-\frac{3}{2}}$, $\mathsf{T}_m = \big((x_\mathrm{User}-x_m)^2+(y_\mathrm{User}-y_m)^2+z^2_{\mathrm{User}}\big)^{-\frac{3}{2}}$, \revise{$\mathsf{S}_m^{\mathrm{Upper}}$ and $\mathsf{T}_m^{\mathrm{Upper}}$ are defined as the right-hand sides of the first rows of \eqref{S^star_m} and \eqref{T^star_m}, respectively. The equality holds when $\sqrt{\mathsf{S}_m}$ and $\sqrt{\mathsf{T}_m}$ are proportional}. Note that the area of each pixel is $d_x d_y$, instead of an ideal point.  Therefore, by using \textit{Riemann sums}\cite{gradshteyn2014table}, we have
\revise{
\begin{align}
    \mathsf{S}_m^{\mathrm{Upper}}=&\int_{x_m-\frac{d_x}{2}}^{x_m+\frac{d_x}{2}}\int_{y_m-\frac{d_y}{2}}^{y_m+\frac{d_y}{2}}\mathsf{S}_m \partial x_m \partial y_m
    \notag \\
    =&\ \mathsf{Q}(t_2,t_4, z_{\mathrm{AP}}) - \mathsf{Q}(t_1,t_4, z_{\mathrm{AP}})
    \notag \\
    -&\  \mathsf{Q}(t_2,t_3, z_{\mathrm{AP}}) + \mathsf{Q}(t_1,t_3, z_{\mathrm{AP}}),
    \label{S^star_m}
\end{align}
}

\noindent where $\mathsf{Q}(s_1, s_2, z)$ is defined as

\begin{equation}
    \mathsf{Q}(s_1, s_2, z) = \frac{1}{z}\arctan\bigg(\frac{s_1 s_2}{z\sqrt{s_1^2+s_2^2+z^2}}\bigg),
\end{equation}

\noindent and $t_1=x_m-\frac{d_x}{2}-x_{\mathrm{AP}}$, $t_2=x_m+\frac{d_x}{2}-x_{\mathrm{AP}}$, $t_3=y_m-\frac{d_y}{2}-y_{\mathrm{AP}}$, and $t_4=y_m+\frac{d_y}{2}-y_{\mathrm{AP}}$. Similarly, we also have
\revise{
\begin{align}
    \mathsf{T}_m^{\mathrm{Upper}}=&\int_{x_m-\frac{d_x}{2}}^{x_m+\frac{d_x}{2}}\int_{y_m-\frac{d_y}{2}}^{y_m+\frac{d_y}{2}}\mathsf{T}_m \partial x_m \partial y_m
    \notag \\
    =&\ \mathsf{Q}(t_6,t_8, z_{\mathrm{User}}) - \mathsf{Q}(t_5,t_8, z_{\mathrm{User}})
    \notag \\
    -&\ \mathsf{Q}(t_6,t_7, z_{\mathrm{User}}) + \mathsf{Q}(t_5,t_7, z_{\mathrm{User}}),
    \label{T^star_m}
\end{align}}

\noindent where $t_5=x_m-\frac{d_x}{2}-x_{\mathrm{User}}$, $t_6=x_m+\frac{d_x}{2}-x_{\mathrm{User}}$, $t_7=y_m-\frac{d_y}{2}-y_{\mathrm{User}}$, and $t_8=y_m+\frac{d_y}{2}-y_{\mathrm{User}}$. For the proofs of \eqref{S^star_m} and \eqref{T^star_m}, please see Appendix \ref{Proof_E_star_F_star}. Consequently, we have

\begin{align}
    \bigg( \sum_{m=1}^M A_{\mathrm{AP}\rightarrow m}A_{m \rightarrow \mathrm{User}}\bigg)^2
   \leq
   \frac{z_{\mathrm{AP}}z_{\mathrm{User}}}{\pi^2}\sum_{m=1}^M\mathsf{S}_m^{\mathrm{Upper}} \sum_{m=1}^M\mathsf{T}_m^{\mathrm{Upper}},
\end{align}

\noindent and the equality holds when the pixel area $d_xd_y$ is small enough. Therefore, we have 
\revise{
\begin{equation}
\mathsf{L}^{\mathrm{Upper}} =  \Gamma_{(\cdot)}(\phi_\mathsf{L}) 
\frac{z_{\mathrm{AP}}z_{\mathrm{User}}}{\pi^2}\sum_{m=1}^M\mathsf{S}_m^{\mathrm{Upper}} \sum_{m=1}^M\mathsf{T}_m^{\mathrm{Upper}}
\label{L^star}
\end{equation}
}

\noindent and 

\begin{equation}
\mathsf{U}^{\mathrm{Upper}} =  \Gamma_{(\cdot)}(\phi_\mathsf{U})
\frac{z_{\mathrm{AP}}z_{\mathrm{User}}}{\pi^2}\sum_{m=1}^M\mathsf{S}_m^{\mathrm{Upper}} \sum_{m=1}^M\mathsf{T}_m^{\mathrm{Upper}}.
\label{U^star}
\end{equation}

\noindent Besides, using the results above, we can obtain $\mathsf{H}$ as

\begin{align}
\mathsf{H} \approx &  \ \bigg( \sum_{m=1}^M A_{\mathrm{AP}\rightarrow m}
\sqrt{\Gamma_{(\cdot)}(\phi_m)}
A_{m \rightarrow \mathrm{User}}\bigg)^2 
\notag \\
\leq &  \ \frac{z_{\mathrm{AP}}z_{\mathrm{User}}}{\pi^2}\sum_{m=1}^M\sqrt{\Gamma_{(\cdot)}(\phi_m)}\mathsf{S}_m^{\mathrm{Upper}} \sum_{m=1}^M\mathsf{T}_m^{\mathrm{Upper}} 
\notag \\
= & \ \mathsf{H}^{\mathrm{Upper}}.
\label{E{|h|^2}^star}
\end{align}

\noindent Finally, we have a theorem as follows.

\textbf{Theorem 1} \big(Lower and upper bounds of the SE\big): \textit{It can be obtained that}

\begin{align}
        \underbrace{\vphantom{\mathsf{SE}_\mathsf{L}^{\mathrm{Upper}}}\mathsf{SE}_\mathsf{L}\overset{(i)}{\le}\mathsf{SE}_{\mathsf{L}}^{\mathrm{Upper}}}_{\mathrm{Contains}\ \phi_\mathsf{L}}
        <       
        \underbrace{\vphantom{\mathsf{SE}_\mathsf{L}^{\mathrm{Upper}}}\mathsf{SE}\overset{(ii)}{\le}\mathsf{SE}^{\mathrm{Upper}}}_{\mathrm{Contains}\ \phi_m}
        <
\underbrace{\vphantom{\mathsf{SE}_\mathsf{L}^{\mathrm{Upper}}}\mathsf{SE}_{\mathsf{U}}\overset{(iii)}{\le}\mathsf{SE}_{\mathsf{U}}^{\mathrm{Upper}}}_{\mathrm{Contains}\ \phi_\mathsf{U}},
\label{theorem1}
\end{align}

\noindent \textit{where $\mathsf{SE}_\mathsf{L} = \log_2 \big(1+\frac{P_{\mathrm{AP}}\mathsf{L}}{\sigma^2}\big)$, $    \mathsf{SE}_{\mathsf{L}}^{\mathrm{Upper}} = \log_2 \big(1+\frac{P_{\mathrm{AP}}\mathsf{L}^{\mathrm{Upper}}}{\sigma^2}\big)$, $\mathsf{SE} \approx\log_2 \big(1+\frac{P_{\mathrm{AP}}\mathsf{H}}{\sigma^2}\big)$, $\mathsf{SE}^{\mathrm{Upper}} \approx\log_2 \big(1+\frac{P_{\mathrm{AP}}\mathsf{H}^{\mathrm{Upper}}}{\sigma^2}\big)$, $\mathsf{SE}_\mathsf{U} = \log_2 \big(1+\frac{P_{\mathrm{AP}}\mathsf{U}}{\sigma^2}\big)$, \textit{and} $\mathsf{SE}_{\mathsf{U}}^{\mathrm{Upper}} = \log_2 \big(1+\frac{P_{\mathrm{AP}}{\mathsf{U}^{\mathrm{Upper}}}}{\sigma^2}\big)$. Note that $(i)$, $(ii)$ and $(iii)$ are satisfied with equal measure when $d_xd_y$ is sufficiently small. Besides, $\mathsf{L}$, $\mathsf{L}^{\mathrm{Upper}}$, $\mathsf{U}$, $\mathsf{U}^{\mathrm{Upper}}$, $\mathsf{H}$, and $\mathsf{H}^{\mathrm{Upper}}$ are defined in \eqref{L}, \eqref{L^star}, \eqref{U}, \eqref{U^star}, and \eqref{E{|h|^2}^star} respectively.}

\textit{Proof}: Consider \eqref{L}, \eqref{L^star}, \eqref{U}, \eqref{U^star}, and \eqref{E{|h|^2}^star}, we have $\mathsf{SE}_\mathsf{L}\leq\mathsf{SE}_{\mathsf{L}}^{\mathrm{Upper}}$, $\mathsf{SE}_\mathsf{U}\leq\mathsf{SE}_{\mathsf{U}}^{\mathrm{Upper}}$ and $\mathsf{SE}\le\mathsf{SE}^{\mathrm{Upper}}$, where the equality holds when $d_x d_y$ is small enough. $\mathsf{SE}_{\mathsf{L}}^{\mathrm{Upper}}<\mathsf{SE}$ and $\mathsf{SE}^{\mathrm{Upper}}<\mathsf{SE}_{\mathsf{U}}$ are also obvious. \ \ \ \ \ \ \ \ \ \ \ \ \ \ \ \  \ \ \ \ \ \ \  \ \ \ \ \ \ \ \ $\blacksquare$

\section{Numerical Evaluations and Discussions}\label{Sec5}

In this section, we present numerical results to evaluate the AN results in Secs. \ref{Sec2}, \ref{Sec3}, and \ref{Sec4}. In particular, $D_\mathrm{AP}=[-20, 15, 8]^\mathsf{T}$ m, $D_\mathrm{User}=[20, 1.5, 8]^\mathsf{T}$ m, the center of RIS with $h_y=10$ m is $D_m=[0, 10, 0 ]^\mathsf{T}$. Besides, $a=1$, $b=0.2$, $c=0.43\pi$, $P_{\mathrm{AP}}=20$ dBm, $\sigma^2=-80$ dBm, and $f_c=2.4$ GHz, the pixel number\footnote{While $M=500$ is sufficient based on Sec.\ref{Sec3}, we choose a larger value to obtain more compelling MC results.} $M=200^2$, the MC realization number is $5000$. Lastly, to evaluate the most severe hardware impairment scenario, we perform system-level simulations using $\Gamma_{(3.10)}^\star$ as the RP expression. Three simulation scenarios are considered for validation. To cover a wider range of boundary conditions, the position coordinates of the RIS, the AP, and the user are allowed to take negative values.

\subsection{When the RIS is Fixed}

We first evaluate the system performance when the RIS is located in the middle of the AP and the user, i.e., $D_m=[0, 10, 0 ]^\mathsf{T}$.

\begin{figure}[htbp]
	\centering
	\begin{subfigure}{0.24\textwidth}
		\centering
		\includegraphics[width=\textwidth]{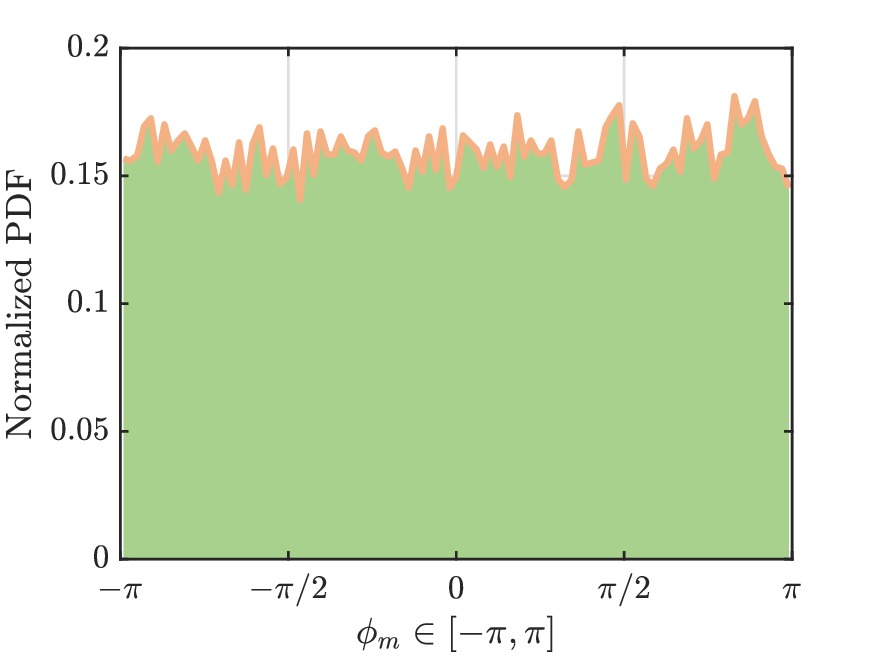}
		\caption{The empirical PDF of $\phi_m$.}
        \label{Fig_Fixed_phi_PDF}
	\end{subfigure}
	\centering
	\begin{subfigure}{0.24\textwidth}
		\centering
		\includegraphics[width=\textwidth]{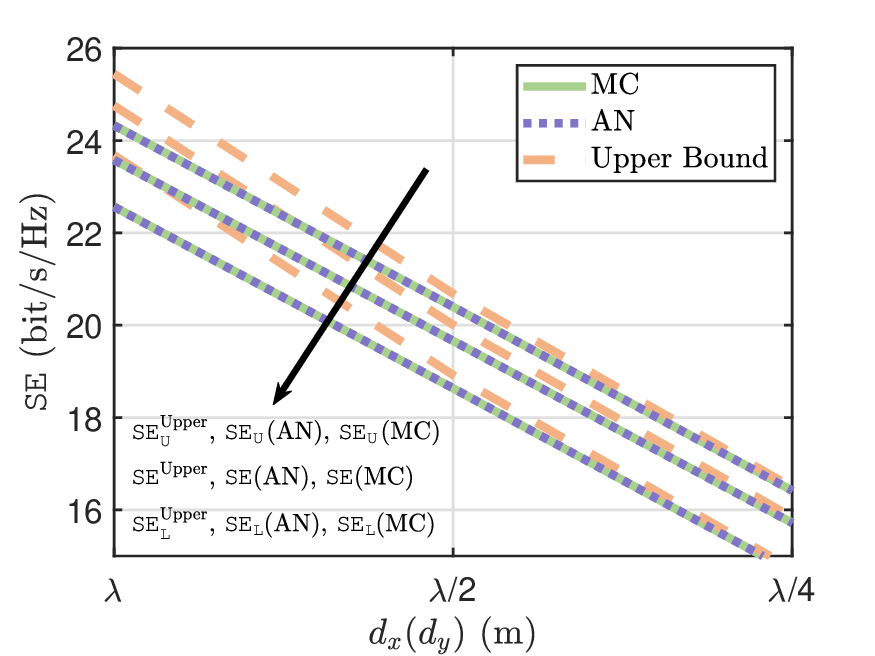}
		\caption{SE on different pixel sizes.}
		\label{Fig_Move_RIS_Diff_dxdy_U}
	\end{subfigure}
    	\centering
	\begin{subfigure}{0.24\textwidth}
		\centering
		\includegraphics[width=\textwidth]{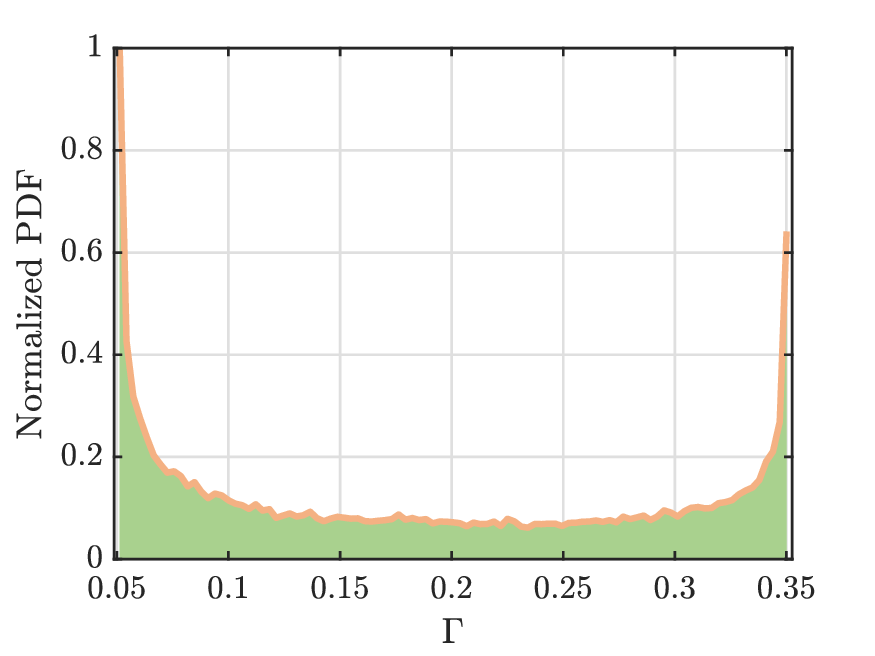}
		\caption{The empirical PDF of $\Gamma$.}
        \label{Fig_Fixed_Gamma_PDF}
	\end{subfigure}
	\centering
	\begin{subfigure}{0.24\textwidth}
		\centering
		\includegraphics[width=\textwidth]{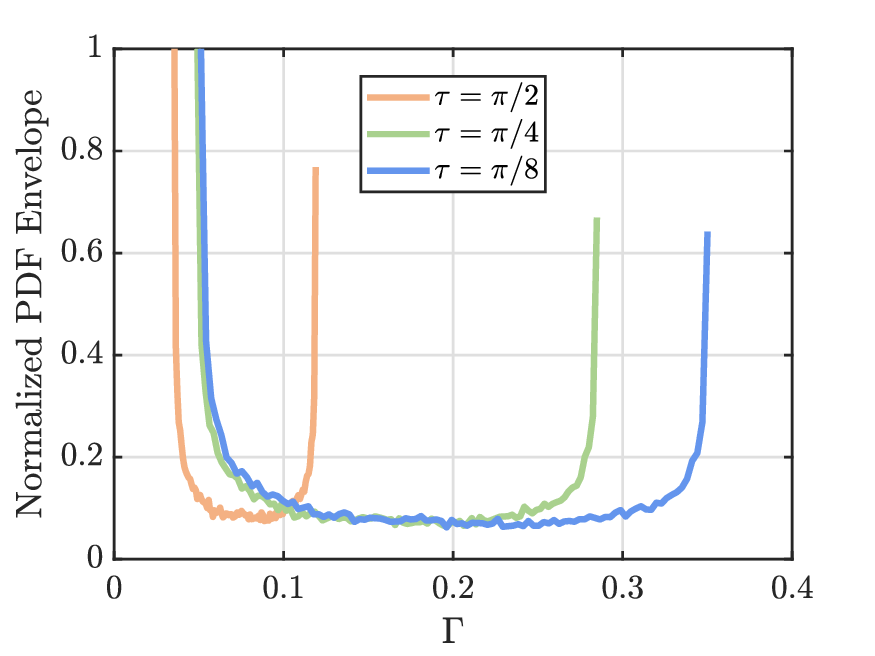}
		\caption{The PDF of $\Gamma$ for different $\tau$.}
		\label{Fig_Fixed_Gamma_PDF_diff_tau}
	\end{subfigure}
	\caption{\textcolor{myColor}{System performance when the RIS is fixed.}}
	\label{Fig_Fixed_PDF}
\end{figure}

\revise{From Fig. \ref{Fig_Fixed_phi_PDF}, it can be observed that the empirical PDF of $\phi_m$ distributed over $[-\pi,\pi]$, this aligns with our assumption in the last step of \eqref{P_RIS_User}. Fig. \ref{Fig_Move_RIS_Diff_dxdy_U} shows that decreasing the area of the pixel (i.e., $d_x d_y$) would diminish the SE. Besides, the upper bound of the SE would approach the simulated results, i.e., \eqref{theorem1} becomes $\mathsf{SE}_\mathsf{L}=\mathsf{SE}_\mathsf{L}^{\mathrm{Upper}}>\mathsf{SE}=\mathsf{SE}^{\mathrm{Upper}}>\mathsf{SE}_\mathsf{U}=\mathsf{SE}_\mathsf{U}^{\mathrm{Upper}}$. This is because the area and the center point of the pixel can be regarded as the same when $d_x=d_y\le \frac{\lambda}{4}$, and shows the correctness of Theorem 1. Although $\phi_m\sim\mathcal{UF}[-\pi, \pi]$, Fig. \ref{Fig_Fixed_Gamma_PDF} shows $\Gamma$ exhibits a U-shaped distribution. This is caused by the nonlinear feature of the PDA. Fig. \ref{Fig_Fixed_Gamma_PDF_diff_tau} illustrates the PDFs for different values of $\tau$, showing broader distributions and reduced peaks as $\tau$ increases, while a similar trend can be observed when $\kappa$ decreases.}

\begin{figure}[htbp]
	\centering
	\begin{subfigure}{0.24\textwidth}
		\centering
		\includegraphics[width=\textwidth]{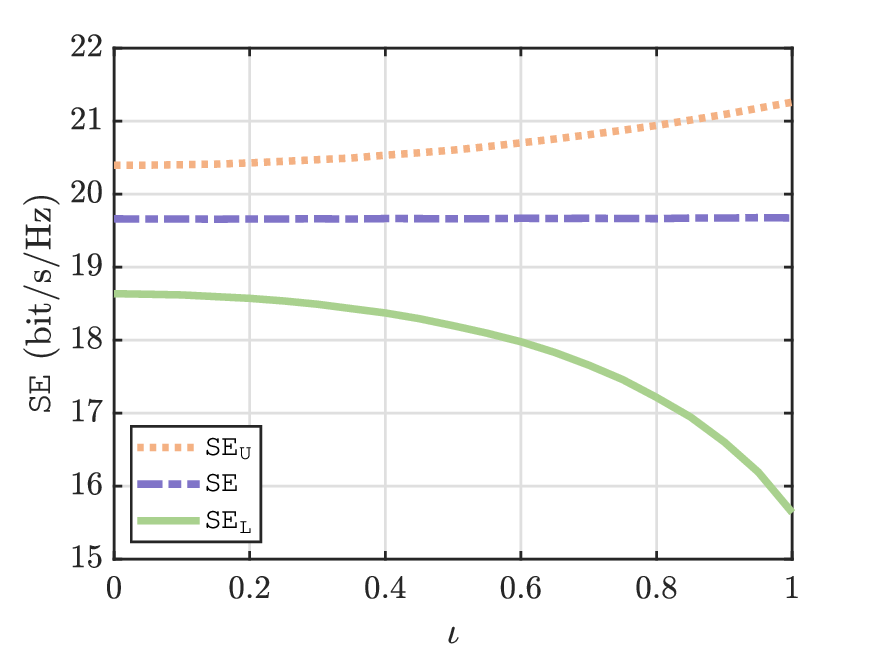}
		\caption{SE for different $\iota$.}
        \label{Fig_Fixed_iota_0to1_1012}
	\end{subfigure}
	\centering
	\begin{subfigure}{0.24\textwidth}
		\centering
		\includegraphics[width=\textwidth]{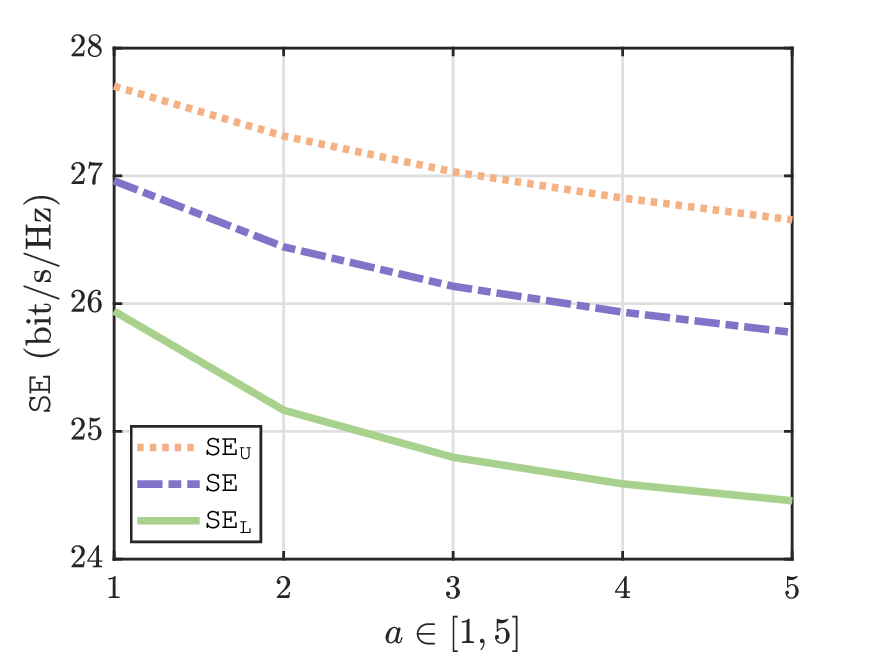}
		\caption{SE for different $a$.}
		\label{Fig_Fixed_vary_a_1012}
	\end{subfigure}
    	\centering
	\begin{subfigure}{0.24\textwidth}
		\centering
		\includegraphics[width=\textwidth]{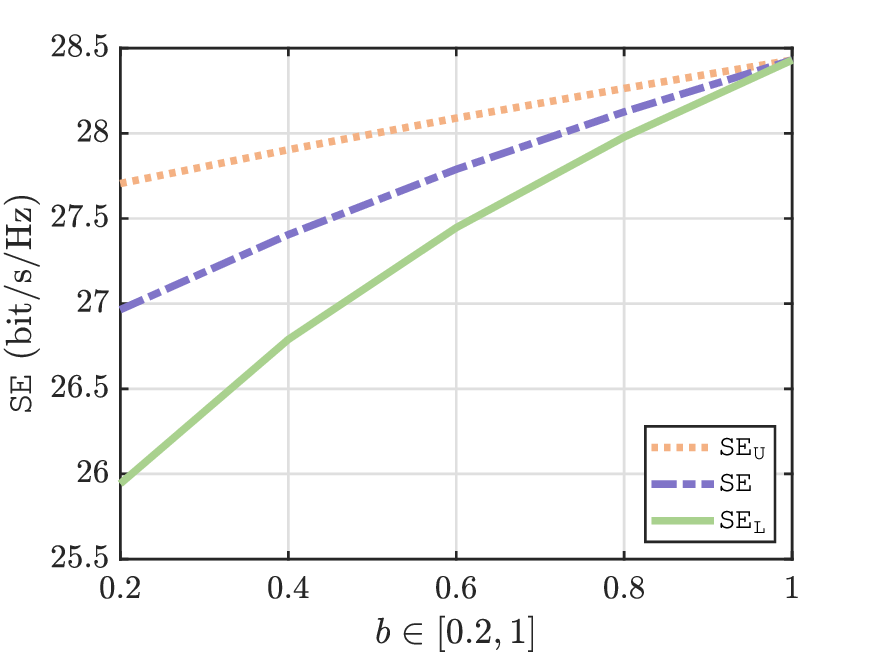}
		\caption{SE for different $b$.}
		\label{Fig_Fixed_vary_b_1012}
	\end{subfigure}
    	\centering
	\begin{subfigure}{0.24\textwidth}
		\centering
		\includegraphics[width=\textwidth]{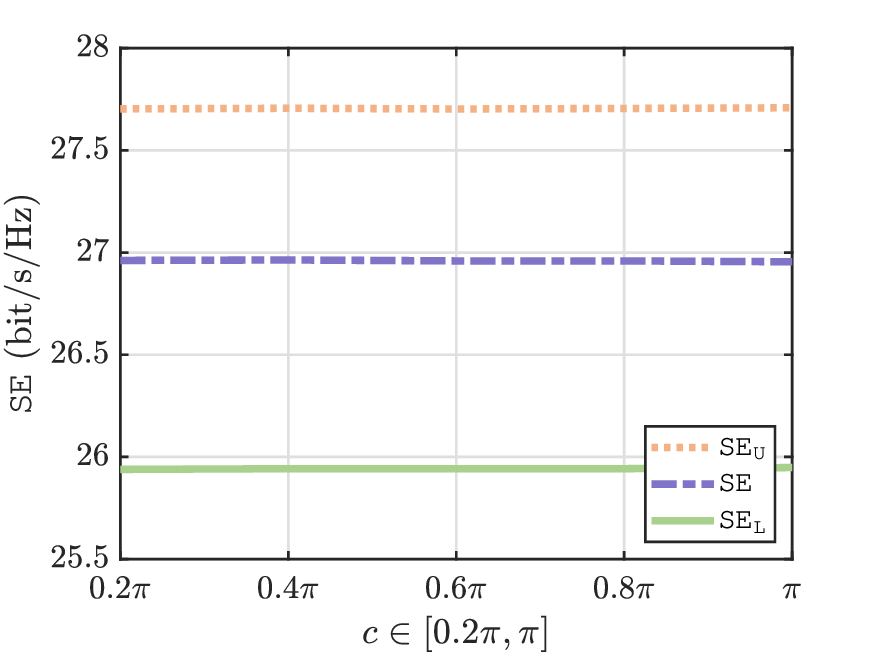}
		\caption{SE for different $c$.}
		\label{Fig_Fixed_vary_c_1012}
	\end{subfigure}
	\caption{\textcolor{myColor}{SE performance for different parameters.}}
	\label{Fig_Fixed_diff_a_b_c_iota}
\end{figure}

\revise{Fig. \ref{Fig_Fixed_diff_a_b_c_iota} illustrates the SE for different $\iota$, $a$, $b$, and $c$. As expected, when $\phi_m = \phi_{\mathsf{L}}$, the SE decreases as $\iota$ increases, and vice versa. This is consistent with Figs.~\ref{Fig_eq8_1004} and~\ref{Remark_iota_vs_Gamma}. Moreover, SE increases with smaller $a$ and decreases with larger $b$. In contrast, $c$ has negligible impact on SE because it only shifts the PDA curve. This result is in good agreement with Fig. \ref{Fig_Feasible_diff_c}.}

\begin{figure}[htbp]
	\centering
	\begin{subfigure}{0.24\textwidth}
		\centering
		\includegraphics[width=\textwidth]{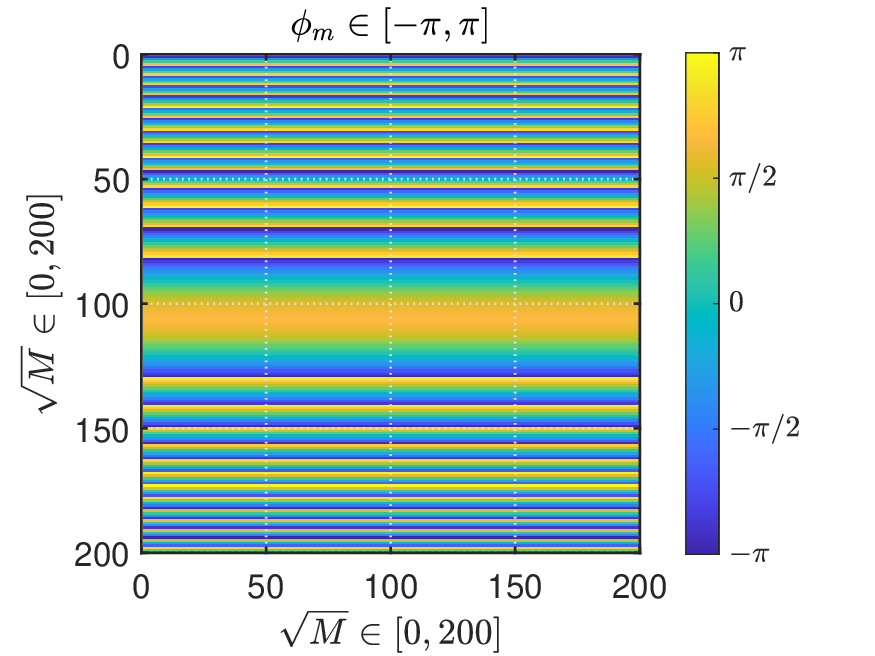}
		\caption{Heatmap of $\phi_m$.}
        \label{Fig_Fixed_phi_HeatMap_1012}
	\end{subfigure}
	\centering
	\begin{subfigure}{0.24\textwidth}
		\centering
		\includegraphics[width=\textwidth]{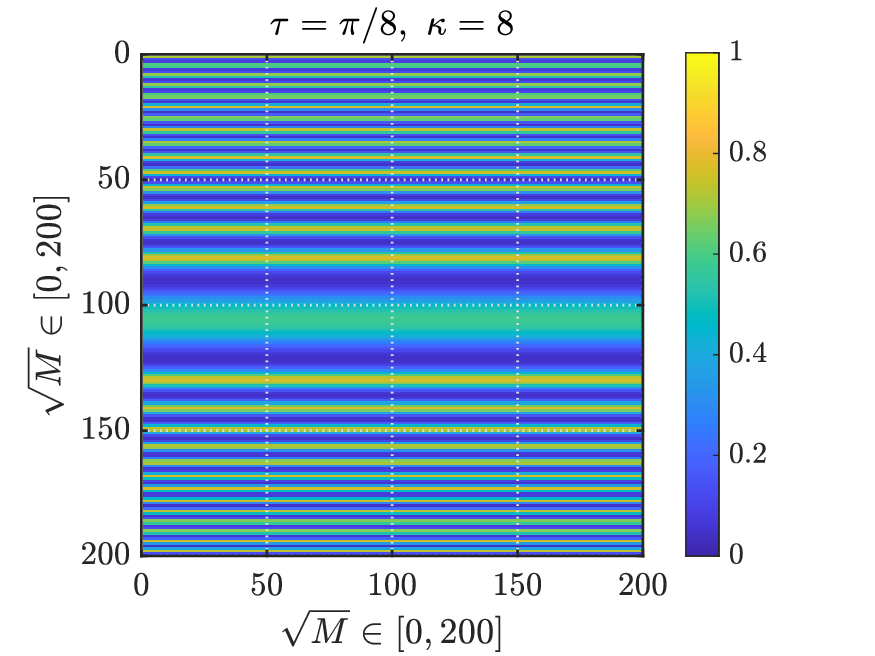}
		\caption{$\tau=\pi/8$ and $\kappa=8$.}
		\label{Fig_Fixed_Gamma_HeatMap_tau8kappa8_1012}
	\end{subfigure}
    	\centering
	\begin{subfigure}{0.24\textwidth}
		\centering
		\includegraphics[width=\textwidth]{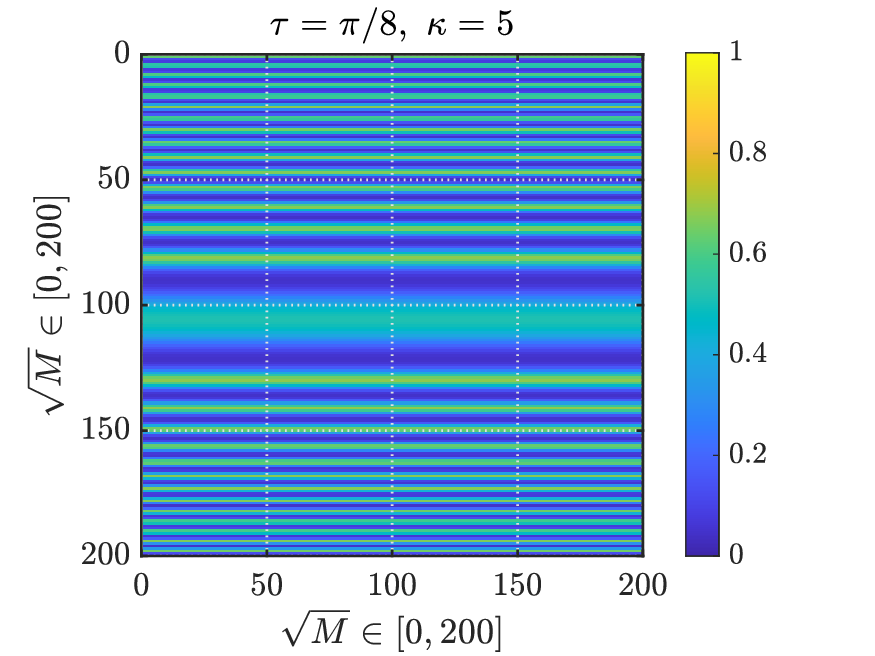}
		\caption{$\tau=\pi/8$ and $\kappa=5$.}
		\label{Fig_Fixed_Gamma_HeatMap_tau8kappa5_1012}
	\end{subfigure}
    	\centering
	\begin{subfigure}{0.24\textwidth}
		\centering
		\includegraphics[width=\textwidth]{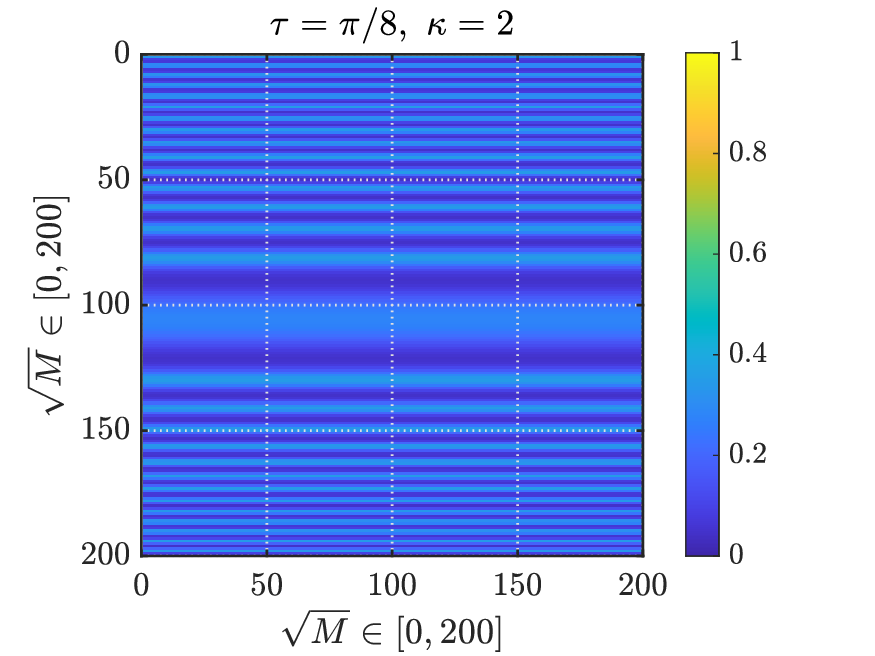}
		\caption{$\tau=\pi/8$ and $\kappa=2$.}
		\label{Fig_Fixed_Gamma_HeatMap_tau8kappa2_1012}
	\end{subfigure}
	\caption{\textcolor{myColor}{Heatmaps of $\phi$ and $\Gamma$ when the RIS is fixed.}}
	\label{Fig_Fixed_HeatMap}
\end{figure}

\revise{Fig. \ref{Fig_Fixed_HeatMap} visualizes $\phi_m$ and $\Gamma$ using heatmaps. With $\tau=\pi/8$ and $\kappa=2, 5, 8$, we can see that $\phi_m$ is approximately uniform over $[-\pi, \pi]$. Besides, as $\kappa$ decreases, phase concentration decreases and both the peak and span of high $\Gamma$ regions shrink, with weak sensitivity to $M$. These trends align with the SE improving as noise decreases. In addition, it is worth emphasizing that the pattern of the phase distribution is related to the positions of the RIS, AP, and user, since the RIS phases are designed based on location information rather than instantaneous channel state information.}

\subsection{When the RIS Moves along \texorpdfstring{$x$}{}-axis}

We further consider a slow movement of the RIS from a position near the AP toward the user. The RIS center is set to $[x_{\mathrm{RIS}}, 10, 0 ]^\mathsf{T}$ m with $x_{\mathrm{RIS}}\in[-8, 8]$ m, and the Doppler effect is neglected.

\begin{figure}[htbp]
	\centering
	\begin{subfigure}{0.24\textwidth}
		\centering
		\includegraphics[width=\textwidth]{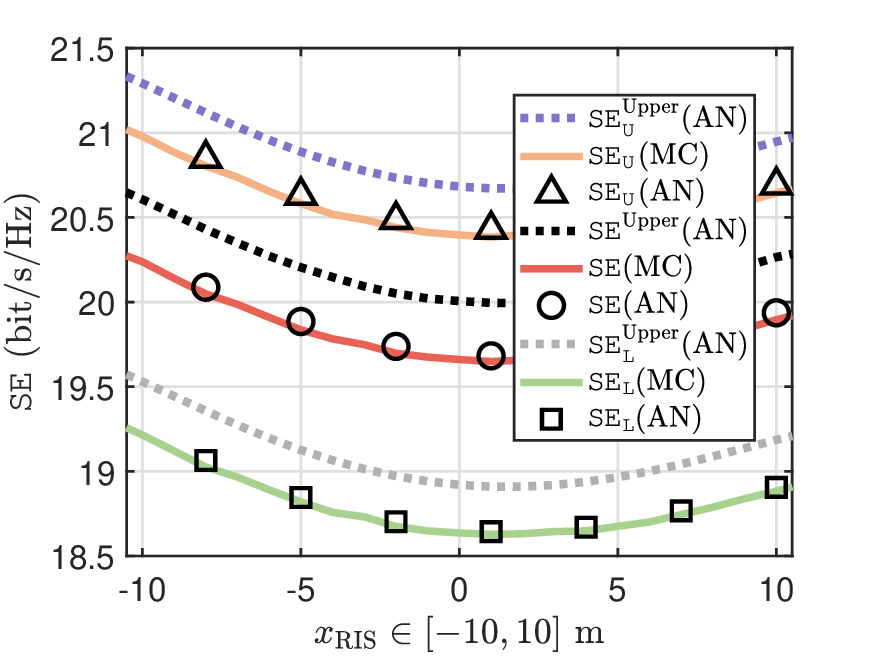}
		\caption{Validation of Theorem 1.}
        \label{Fig_Move_xAxis_Prop310_05lambda_1006}
	\end{subfigure}
	\centering
	\begin{subfigure}{0.24\textwidth}
		\centering
		\includegraphics[width=\textwidth]{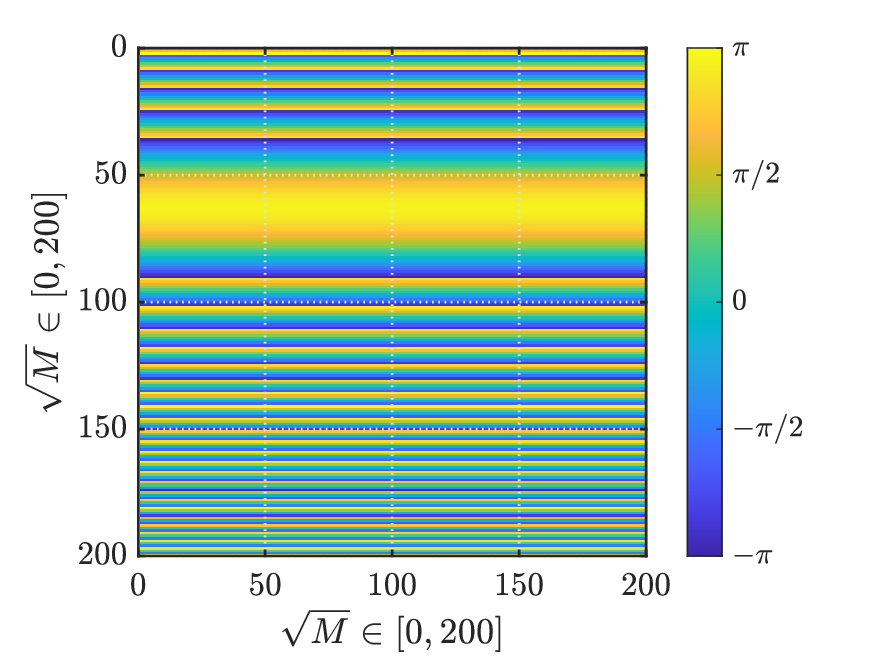}
		\caption{Heatmap of $\phi_m$, $x_\mathrm{RIS}=8$.}
		\label{Fig_HotMap_Phi_xRIS8_1013}
	\end{subfigure}
    	\centering
	\begin{subfigure}{0.24\textwidth}
		\centering
		\includegraphics[width=\textwidth]{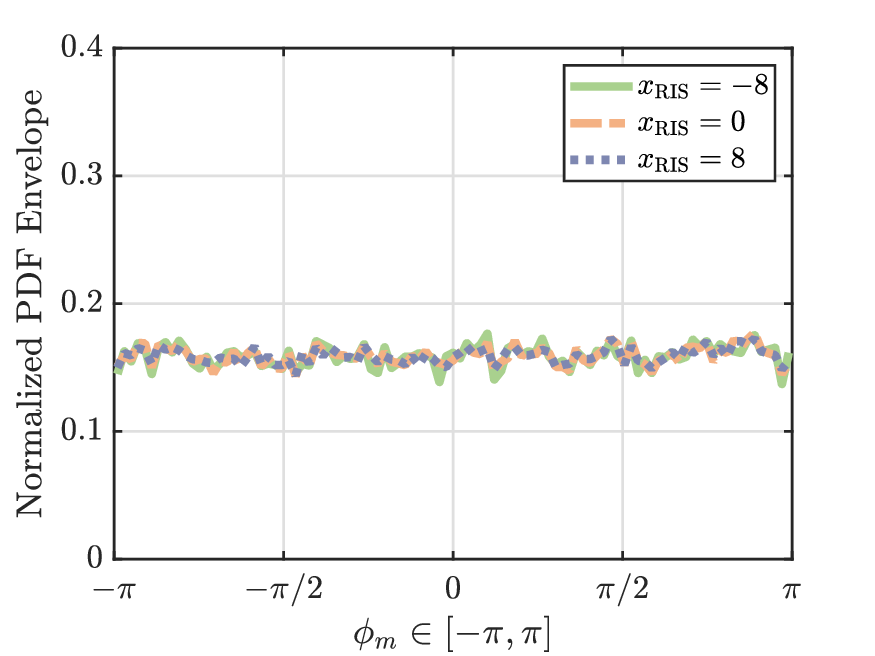}
		\caption{The empirical PDF of $\phi_m$.}
		\label{Fig_Move_PDF_phi_m_1012}
	\end{subfigure}
    	\centering
	\begin{subfigure}{0.24\textwidth}
		\centering
		\includegraphics[width=\textwidth]{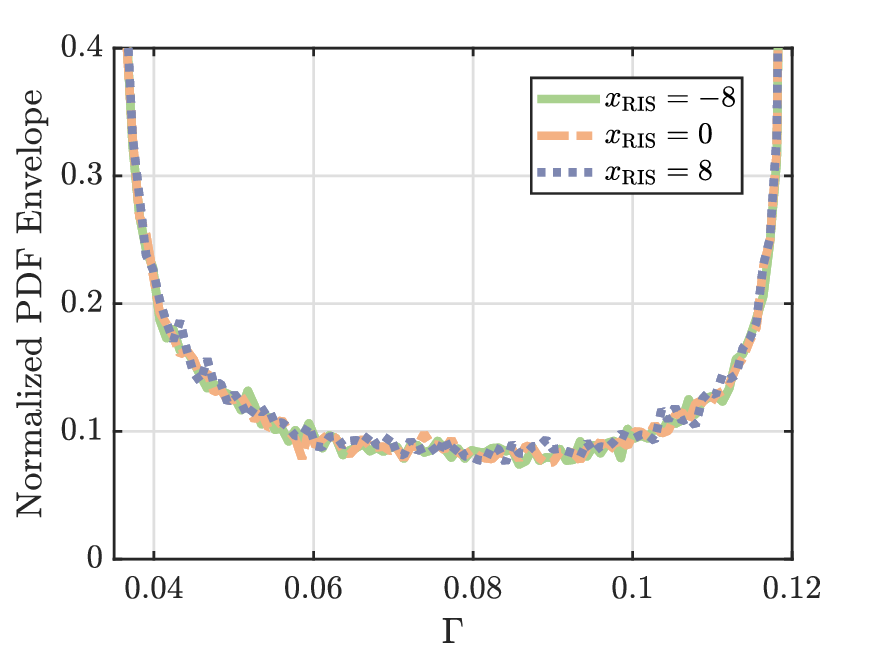}
		\caption{The empirical PDF of $\Gamma$.}
		\label{Fig_Move_PDF_Gamma_1012}
	\end{subfigure}
	\caption{\textcolor{myColor}{System performance for the RIS $x$-axis movement.}}
	\label{Fig_Move_x_axis}
\end{figure}

\revise{Fig. \ref{Fig_Move_xAxis_Prop310_05lambda_1006} verifies the correctness of Theorem 1 in Sec. \ref{Sec4}, as MC and AN results match well. Moreover, SE first decreases and then increases, as the RIS performs worst when it is deployed in the middle of the AP and the user\cite{emil20howlarge}. Fig. \ref{Fig_HotMap_Phi_xRIS8_1013} shows heatmap of $\phi_m$ when $x_\mathrm{RIS}=8$. Compared to Fig. \ref{Fig_Fixed_phi_HeatMap_1012} with $x_\mathrm{RIS}=0$, although the overall periodic pattern remains similar, slight displacements in the central region and phase gradient occur, indicating that the horizontal shift of the RIS introduces spatial variations in the phase distribution. Figs. \ref{Fig_Move_PDF_phi_m_1012} and \ref{Fig_Move_PDF_Gamma_1012} demonstrate that the statistical distributions of $\phi_m$ and $\Gamma$ are independent of the RIS's position, as the curves for different $x_{\mathrm{RIS}}$ are nearly identical.}

\subsection{When the RIS Moves along \texorpdfstring{$z$}{}-axis}

\begin{figure}[htbp]
	\centering
	\begin{subfigure}{0.24\textwidth}
		\centering
		\includegraphics[width=\textwidth]{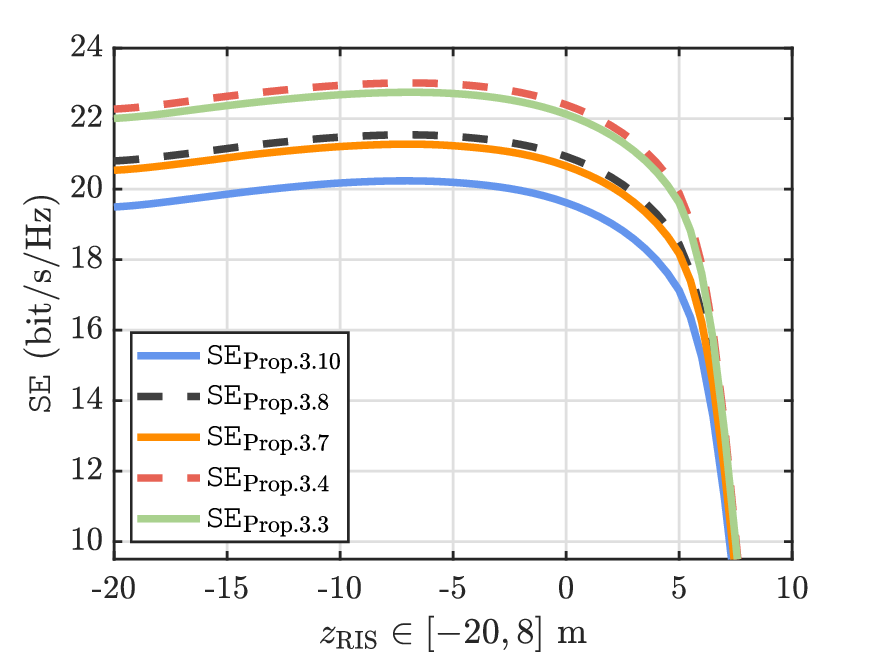}
		\caption{SE with different $\Gamma$.}
        \label{Fig_Move_zAxis_P31038373433_1012}
	\end{subfigure}
	\centering
	\begin{subfigure}{0.24\textwidth}
		\centering
		\includegraphics[width=\textwidth]{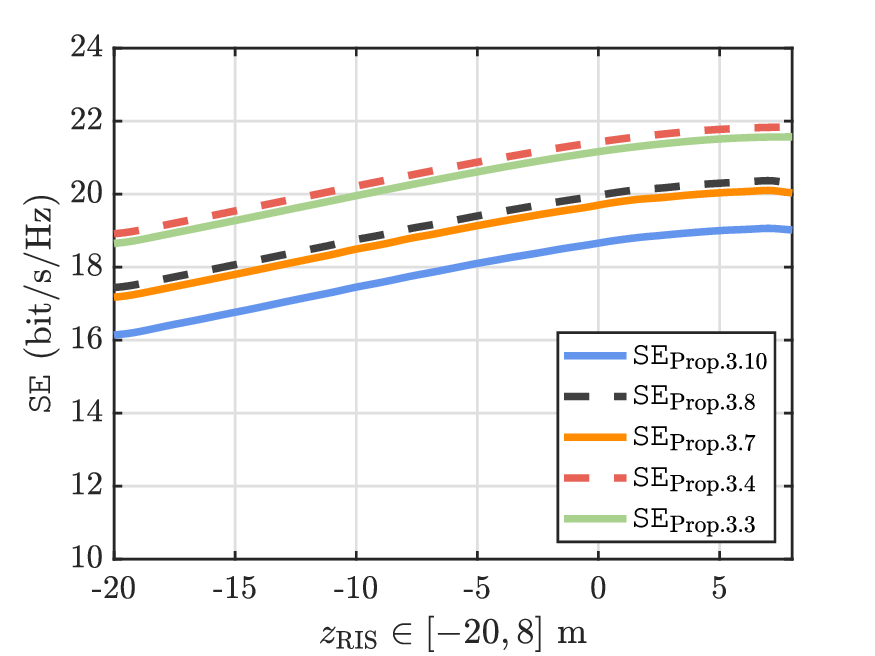}
		\caption{Without project aperture.}
		\label{Fig_Move_zAxis_P31038373433_1012_iso}
	\end{subfigure}
    \caption{\textcolor{myColor}{System performance for the RIS $z$-axis movement.}}
	\label{Fig_Move_z_axis}
\end{figure}

\revise{Next, we consider the RIS center is set to $[0, 10, z_{\mathrm{RIS}}]^\mathsf{T}$ m with $z_{\mathrm{RIS}}\in[-20, 10]$ m, and the Doppler effect is neglected. Fig.~\ref{Fig_Move_zAxis_P31038373433_1012} illustrates that as $z_{\mathrm{RIS}}$ increases, the SE first rises due to the reduced propagation distance, but then falls as the projected aperture diminishes to nearly zero. However, as shown in Fig.~\ref{Fig_Move_zAxis_P31038373433_1012_iso}, without the projected aperture, the SE is simply inversely proportional to the distance between the transmitter and receiver. Besides, $\mathsf{SE}_{\mathrm{Prop. 3.10}}$ is smallest, as expected.}

\subsection{Optimal Phase Feasible Set Comparison}

\begin{figure}[htbp]
	\begin{subfigure}{0.24\textwidth}
		\centering
		\includegraphics[width=\textwidth]{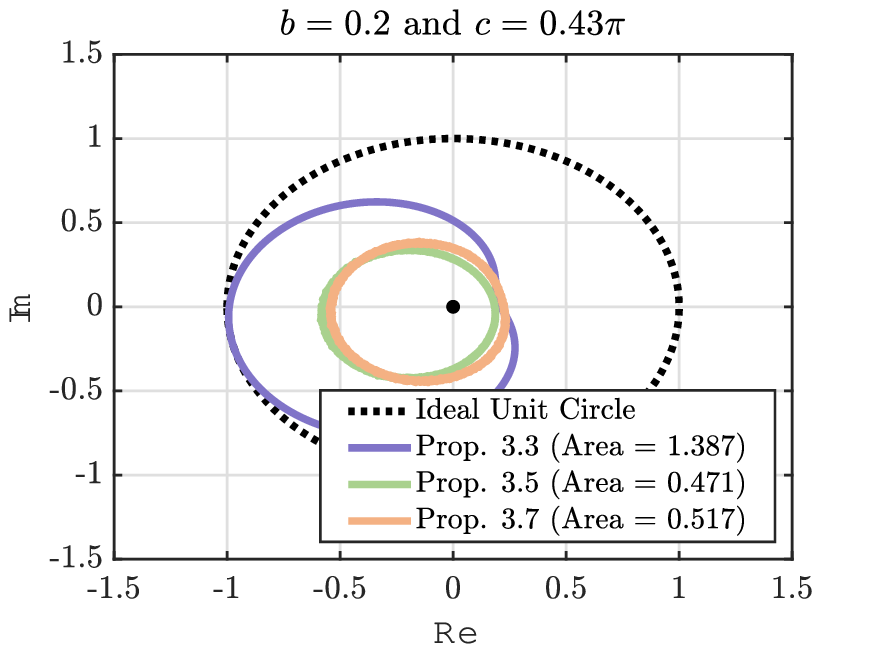}
		\caption{Props. 3.3, 3.5, and 3.7.}
        \label{Fig_Feasible_Prop333537_UF}
	\end{subfigure}
	\centering
	\begin{subfigure}{0.24\textwidth}
		\centering
		\includegraphics[width=\textwidth]{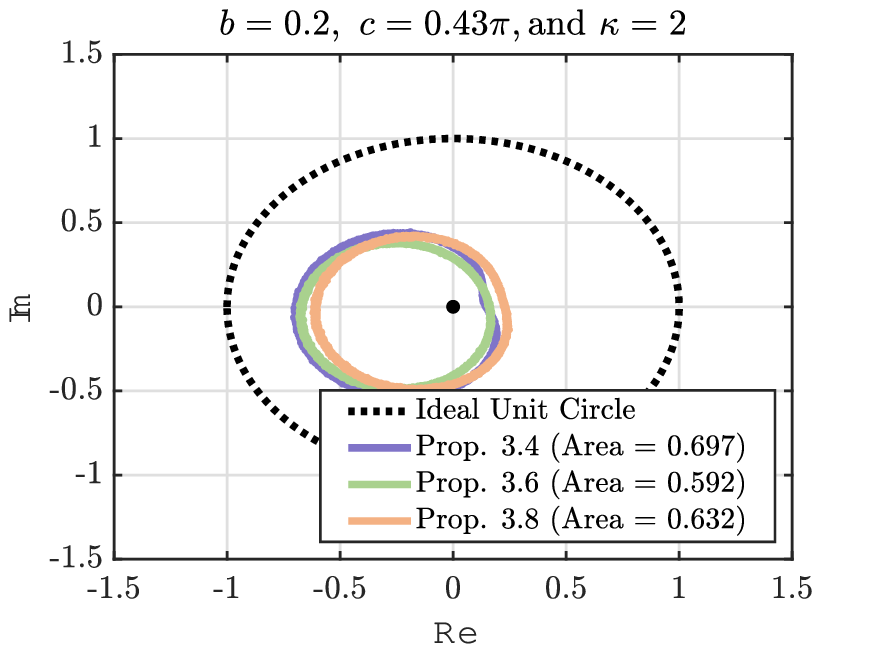}
		\caption{Props. 3.4, 3.6, and 3.8.}
		\label{Fig_Feasible_Prop343638_VM}
	\end{subfigure}
	\centering
	\begin{subfigure}{0.24\textwidth}
		\centering
		\includegraphics[width=\textwidth]{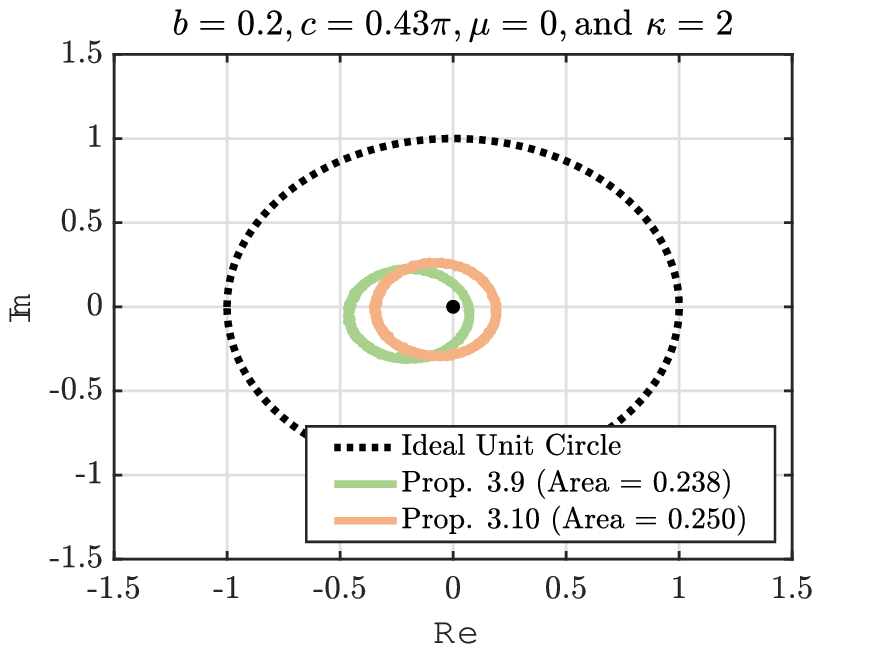}
		\caption{Props. 3.9 and 3.10.}
        \label{Fig_Feasible_Prop39310}
	\end{subfigure}
	\centering
	\begin{subfigure}{0.24\textwidth}
		\centering
		\includegraphics[width=\textwidth]{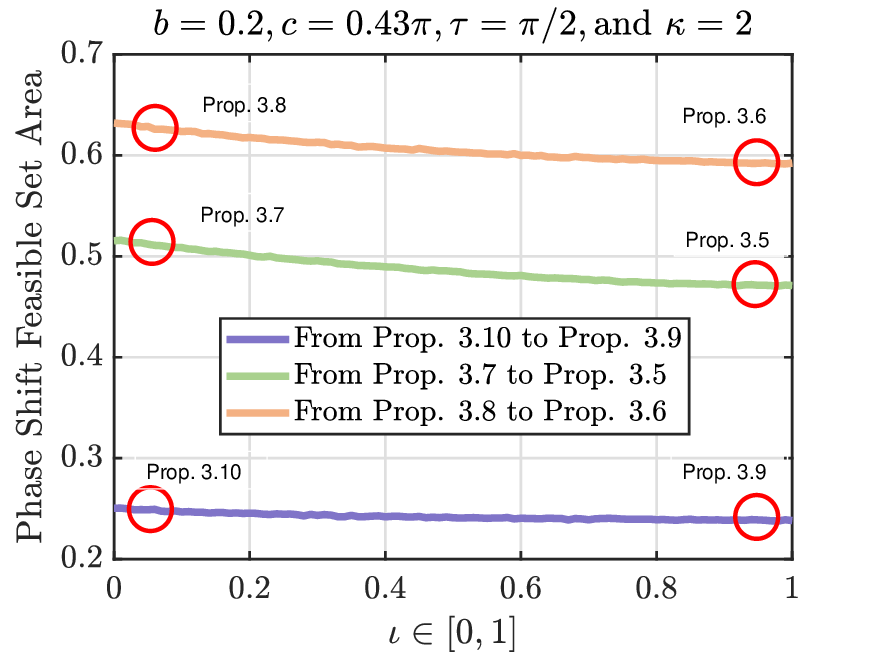}
		\caption{Feasible set areas vs. $\iota$.}
		\label{Fig_Feasible_iota01}
	\end{subfigure}
    \caption{\textcolor{myColor}{Feasible set and areas.}}
	\label{Fig_Feasible_Prop}
\end{figure}

\revise{Figs.~\ref{Fig_Feasible_Prop} show the feasible sets for designed RIS phases in the complex plane and their corresponding areas for the different propositions in Sec.~\ref{Sec3}. As expected, greater uncertainties lead to smaller feasible set areas. Moreover, the areas and $\iota$ are inversely related. This suggests that RIS designs should be optimized to minimize noise coupling. Specifically, during the initial deployment of the RIS (i.e., $\iota=1$), an optimization constraint can be formulated to ensure that the optimal phase remains close to the phase boundaries. However, after prolonged operation (i.e., $\iota=0$), this constraint should be reformulated to keep the optimal phase near the middle of the phase range.}

\section{Conclusions and Future Work}\label{Sec6}

\revise{This paper has investigated the impact of mutual coupling and hardware imperfections in the RIS-aided system. A practical reflection framework of the RIS pixel has been established by incorporating the PSE and the PDA, and four unified models have been proposed to characterize phase and amplitude distortions. A new metric, i.e., the RP, with asymptotic convergence has been derived to quantify reflection performance, providing a rigorous theoretical basis for system analysis. Moreover, a general NF LoS channel model has been developed, and analytical upper and lower bounds on the SE have been obtained. The results have demonstrated that the interplay between the PSE and the PDA significantly affects the RIS response.}

\revise{Future work will focus on RIS phase optimization considering the coupling between the PDA and PE. For active RISs, where the PDA is more complex \cite{panagiotis25}, it is also important to investigate imperfections in the relationship between the PDA and the PE of each active pixel. Another research direction is how to quantitatively characterize the phase and amplitude of an RIS pixel as it transitions from fully coupled to fully independent, which, based on known results, is closely related to hardware aging that occurs over time during operation\cite{ken23vtcF}. Furthermore, we will extend these studies to multiuser scenarios to analyze the impact of the coupling on beamforming and interference control.}

\appendices
\section{Proof of Proposition 3.2}\label{Proof_Prop_31_32}

\revise{We study $|\mathbb{E}\{\beta\exp{(-\jmath\gamma_m)}\}|^2$ where $\gamma_m\sim\mathcal{VM}(0,\kappa)$. Since $\mathbb{E}\{\beta\exp{(-\jmath\gamma_m)}\}=\beta\mathbb{E}\{\exp{(\jmath\gamma_m)}\}$, then we have}

\begin{align}
\mathbb{E}\{\exp{(\jmath\gamma_m)}\} =& \int_{-\pi}^{\pi}\exp{(\jmath\gamma_m)}f_{\kappa}(\gamma_m)\partial\gamma_m
    \notag \\
    = &\int_{-\pi}^{\pi}\exp{(\jmath\gamma_m)}\frac{\exp(\kappa\cos(\gamma_m))}{2\pi I_0(\kappa)}\partial\gamma_m
    \notag \\
    =&\frac{1}{2\pi I_0(\kappa)}\int_{-\pi}^{\pi}\exp(\jmath\gamma_m+\kappa\cos(\gamma_m))\partial\gamma_m
    \notag \\
    =&\frac{\frac{1}{2\pi}\int_{-\pi}^{\pi}\exp(\jmath\gamma_m+\kappa\cos(\gamma_m))\partial\gamma_m}{I_0(\kappa)}.
    \label{exp_jmath_gamma_m}
\end{align}

\noindent Note that $I_n(\kappa)=\jmath^{-n}J_n(\jmath\kappa)=\jmath^{-n}\frac{1}{2\pi}\int_{-\pi}^{\pi}\exp(\jmath n \gamma_m+\kappa\sin(\gamma_m))\partial\gamma_m$, \revise{where $J_n$ is the Bessel function of the first kind of order $n$.} Let $n=1$, then $I_1(\kappa)=\frac{1}{2\pi}\int_{-\pi}^{\pi}\exp(\jmath\gamma_m+\kappa\cos(\gamma_m))\partial\gamma_m$. Thus we have $\mathbb{E}\{\exp{(-\jmath\gamma_m)}\}=\mathbb{E}\{\exp{(\jmath\gamma_m)}\} ={I_1(\kappa)}/{I_0(\kappa)}$.
\noindent Since the power series expansion of $I_n(\kappa)$ is \cite{mardia2009directionalbook}
\revise{
\begin{equation}
I_n(\kappa)=\sum_{r=0}^{\infty}\frac{1}{r!\,(n+r)!}\left(\frac{\kappa}{2}\right)^{n+2r},
\end{equation}}

\noindent and after some manipulations, $\rho={I_1(\kappa)}/{I_0(\kappa)}$ can be obtained as \eqref{rho}. Besides, for small $\kappa$, let $\frac{\partial \rho^2}{\partial x}\ge 0$, we have $\kappa\in[-2.6, -1.81]\cup[0,1.81]$. For large $\kappa$, let $\frac{\partial \rho^2}{\partial x}\ge 0$, we have $\kappa\in(-\infty, 0)\cup[0.83, +\infty)$. For small $\kappa$, we have $\kappa\in[1,1.6]\subset[0,1.81]$; for large $\kappa$, we have $\kappa\in[1.6,+\infty)\subset[0.83, +\infty)$, thus we finish the proof.

\section{Proof of  \texorpdfstring{$\Gamma(0)|_{\iota=0}>\Gamma(0)|_{\iota=1}$}{} in Remark II}\label{Proof_Remark_II}

\revise{Let $\phi_m=0$, then we have $\Gamma(0)=|\mathbb{E}\{\beta(\bar{\Delta})\exp(-\jmath\Delta)\}|^2=(\mathbb{E}\{\beta(\bar{\Delta})\cos(\Delta)\})^2 + (\mathbb{E}\{\beta(\bar{\Delta})\sin(\Delta)\})^2$. Since $\mathcal{UF}$ and zero-mean $\mathcal{VM}$ distributions are all symmetry, $\mathbb{E}\{\sin(\Delta)\}=\mathbb{E}\{\sin(\bar{\Delta})\}=0$, and $\mathbb{E}\{\beta(\bar{\Delta})\sin(\Delta)\}\approx 0$. So we have $\Gamma(0) = (\mathbb{E}\{\beta(\bar{\Delta})\cos(\Delta)\})^2=(\mathsf{P}(\bar{\Delta},\Delta))^2$. Therefore, $\mathsf{P}(\bar{\Delta},\Delta) = \mathbb{E}\{\beta(\bar{\Delta})\cos(\Delta)\} =\mathbb{E}\{\beta(\bar{\Delta})\} \mathbb{E}\{\cos(\Delta)\}+\mathsf{Cov}(\beta(\bar{\Delta}), \cos(\Delta))$. Consequently, when $\iota=0$, $\bar{\Delta}$ and $\Delta$ are i.i.d., $\mathsf{Cov}(\beta(\bar{\Delta}), \cos(\Delta))=0$, so $\Gamma(0)|_{\iota=0} = (\mathsf{P}(\bar{\Delta},\Delta)|_{\iota=0})^2=(\mathbb{E}\{\beta(\bar{\Delta})\} \mathbb{E}\{\cos(\Delta)\})^2$. Similarly, when $\iota=1$, $\bar{\Delta}=\Delta$, $\Gamma(0)|_{\iota=1} = (\mathsf{P}(\bar{\Delta},\Delta)|_{\iota=1})^2=(\mathbb{E}\{\beta(\bar{\Delta})\} \mathbb{E}\{\cos(\Delta)\}+\mathsf{Cov}(\beta(\bar{\Delta}), \cos(\Delta)))^2$. Now, the target is to show $\mathsf{Cov}(\beta(\bar{\Delta}), \cos(\Delta))=\mathsf{Cov}(\beta(\Delta), \cos(\Delta))<0$.}

\revise{Consider Taylor expansion $\beta(\Delta) = \beta(0)+\frac{\partial\beta}{\partial\Delta}|_{\Delta=0}\Delta+\frac{1}{2}\frac{\partial^2\beta}{\partial\Delta^2}|_{\Delta=0}\Delta^2+\mathcal{O}(\Delta^3)$ and $\cos{(\Delta)}=1-\frac{1}{2}\Delta^2+\mathcal{O}(\Delta^4)$, we can obtain that $\mathsf{Cov}(\beta(\Delta), \cos(\Delta))\approx -\frac{1}{4}\frac{\partial^2\beta}{\partial\Delta^2}|_{\Delta=0}\mathsf{Var}(\Delta^2)$. Note that $\frac{\partial^2\beta}{\partial\Delta^2}|_{\Delta=0}=\frac{1-b}{2}\sin(c)>0$ when $c\in (0, \pi/2]$ and $b\in (0,1)$. Besides, $\mathsf{Var}(\Delta^2)>0$, we then finish the proof.}

\section{Proof of Proposition 3.9}\label{Proof_Prop_39}

Note that $\sin(\phi_m-c+\delta_m+\gamma_m)\exp(-\jmath\delta)\exp(-\jmath\gamma_m)=\sin(\phi_m-c)\cos(\delta_m)\cos(\gamma_m)\exp(-\jmath\delta_m)\exp(-\jmath\gamma_m)+\cos(\phi_m-c)\sin(\delta_m)\cos(\gamma_m)\exp(-\jmath\delta_m)\exp(-\jmath\gamma_m)+\cos(\phi_m-c)\cos(\delta_m)\sin(\gamma_m)\exp(-\jmath\delta_m)\exp(-\jmath\gamma_m)-\sin(\phi_m-c)\sin(\delta_m)\sin(\gamma_m)\exp(-\jmath\delta_m)\exp(-\jmath\gamma_m)$, thus $\mathbb{E}\{\sin(\phi_m-c+\delta_m+\gamma_m)\exp(-\jmath\delta_m)\exp(-\jmath\gamma_m)\}=\sin(\phi_m-c)\mathbb{E}\{\cos(\delta_m)\exp(-\jmath\delta_m)\}\mathbb{E}\{\cos(\gamma_m)\exp(-\jmath\gamma_m)\}+\cos(\phi_m-c)\mathbb{E}\{\sin(\delta_m)\exp(-\jmath\delta_m)\}\mathbb{E}\{\cos(\gamma_m)\exp(-\jmath\gamma_m)\}+\cos(\phi_m-c)\mathbb{E}\{\cos(\delta_m)\exp(-\jmath\delta_m)\}\mathbb{E}\{\sin(\gamma_m)\exp(-\jmath\gamma_m)\}-\sin(\phi_m-c)\mathbb{E}\{\sin(\delta_m)\exp(-\jmath\delta_m)\}\mathbb{E}\{\sin(\gamma_m)\exp(-\jmath\gamma_m)\}$. Considering $\mathbb{E}\{\cos(\delta_m)\exp(-\jmath\delta_m)\}=1 - \frac{\tau^2}{3} + \frac{\tau^4}{15} - \frac{2\tau^6}{315} + \mathcal{O}(\tau^8)$, $\mathbb{E}\{\cos(\gamma_m)\exp(-\jmath\gamma_m)\}= 1 - \frac{1}{2\kappa}-\frac{1}{8\kappa^2}-\frac{1}{16\kappa^3} + \mathcal{O}(\kappa^{-4})$, $\mathbb{E}\{\sin(\delta_m)\exp(-\jmath\delta_m)\}=-\jmath \left( \frac{\tau^2}{3} - \frac{\tau^4}{15} + \frac{2\tau^6}{315} + \mathcal{O}(\tau^8) \right)$, and $\mathbb{E}\{\sin(\gamma_m)\exp(-\jmath\gamma_m)\}=-\jmath \left( \frac{1}{2\kappa} + \frac{1}{8\kappa^2} + \frac{1}{16\kappa^3} + \mathcal{O}(\kappa^{-4}) \right)$, the proof is completed.

\section{Proof of Proposition 3.10}\label{Proof_Prop_310}

Note that $\sin(\phi_m-c+\bar{\delta}_m+\bar{\gamma}_m)=\big(\sin(\phi_m-c)\cos(\bar{\delta}_m)+\cos(\phi_m-c)\sin(\bar{\delta}_m)\big)\cos(\bar{\gamma}_m)+\big(\cos(\phi_m-c)\cos(\bar{\delta}_m)-\sin(\phi_m-c)\sin(\bar{\delta}_m)\big)\sin(\bar{\gamma}_m)$, $\mathbb{E}\{\sin(\bar{\gamma}_m)\}=\mathbb{E}\{\sin(\gamma_m)\}=\mathbb{E}\{\sin(\bar{\delta}_m)\}=\mathbb{E}\{\sin(\delta_m)\}=0$, $\mathbb{E}\{\cos(\bar{\gamma}_m)\}=\mathbb{E}\{\cos(\gamma_m)\}=\rho$, and $\mathbb{E}\{\cos(\bar{\delta}_m)\}=\mathbb{E}\{\cos(\delta_m)\}\approx 1-\frac{1}{6}\tau^2+\frac{1}{120}\tau^4$, then the proof is completed.

\section{Proofs of \texorpdfstring{\eqref{S^star_m}}{} and \texorpdfstring{\eqref{T^star_m}}{}}\label{Proof_E_star_F_star}

First, we have 

\begin{align}
    &\mathsf{S}_m^{\mathrm{Upper}}=\int_{x_m-\frac{\lambda}{4}}^{x_m+\frac{\lambda}{4}}\int_{y_m-\frac{\lambda}{4}}^{y_m+\frac{\lambda}{4}}\mathsf{S}_m \partial x_m \partial y_m
    \notag \\
    =&\int_{x_m-\frac{\lambda}{4}}^{x_m+\frac{\lambda}{4}}\int_{y_m-\frac{\lambda}{4}}^{y_m+\frac{\lambda}{4}}\frac{\partial x_m \partial y_m}{\big((x_m-x_{\mathrm{AP}})^2+(y_m-y_{\mathrm{AP}})^2+z^2_{\mathrm{AP}}\big)^{\frac{3}{2}}}.
\end{align}
\revise{
\noindent Let $x_m-x_{\mathrm{AP}}=\mathsf{v}_m$, $y_m-y_{\mathrm{AP}}=\mathsf{w}_m$, then we have
\begin{align}
    \mathsf{S}_m^{\mathrm{Upper}}=\ &\int_{x_m-\frac{\lambda}{4}}^{x_m+\frac{\lambda}{4}}\int_{y_m-\frac{\lambda}{4}}^{y_m+\frac{\lambda}{4}}\mathsf{S}_m \partial x_m \partial y_m
    \notag \\
    =\ &\int_{x_m-\frac{\lambda}{4}-x_{\mathrm{AP}}}^{x_m+\frac{\lambda}{4}-x_{\mathrm{AP}}}\int_{y_m-\frac{\lambda}{4}-y_{\mathrm{AP}}}^{y_m+\frac{\lambda}{4}-y_{\mathrm{AP}}}\frac{\partial \mathsf{v}_m \partial \mathsf{w}_m}{\big(\mathsf{v}_m^2+\mathsf{w}_m^2+z^2_{\mathrm{AP}}\big)^{\frac{3}{2}}}
    \notag \\
    \overset{(i)}=&\int_{t_1}^{t_2}
    \bigg[\frac{\mathsf{w}_m}{(\mathsf{v}_m^2+z^2_{\mathrm{AP}})\sqrt{\mathsf{v}_m^2+\mathsf{w}_m^2+z^2_{\mathrm{AP}}}}\bigg]_{t_3}^{t_4}\partial \mathsf{v}_m
    \notag \\
    \overset{(ii)}=& \mathsf{Q}(t_2,t_4, z_{\mathrm{AP}}) - \mathsf{Q}(t_1,t_4, z_{\mathrm{AP}})
    \notag \\
    -\ & \mathsf{Q}(t_2,t_3, z_{\mathrm{AP}}) + \mathsf{Q}(t_1,t_3, z_{\mathrm{AP}}),
\end{align}
\noindent where $\mathsf{Q}(s_1, s_2, z)$ is}

\begin{equation}
    \mathsf{Q}(s_1, s_2, z) = \frac{1}{z}\arctan\bigg(\frac{s_1 s_2}{z\sqrt{s_1^2+s_2^2+z^2}}\bigg),
\end{equation}

\noindent and $t_1=x_m-\frac{\lambda}{4}-x_{\mathrm{AP}}$, $t_2=x_m+\frac{\lambda}{4}-x_{\mathrm{AP}}$, $t_3=y_m-\frac{\lambda}{4}-y_{\mathrm{AP}}$, and $t_4=y_m+\frac{\lambda}{4}-y_{\mathrm{AP}}$. Besides, $(i)$ uses $\int\frac{\partial \mathsf{v}}{(\mathsf{v}^2+\mathsf{u})^{\frac{3}{2}}} = \frac{\mathsf{v}}{\mathsf{u}\sqrt{\mathsf{v}^2+\mathsf{u}}}$ \cite{gradshteyn2014table} and the constant term is ignored, $(ii)$ uses $\int\frac{\mathsf{u}\partial \mathsf{v}}{(\mathsf{v}^2+\mathsf{w})\sqrt{\mathsf{v}^2+\mathsf{u}+\mathsf{w}}}=\frac{1}{\sqrt{\mathsf{w}}}\mathrm{arctan}\big(\frac{\sqrt{\mathsf{u}}\mathsf{v}}{\sqrt{\mathsf{w}}\sqrt{\mathsf{v}^2+\mathsf{u}+\mathsf{w}}}\big)$ \cite{gradshteyn2014table} and the constant term is ignored. Thus \eqref{S^star_m} is obtained. Similarly,
let $t_5=x_m-\frac{\lambda}{4}-x_{\mathrm{User}}$, $t_6=x_m+\frac{\lambda}{4}-x_{\mathrm{User}}$, $t_7=y_m-\frac{\lambda}{4}-y_{\mathrm{User}}$, and $t_8=y_m+\frac{\lambda}{4}-y_{\mathrm{User}}$, then we have $\mathsf{T}^\mathrm{Upper}_m = \mathsf{Q}(t_6,t_8, z_{\mathrm{User}}) - \mathsf{Q}(t_5,t_8, z_{\mathrm{User}}) - \mathsf{Q}(t_6,t_7, z_{\mathrm{User}}) + \mathsf{Q}(t_5,t_7, z_{\mathrm{User}})$. Therefore, \eqref{T^star_m} is achieved, concluding the proof.

\bibliographystyle{./bibliography/IEEEtran}
\bibliography{./bibliography/IEEEabrv,./bibliography/ref}

\end{document}